%% file: dissertation.tex
\documentclass[dsingle]{Dissertate}

\usepackage{varioref, cleveref}

\crefname{pluralequation}{eqs.}{eqs.}
\Crefname{pluralequation}{Eqs.}{Eqs.}
\crefformat{pluralequation}{eqs.~(#2#1#3)}
\Crefformat{pluralequation}{Eqs.~(#2#1#3)}

\newcommand{\dsum}{\displaystyle\sum}

\renewcommand{\d}{\partial}

\newcommand{\overbar}[1]{\mkern 2mu\overline{\mkern-2mu#1\mkern-2mu}\mkern 2mu}
\newcommand{\C}{\mathbb{C}}
\newcommand{\R}{\mathbb{R}}

\newcommand{\Z}{\mathbb{Z}}

\newcommand\mperiod[1][\rlap]{#1{\ .}}	
\newcommand\mcomma[1][\rlap]{#1{\ ,}}

\crefname{table}{table}{tables}
\Crefname{table}{Table}{Tables}
\crefname{figure}{figure}{figures}
\Crefname{figure}{Figure}{Figures}

\newenvironment{eq}
    {\begin{equation}
    \begin{aligned}
    }
    { 
    \end{aligned}
    \end{equation}
    \ignorespacesafterend
    }

\usepackage{subcaption}

\begin{document}

\input{frontmatter/personalize}
\frontmatter
\setstretch{\dnormalspacing}


\include{chapters/introduction}
\include{chapters/chapter1}
\include{chapters/chapter2}
\include{chapters/conclusion}

\setstretch{\dnormalspacing}

\backmatter



\end{document}

%% file: frontmatter/personalize.tex
\title{Spacetime aspects of \\ non-supersymmetric strings}
\author{Salvatore Raucci}

\advisor{Augusto Sagnotti}




\degree{Doctor of Philosophy}
\field{Physics}
\degreeyear{2024}
\degreemonth{September}
\department{Classe di Scienze}



%% file: chapters/introduction.tex

\chapter{Introduction}
\label{introduction}

We do not understand why apples fall. If apples were BPS states, static and in stable equilibrium in a supersymmetric universe, string theory would be a compelling candidate to describe gravitational interactions in nature.

I intend to introduce the contents of this thesis using this provocative statement. Apples fall, and we currently lack a satisfactory understanding and a consistent mathematical framework for the mechanism by which they fall: gravity. Explicitly, the missing tile is the absence of spacetime supersymmetry in a quantum theory of gravity. 
For BPS apples, string theory would capture the physics at least in some regions of the moduli space, and the known dualities would complete the picture. 
Non-BPS apples in supersymmetric spacetimes would still be fine, provided that the corrections induced by their non-BPS nature are small and under control. The setting that remains shrouded in mystery is non-BPS apples in non-supersymmetric spacetimes, which is precisely the scenario presented by our universe.

The key point is that we understand and control string theory only in the presence of protecting principles that prevent wild quantum corrections from becoming relevant.
The known principles can be separated into two classes: topological or dynamical.
Supersymmetry\footnote{In this thesis, when I write \emph{supersymmetry} I mean \emph{local supersymmetry}. The overscrupulous reader can replace it with \emph{supergravity} throughout the text.} is the leading actor of the latter type of principle.
Unsurprisingly, as a dynamical statement, supersymmetry is typically synonymous with stability. Yet, in a theory that can approximate our universe, it cannot be the ultimate stability proxy.
After all, it is quantum mechanics that makes atoms stable.

In a quantum theory, quantum corrections are expected, which appears to suggest that understanding non-supersymmetric setups in string theory is most likely a technical problem. 
Perhaps surprisingly, in the next chapter, I will argue that the absence of spacetime supersymmetry has dramatic consequences for string theory, demanding a complete change of perspective.

This thesis is based on~\cite{Raucci:2022jgw,Raucci:2022bjw,Basile:2022ypo,Raucci:2023xgx,Mourad:2024dur,Mourad:2024mpg}, and focuses on specific non-supersymmetric string theory settings. These are the only three (known) ten-dimensional tachyon-free string theories, that are introduced in this chapter. I shall colloquially refer to these models as non-supersymmetric strings. 
Clearly, this is not the only way to study non-BPS apples in non-supersymmetric universes. String theory provides various options starting from the five spacetime supersymmetric models or from M theory. For instance, one could consider compactifications on manifolds or orbifolds that break all supersymmetries. While such choices might lead to more attractive effective field theories, possibly closer to the standard model of particle physics, non-supersymmetric strings are the most basic playground for understanding the gravitational effects of the absence of supersymmetry while remaining in the worldsheet regime of string theory. 
Indeed, they are the most basic perturbative string theory models in which one can address these issues, with the only requirements being consistency, criticality, and the absence of tachyons in the spectrum. The next step is to extrapolate the generic considerations of the spacetime consequences and apply them to more complicated settings, such as the compactifications of supersymmetric strings.

Recently, there has been growing interest in lower-dimensional string models without spacetime supersymmetry~\cite{Fraiman:2023cpa,DeFreitas:2024ztt,Baykara:2024tjr,Angelantonj:2024jtu}. The considerations of this thesis apply to these models without much difference; however, here I focus on the ten-dimensional cases for definiteness. The works~\cite{Raucci:2022jgw,Basile:2022ypo,Mourad:2024dur,Mourad:2024mpg} contain comments and explicit generalizations to $D$ spacetime dimensions, which apply to lower-dimensional examples with appropriate changes to the scalar potentials.

The structure of this thesis rests on two lines of research, after a rapid review of the worldsheet construction of the models. The first, which is the subject of \cref{chapter:tadpoles}, contains the paradigm shift that string theory requires without spacetime supersymmetry and the immediate consequences for the vacuum solutions of non-supersymmetric strings. The second, which is the subject of \cref{chapter:landscape}, is an attempt to understand the content of these string models both in terms of their landscape of vacua and in terms of deformations of brane-like solutions.

\section{Conformal invariance}

This section is not an attempt to review string theory. Rather, I focus on the single aspect that is most relevant to the discussion of non-supersymmetric strings: conformal invariance.

Conformal invariance is the defining feature of string theory. The sum over two-dimensional geometries, which characterizes the worldsheet formulation, can be understood only when conformal invariance holds.
Conformal invariance is the guiding principle for the spacetime backreaction of non-supersymmetric strings; therefore, its role is worthy of recalling. 

A possible way to characterize string theory is to view it as a two-dimensional quantum theory of gravity in a naive sense, with a schematic partition function
\begin{eq}
    \mathcal{Z} \sim \int \mathcal{D}[g]\mathcal{D}[\phi] e^{-S[g,\phi]}\mcomma
\end{eq}
where $\phi$ represents, for the moment, generic matter fields, and $g$ is the two-dimensional metric. Specifying what is meant by
\begin{eq}
    \mathcal{D}[g]
\end{eq}
is typically the main issue in this perspective. In two-dimensions, reparametrization invariance removes two degrees of freedom from the metric tensor, and one is left with a single scalar degree of freedom, for instance the conformal mode. However, the space of metrics is too broad, and it is even less clear whether $\mathcal{D}[g]$ should include contributions from Riemannian surfaces with different topological invariants.
In fact, string theory is not a two-dimensional quantum field theory of Riemannian surfaces. Rather, it is a two-dimensional quantum field theory of conformal classes of Riemanninan surfaces, or equivalently, a two-dimensional quantum field theory of Riemann surfaces.

The equivalence between the two characterizations rests on two interchangeable definitions of Riemann surfaces: complex manifolds with complex dimension one, or equivalence classes of oriented real two-dimensional Riemannian manifolds under the action of conformal transformations. Here, I follow the second perspective. 
In this language, the functional integral $\mathcal{D}[g]$ of string perturbation theory instructs us to sum over conformal structures of Riemannian surfaces, so that conformal invariance is part of the consistency requirements of the theory.
Using the mathematical equivalence to Riemann surfaces, one can then interpret the sum as a sum over Riemann surfaces, which afford a topological classification in terms of their genus (if the surfaces are oriented and without boundaries. I will comment later on the addition of boundaries and crosscaps).
At the practical level, Weyl invariance, 
\begin{eq}\label{eq:conformal_mode}
    g_{ab}\sim e^\rho g_{ab} \mcomma
\end{eq}
demands that the conformal mode of the two-dimensional metric be decoupled. The immediate consequence is that not any matter sector $\phi$ can be added to the two-dimensional action. It must be conformally invariant, hence a CFT.

The requirement of conformal invariance has a neat physical interpretation: it makes string theory an on-shell formulation of physics. This point will be particularly relevant later, for linking the two-dimensional picture with the emergence of spacetime.
To date, we lack an off-shell formulation of string theory that is under reasonable control, although string field theory has long been considered a viable candidate.
The geometric characterization of string theory, which I adopt in this thesis, is therefore necessarily incomplete. In addition to being only on-shell, it is also a first-quantized formulation, which can be easily understood by energy considerations. The energy of a constant-tension string is the product of the tension and length of the string. Therefore, strings with long spikes, which are the seeds for splitting, are suppressed.
Taking two-dimensional geometry and conformal invariance as our guides for the theory of gravity, we are bound to find limitations in terms of predictive power, particularly when discussing non-supersymmetric setups. 

At any rate, trusting the on-shell and first-quantized string theory defined through Riemann surfaces, no local degree of freedom is left in the two-dimensional metric, leaving $\mathcal{D}[g]$ as a placeholder for the sum over the conformal classes of Riemannian surfaces. 
The required gauge-fixing procedure for the metric introduces a reparametrization-ghost sector with a negative central charge.
Conformal invariance takes the lead again, and demands that the total central charge vanishes. This, with its simplicity and elegance, is the starting point of string theory as a consistent formulation of gravity.

\subsection{Spacetime geometry}

Spacetime emerges in the two-dimensional string theory of Riemann surfaces as the target space of the 2D matter CFT that is coupled to the worldsheet gravity.
In its two-dimensional formulation, the structure of string theory does not require the existence of a geometric spacetime: the matter CFT can be any abstract CFT. 
However, a simple CFT with a clear physical interpretation leads to a suitable geometric spacetime in its target space: free two-dimensional scalars that capture the position of the string in the target space.
For example, one can consider $n$ free real scalars with target space $\R^n$, which reconstruct the spacetime embedding of the string worldsheet. 
The impact of conformal invariance is already memorable in this most basic case: Weyl anomaly cancellation restricts the number of scalars and hence the number of spacetime dimensions to $n=26$. This number can be understood by considering the universal contribution of 2D reparametrization ghosts, which have central charge $c=-26$. Requiring a non-anomalous Weyl symmetry demands the total central charge
\begin{eq}
    c_{tot}=0\mcomma
\end{eq}
and since each free boson adds one unit of central charge one finally obtains the infamous $n=26$.
The importance of this simplest matter CFT coupled to two-dimensional gravity lies in its feature of containing a massless spin-two field in the spacetime spectrum, thus elevating string theory to a consistent theory of gravity in spacetime~\cite{Scherk:1974ca}. 
Yet, there is a major price to pay for simplicity: this theory is tachyonic, which signals that the free-scalar description is inadequate and that the true vacuum of the theory lies elsewhere, if it exists. I embarked on this discussion to emphasize that conformal invariance, albeit restrictive, is not sufficient to have a non-pathological string theory.

Other two-dimensional matter sectors can be added to build more interesting, possibly tachyon-free, models. The simplest option is to include free fermions, which leads to the RNS string, the starting point of superstrings. In this case, appropriate GSO projections~\cite{Gliozzi:1976jf,Gliozzi:1976qd} remove the tachyonic excitations of the free-scalar sector and stabilize the theory of gravity.
I shall comment more on these setups when introducing the three non-supersymmetric strings.
Here, I want to highlight alternative scenarios: WZW models, minimal models, or more general rational CFTs can be part of the worldsheet matter content and can lead to interesting possibilities that do not necessarily afford a geometric spacetime interpretation.
The only requirement is conformal invariance, which translates into the vanishing of the total central charge of the two-dimensional worldsheet CFT. 
Given this variety of options, a compelling question arises: can any string theory, possibly completely non-geometric, be connected to a geometric model through perturbative or non-perturbative steps? Recent works~\cite{McNamara:2019rup,McNamara:2022xkg,Aoufia:2024awo} point to a positive answer, which would elevate the role played by the simple free-boson sectors, although a complete understanding of the problem is still beyond our current knowledge. 

\subsection{Sigma models}
\label{sec:sigma_models}

Much of what we know about string theory comes from the cases of free bosons and free fermions as two-dimensional CFTs.
In their simplest formats, they describe the propagation of a string, together with its excitations, in a flat spacetime background. The generalization to curved spacetime backgrounds, which will be the main theme of this work, involves non-linear sigma models as two-dimensional CFTs, which are the geometric avatars of interacting CFTs. The most basic sigma model, which captures nothing but the motion of a string in a curved target space, is given by the Polyakov action
\begin{eq} \label{eq:Polyakov_action}
    S=\frac{1}{4\pi\alpha'}\int d^2 \sigma \sqrt{g}  \,  g^{ab} G_{\mu\nu}(X)  \d_a X^\mu \d_b X^\nu\mcomma
\end{eq}
where $g_{ab}$ is the worldsheet metric, $G_{\mu\nu}$ is the target space metric, $X^\mu$ are the scalar fields, and $l_s=\sqrt{\alpha'}$ is the only dimensionful parameter in string theory, the string length.
The spacetime metric $G_{\mu\nu}$ is built by a coherent state of the spin-two excitations of the string, so that the string modes generate the spacetime background. This is the background independence of string theory.
Similar considerations apply to the universal Kalb-Ramond field $B$ and the dilaton $\phi$, although they are not discussed here.

Not all sigma models are good worldsheet candidates: they must be CFTs.
Here, the role of conformal invariance is even more significant because it is the heart of the spacetime geometric description. The sigma model in \cref{eq:Polyakov_action} is classically invariant under conformal transformations of the two-dimensional metric. Quantum conformal invariance, namely the vanishing of Weyl anomalies, translates into the vanishing of the beta functional:
\begin{eq}
    \mu \frac{\d G_{\mu\nu}}{\d \mu} = 0 \mperiod
\end{eq}
Computing the beta functional at one loop leads to the far-reaching result
\begin{eq}
    \alpha' R_{\mu\nu}=0\mcomma
\end{eq}
where $R_{\mu\nu}$ is the Ricci tensor of the spacetime metric $G_{\mu\nu}$. Weyl invariance becomes equations of motion for spacetime fields~\cite{Callan:1985ia}, which again signals that string theory is an on-shell formulation of physics. One can add the Kalb-Ramond field and the dilaton in a similar fashion to obtain the equations of motion for $G_{\mu\nu}$, $B_{\mu\nu}$ and $\phi$ in a worldsheet-loop expansion, which is weighted by 
\begin{eq}
    \frac{\alpha'}{r^2}\mcomma
\end{eq}
where $r$ is the typical spacetime curvature radius.
This is how spacetime QFTs and the geometrical nature of gravity emerge from the two-dimensional worldsheet.

A fundamental feature of string theory is the double expansion in $g_s$ and $\alpha'$, as measures of the string perturbative expansion and the worldsheet-loop renormalization.
A comment is in order on the interpretation of this double expansion. The most naive viewpoint is to work order by order in string perturbation theory, which requires conformal invariance for each worldsheet topology. In this way, Weyl anomalies must separately vanish on each Riemann surface, and the $\alpha'$ expansion is performed at any fixed $g_s$ order.
As I explain in \cref{sec:FS_mechanism}, this is not true in the absence of protecting principles, and in particular, without spacetime supersymmetry. A more involved type of Weyl anomaly cancellation is the key to understand the physics of such cases.

In view of the above considerations, it is difficult to underestimate the relevance of sigma models. The latter are the true heralds of spacetime geometry and Einstein gravity in string theory.
Again, the cases that are under control are only small drops in the ocean of possibilities. More exotic interacting sigma models may hide phenomenologically attractive corners, such as de Sitter spacetimes~\cite{Polyakov:2007mm}.

Before moving to the explicit constructions of the relevant superstring theories, I want to stress a point that is not usually emphasized.
The coupling to two-dimensional quantum gravity in string perturbation theory, namely the sum over Riemann surfaces, requires matter CFTs to exist on all such surfaces. Typically, the focus is on the sphere topology, which is the first contribution to the string perturbative expansion; however, the existence requirement on all topologies is a consistency prerequisite for the formulation.
It turns out~\cite{Moore:1988qv} that crossing symmetry and modular invariance are sufficient conditions to guarantee the existence of a CFT on any Riemann surface. 
The reason is that any Riemann surface can be obtained, in a non-unique way, by sewing three-holed spheres, and consistency amounts to the independence on the chosen sewing. A consequence of~\cite{Moore:1988qv} is that consistency of a CFT on both the sphere and the torus implies consistency on all Riemann surfaces.

\section{Vacuum amplitudes}

The aim of this and the following section is to introduce the non-supersymmetric string theories of interest. 
The intent is not to be exhaustive but to provide some context on the general constructions of string theories. The interested reader can consult the reviews~\cite{Angelantonj:2002ct,Mourad:2017rrl,Basile:2020xwi,Basile:2021mkd,Basile:2021vxh}.

The worldsheet formulation is a geometrical generalization of the Feynman amplitudes for point particles, where singular corner points are smoothed out by the two-dimensional geometry and are recovered in particular limits in the moduli spaces of Riemann surfaces. A single contribution from each order in string perturbation theory enters the scattering amplitudes for closed oriented strings~\cite{Polyakov:1981rd,Polyakov:1981re} (or a few contributions for open and unoriented strings), 
\begin{eq}
    \mathcal{A}=\sum_{n=0}^{\infty} g_s^{2(n-1)} \mathcal{A}_n \mcomma
\end{eq}
where $\mathcal{A}_n$ are computed by integrals over moduli spaces of genus-$n$ Riemann surfaces.
A special amplitude stands out for its physical relevance: the one-loop vacuum amplitude.
In quantum field theory, it encodes the masses of the finite number of fields that constitute the model, and is fully determined by the free spectrum. Likewise, in string theory, since the spectrum contains infinitely many excitations, one-loop vacuum amplitudes are efficient tools for determining the complete perturbative spectrum, at least if the string is quantized in its true vacuum.
It is instructive to compare the geometric interpretations of the two cases of a point particle and a string to understand how the worldsheet generalizes the Feynman amplitudes. 

For a point particle with worldline (Schwinger) parameter $t$, the one-loop vacuum amplitude for a free scalar is given by
\begin{eq}
    \frac{1}{2}\int_0^\infty \frac{dt}{t} \Tr e^{- t H} = -\frac{1}{2}\log\det\Delta\mcomma
\end{eq}
where the measure contains a combinatorial factor $\frac{1}{t}$ that comes from the action of the isometry group of the circle, a factor $\frac{1}{2}$ that takes into account the two possible orientations, and where the inverse propagator $\Delta$ is 
\begin{eq}\label{eq:propagator}
    \frac{1}{\Delta}=\int_0^\infty dt e^{- t H} \mperiod
\end{eq}
In general, considering the bosonic and fermionic nature of fields and performing the Gaussian momentum integrals, the one-loop vacuum amplitude in $D$ spacetime dimensions takes the compact form
\begin{eq} \label{eq:1_loop_amplitude_particle}
    \Gamma=-\frac{1}{2(4\pi)^{\frac{D}{2}}}\int_0^\infty \frac{dt}{t^{1+\frac{D}{2}}} \text{Str}\left(e^{- t M^2}\right)\mcomma
\end{eq}
where one must insert a UV cutoff and cut the integration at some $\epsilon>0$.

For closed and oriented strings one-loop translates into genus-one, and in fact, a circle can be seen as a torus with a shrunk cycle, which physically would be the spatial length of the string shrinking to a point.
The expression that replaces \cref{eq:1_loop_amplitude_particle} is (note that the signs of $\Gamma$ and $\mathcal{T}$ are opposite)
\begin{eq} \label{eq:1_loop_amplitude_string}
    \Gamma \coloneq - \frac{\mathcal{T}}{2 (4\pi^2\alpha')^{\frac{D}{2}}}  = - \frac{1}{2 (4\pi^2\alpha')^{\frac{D}{2}}}\int_{\mathcal{F}} \frac{d^2\tau}{\tau_2^{1+\frac{D}{2}}} \Tr q^{L_0-\frac{c}{24}}\bar{q}^{\bar{L}_0-\frac{\bar{c}}{24}}\mcomma
\end{eq}
where the Virasoro operators $L_0$ and $\bar{L}_0$ minus the vacuum Casimir energies (central charges $c$ and $\bar{c}$) replace the mass, $\tau=\tau_1+ i \tau_2$ is the modular parameter of the torus, $q=e^{2\pi i \tau}$, and the integral is over a single fundamental domain $\mathcal{F}$.
The combinatorial factors in this amplitude are dictated by the SL(2,$\Z$) modular invariance of the torus; in particular, the measure is the modular invariant combination 
\begin{eq}
    \frac{d^2\tau}{\tau_2^2} \mcomma
\end{eq}
where $\tau_2$ replaces the Schwinger parameter $t$ of \cref{eq:1_loop_amplitude_particle}. The SL(2,$\Z$) identification is demanded by the two-dimensional large-diffeomorphism invariance, which would otherwise result in a global worldsheet gravitational anomaly~\cite{Witten:1985mj}, thus making the two-dimensional formulation inconsistent. This is a consistency requirement of the initial definition of the theory.
It is striking that it carries perhaps the most important physical consequence of string theory, implementing the intuitive idea that the corner points of Feynman amplitudes are smoothed out: the singular UV $t\to0$ point of \cref{eq:1_loop_amplitude_particle} does not belong to the integration domain. 

Relaxing the string-derived restriction of integrating \cref{eq:1_loop_amplitude_string} over a single fundamental domain and performing the $\tau_1$ integral over the full range, from $\tau_1=-\frac{1}{2}$ to $\tau_1=\frac{1}{2}$, implements level matching; thus, after performing the $\tau_1$ integral, one is left with the sum over the physical states:
\begin{eq}\label{eq:level_matched_1_loop_amplitude}
    \mathcal{T}\propto \sum_{\text{phys}} (-1)^F e^{-\pi\alpha' m^2 \tau_2} \mcomma
\end{eq}
where $F$ is the spacetime fermion number.
\Cref{eq:level_matched_1_loop_amplitude} reduces to the quantum field theory supertrace, counting the degeneracy of states with a given mass, albeit with infinitely many states, so that the analogy between the one-loop vacuum amplitude of particles and that of strings is now complete.

All closed-string theories that are relevant for this discussion derive from two-dimensional SCFTs of free bosons and fermions, which are therefore the building blocks of one-loop vacuum amplitudes.
Free bosons, corresponding to the worldsheet-periodic scalar fields denoting the position of the string in the target space, lead to copies of the contribution of a single scalar, which is proportional to
\begin{eq}
    \frac{1}{\sqrt{\alpha' \tau_2}} \frac{1}{|\eta(\tau)|^2}\mcomma
\end{eq}
where $\eta$ is the Dedekind $\eta$ function
\begin{eq}
    \eta(\tau)=q^{\frac{1}{24}}\prod_{n=1}^\infty \left(1-q^n\right)\mperiod
\end{eq}
On the other hand, for free fermions, one has the freedom to choose periodic and antiperiodic boundary conditions around the two cycles of the torus, and the four possibilities are usually denoted $(A,A)$, $(P,A)$, $(A,P)$, and $(P,P)$. String perturbation theory does not select a privileged option: modular invariance requires the sum of all possibilities. Indeed, the first three spin structures are in the same orbit under modular transformations, which also fixes the relative phases in the summation. The remaining $(P,P)$ spin structure is modular invariant by itself, and the single phase ambiguity in adding it to the other three precisely amounts to the available GSO projections.
A similar structure persists for genus-$g$ surfaces: the modular group has two orbits, one with $2^{g-1}(2^g - 1)$ spin structures and the other with $2^{g-1}(2^g + 1)$ spin structures, which are distinguished by the different numbers of zero modes of the Dirac operator.
The two orbits do not mix at genus-$g$, and consistency with higher-genus amplitudes only leads to a sign ambiguity in the relative phases of the two sectors~\cite{Seiberg:1986by}.

The building block of the fermionic one-loop vacuum amplitude is the determinant of the chiral Dirac operator. If fermions change by arbitrary phases along the two cycles of the torus,
\begin{eq}
    \psi(\sigma_1+2\pi,\sigma_2)= - e^{\pi i \alpha}\psi(\sigma_1,\sigma_2) \mcomma \qquad \psi(\sigma_1,\sigma_2+2\pi)= - e^{\pi i \beta}\psi(\sigma_1,\sigma_2)\mcomma
\end{eq}
the Dirac determinant is
\begin{eq}
    \det (\alpha,\beta)=\frac{\theta\smqty[\alpha\\ \beta](0|\tau)}{\eta(\tau)} \mcomma
\end{eq}
where $\theta\smqty[\alpha\\ \beta](0|\tau)$ denotes the Jacobi theta $\theta$ functions with characteristics, 
\begin{eq}
    \theta\smqty[\alpha\\ \beta](z|\tau)=\sum_{n\in\Z}q^{\frac{1}{2}\left(n+\frac{\alpha}{2}\right)^2}e^{2\pi i \left(n+\frac{\alpha}{2}\right)\left(z+\frac{\beta}{2}\right)}\mperiod 
\end{eq}
The two relevant cases correspond to $\alpha,\beta=0$ for antiperiodic boundary conditions and $\alpha,\beta=1$ for periodic boundary conditions, and the modular-invariant combination that reconstructs the SL(2,$\Z$) orbit is 
\begin{eq}
    \theta^4\smqty[0\\0]-\theta^4\smqty[1\\0]-\theta^4\smqty[0\\1] \mcomma
\end{eq}
which vanishes numerically. The remaining $\theta\smqty[1\\ 1]$, which is also modular invariant by itself, also vanishes because the fermion constant zero mode is compatible with the periodic-periodic spin structure.

In order to discuss ten-dimensional superstrings, it is convenient to introduce specific combinations of the $\theta$ functions, relying on the isometries of the target space. These are the characters of level-one $\mathfrak{so}(8)$ current algebras, which allow the clear identification of Lorentz-invariant massless spectra in the one-loop vacuum amplitudes.
This is a special case of a more general strategy, which applies to rational CFTs, for which characters are the generating functions of the degeneracy of states at all levels,
\begin{eq}\label{eq:characters}
    \chi_i(\tau)=\Tr_i q^{L_0-\frac{c}{24}}=q^{h_i -\frac{c}{24}}\sum_{n=0}^\infty d(n)q^n \mperiod
\end{eq}
Once the appropriate characters are selected by the symmetries of the setup, one builds all modular invariant combinations of them, and the classification of (closed) string theories reduces to finding matrices $M_{ij}$ such that
\begin{eq}\label{eq:torus_amplitude_generic}
    \mathcal{T}=\dsum_{ij}\chi_i(\tau)M_{ij}\chi_j(\bar{\tau})
\end{eq}
is a modular invariant vacuum amplitude.
For ten-dimensional superstrings, the $\mathfrak{so}(8)$ characters are special cases of $\mathfrak{so}(2n)$ characters, that will be of use later, which read
\begin{eq}
    O_{2n}&=\frac{\theta^n\smqty[0\\0](0|\tau)+\theta^n\smqty[0\\1](0|\tau)}{2\eta^n(\tau)}\mcomma \qquad S_{2n}=\frac{\theta^n\smqty[1\\0](0|\tau)+i^{-n}\theta^n\smqty[1\\1](0|\tau)}{2\eta^n(\tau)}\mcomma\\
    V_{2n}&=\frac{\theta^n\smqty[0\\0](0|\tau)-\theta^n\smqty[0\\1](0|\tau)}{2\eta^n(\tau)}\mcomma \qquad C_{2n}=\frac{\theta^n\smqty[1\\0](0|\tau)-i^{-n}\theta^n\smqty[1\\1](0|\tau)}{2\eta^n(\tau)}\mperiod
\end{eq}
$T$ and $S$ modular transformations act on them as
\begin{eq}
    T=e^{-\frac{i n\pi}{12}}\begin{pmatrix}1 & 0 & 0 & 0 \\ 0 & -1 & 0 & 0 \\ 0 & 0 & e^{\frac{i n\pi}{4}} & 0 \\ 0 & 0 & 0 & e^{\frac{i n \pi}{4}}\end{pmatrix}\mcomma \qquad S=\frac{1}{2}\begin{pmatrix}1 & 1 & 1 & 1 \\ 1 & 1 & -1 & -1 \\ 1 & -1 & i^{-n} & -i^{-n} \\ 1 & -1 & -i^{-n} & i^{-n}\end{pmatrix}\mcomma
\end{eq}
and on the Dedekind $\eta$ as
\begin{eq}
    \eta(\tau+1)=e^{\frac{i\pi}{12}}\eta(\tau)\mcomma \qquad \eta(-1/\tau)=(-i\tau)^{\frac{1}{2}}\eta(\tau)\mperiod
\end{eq}
Note that numerically $\theta\smqty[1 \\ 1](0|\tau)$ vanishes, and therefore $S_{2n}=C_{2n}$. However, these characters label the two different spinor chiralities in ten dimensions, and separating them makes it possible to include information on spacetime chirality in the vacuum amplitude.
In detail, the spectrum encoded in $O_{2n}$ contains states with integer level, starting with the NS vacuum at level zero, while $V_{2n}$ begins with one NS oscillator and therefore corresponds to half-integer levels starting from $\frac{1}{2}$, and finally the two spinorial characters $S_{2n}$ and $C_{2n}$ contain integer-spaced levels starting from $\frac{n}{8}$.

These basic constituents are sufficient for discussing the one-loop vacuum amplitudes of the string theories of interest.
Before moving on to the three tachyon-free non-supersymmetric strings, it is instructive to examine the other ten-dimensional superstrings. Left- and right-moving RNS closed superstrings assemble type IIA, IIB, 0A, and 0B string theories~\cite{Seiberg:1986by}, whose torus vacuum amplitudes are
\begin{eq}\label{eq:torus_of_closed_superstrings}
    {\cal T}_{\text{IIA}} & = \int_{{\cal F}}\frac{d^2\tau}{\tau_2^2}\frac{(V_8-S_8)(\bar{V}_8-\bar{C}_8)}{\tau_2^4 \eta^8\bar{\eta}^8} \mcomma \\
    {\cal T}_{\text{IIB}} & = \int_{{\cal F}}\frac{d^2\tau}{\tau_2^2}\frac{(V_8-S_8)(\bar{V}_8-\bar{S}_8)}{\tau_2^4 \eta^8\bar{\eta}^8} \mcomma \\
    {\cal T}_{\text{0A}} & = \int_{{\cal F}}\frac{d^2\tau}{\tau_2^2}\frac{O_8\bar{O}_8+V_8 \bar{V}_8+ S_8\bar{C}_8+ C_8\bar{S}_8}{\tau_2^4 \eta^8\bar{\eta}^8} \mcomma \\
    {\cal T}_{\text{0B}} & = \int_{{\cal F}}\frac{d^2\tau}{\tau_2^2}\frac{O_8\bar{O}_8+V_8 \bar{V}_8+ S_8\bar{S}_8+ C_8\bar{C}_8}{\tau_2^4 \eta^8\bar{\eta}^8} \mperiod
\end{eq}
The first two are spacetime supersymmetric, while the other two only contain spacetime bosons and are tachyonic. The two type B theories are left-right symmetric on the worldsheet and are the starting points for constructing two of the non-supersymmetric strings.
Type IIA, IIB, 0A, and 0B are related by orbifolds in ten dimensions: type A and type B theories are interchanged by a $(-1)^{f_R}$ orbifold, where $f_R$ is the right-moving worldsheet fermion number, and type 0 theories are obtained from the corresponding type II theories after a $(-1)^F$ orbifold, where $F$ is the spacetime fermion number.
The other class of known construction, heterotic strings, rests on the independence of left- and right-moving sectors and combines an RNS string with a bosonic string. 
Closed-string theories of this type can be completely classified in terms of several equivalent languages~\cite{Dixon:1986iz,Kawai:1986vd,BoyleSmith:2023xkd}, for instance, using the inequivalent compactifications of the right-moving bosonic strings on $T^{16}$, or using the fermionized version of heterotic strings with 32 right-moving Majorana-Weyl spinors and taking into account all the consistent sums over spin structures. More recently, another equivalent perspective~\cite{BoyleSmith:2023xkd} has re-expressed the classification in terms of chiral CFTs with central charge $(0,16)$, relying on the interpretation of ten-dimensional heterotic strings as $\mathcal{N}=(1,0)$ ten-dimensional sigma models times chiral CFTs with central charge $(0,16)$. 
At any rate, the result is that two of these heterotic theories are spacetime supersymmetric, and these are the ones for which the right-moving chiral CFT is a bosonic CFT. Their one-loop vacuum amplitudes can be written as
\begin{eq}
    {\cal T}_{\text{HE}} & = \int_{\cal F}\frac{d^2\tau}{\tau_2^2}\frac{(V_8-S_8)(\bar{O}_{16}+\bar{S}_{16})^2}{\tau_2^4 \eta^8\bar{\eta}^8} \mcomma \\
    {\cal T}_{\text{HO}} & = \int_{\cal F}\frac{d^2\tau}{\tau_2^2}\frac{(V_8-S_8)(\bar{O}_{32}+\bar{S}_{32})}{\tau_2^4 \eta^8\bar{\eta}^8} \mperiod
\end{eq}
The former becomes the latter after a $(-1)^{f_R}$ orbifold, while the HE theory can be obtained from the HO theory by decomposing SO$(32)$ representations into SO$(16)$$\times$SO$(16)$ ones and modding out by a sign change of the vector and conjugate spinor representations of the first SO$(16)$.
I shall return to the complete classification of heterotic models when discussing the non-supersymmetric strings.

The last basic ingredient is the construction of open-string theories with the orientifold projection~\cite{Sagnotti:1987tw,Pradisi:1988xd,Horava:1989vt,Horava:1989ga,Bianchi:1990yu,Bianchi:1990tb,Bianchi:1991eu,Sagnotti:1992qw}. This fits nicely with the general idea behind the classifications of ten-dimensional closed-string theories: one builds some parent models in terms of simple building blocks, such as free bosons and fermions, and then considers quotients (CFT orbifolds) under all the automorphisms of the parent models to generate new ones.
Open strings can be interpreted along these lines as worldsheet orbifolds of left-right symmetric closed-string theories under the action of an automorphism that exchanges left- and right-movers on the worldsheet. 
An alternative perspective on the orientifold projection is to view it as a generalization of the sum over topologies of string theory, namely of $\mathcal{D}[g]$, allowing for reflection orbifolds as two-dimensional surfaces. In this fashion, the orientifold projection introduces unoriented surfaces and, in the twisted sectors, open strings. 
The reader can find a comprehensive explanation of the one-loop vacuum amplitudes in the review~\cite{Angelantonj:2002ct}. Here, I only sketch it with the aim of fixing the notation.
The starting point is a left-right symmetric torus amplitude as in \cref{eq:torus_amplitude_generic}, to which one adds the Klein bottle amplitude
\begin{eq} \label{eq:Klein}
    {\cal K}=\frac{1}{2}\sum_{i}\sigma_i K_{i}\chi_i \mcomma
\end{eq}
where $K_{i}=M_{ii}$ and $\sigma_i$ are signs that can arise whenever the parent theory has non-trivial automorphisms~\cite{Fioravanti:1993hf,Pradisi:1995pp}.
The twisted sector of the orientifold procedure at one loop corresponds to the addition of the annulus and M\"obius strip amplitudes. The former is constrained by its transverse channel interpretation as a closed string that annihilates at two boundaries, and takes the generic form 
\begin{eq}
    {\cal A}= \frac{1}{2}\sum_{k;i,j} A^k_{ij}n^i n^j \chi_k \mcomma 
\end{eq}
where $A^k_{ij}$ is related to the one-point function of bulk conformal fields in front of a boundary. $n^i$ are Chan-Paton charges that can be assigned to the boundaries, or equivalently labels for D branes on which open strings end, with the constraint that the boundary must act as a mirror, so that states can only be paired with their GSO conjugates.
The last contribution, the M\"obius strip, generically reads
\begin{eq}
    {\cal M}\pm \frac{1}{2}\sum_{k,i}M^k_i n^i \hat{\chi}_k \mcomma 
\end{eq}
where the use of hatted characters is explained in~\cite{Angelantonj:2002ct} and corresponds to a phase difference with respect to the definition in \cref{eq:characters}, due to the shifted argument:
\begin{eq}
    \hat{\chi}_i \left(\frac{1}{2}+i\frac{\tau_2}{2}\right)=e^{-\pi i \left(h_i-\frac{c}{24}\right)} \chi_i \left(\frac{1}{2}+i\frac{\tau_2}{2}\right) \mperiod
\end{eq}
The simplest orientifold of ten-dimensional superstrings is type I string theory, which is type IIB modded out by the worldsheet reflection automorphism $\Omega$.
The four contributions to the one-loop vacuum amplitude read
\begin{eq}\label[pluralequation]{eq:vacuum_ampl_type_I}
    {\cal T}_{\text{I}} & = \frac{1}{2}{\cal T}_{\text{IIB}} \mcomma \\
    {\cal K}_{\text{I}} & = \frac{1}{2}\int_0^\infty \frac{d\tau_2}{\tau_2^2}\frac{V_8-S_8}{\tau_2^4 \eta^8}(2i\tau_2)\mcomma \\
    {\cal A}_{\text{I}} & = \frac{N^2}{2}\int_0^\infty \frac{d\tau_2}{\tau_2^2}\frac{V_8-S_8}{\tau_2^4 \eta^8}\left(i\frac{\tau_2}{2}\right)\mcomma \\
    {\cal M}_{\text{I}} & = -\frac{N}{2}\int_0^\infty \frac{d\tau_2}{\tau_2^2}\frac{\hat{V}_8-\hat{S}_8}{\tau_2^4 \hat{\eta}^8}\left(\frac{1}{2}+i\frac{\tau_2}{2}\right)\mcomma
\end{eq}
where the dependence on the complex structures is explicitly given for each case, and where the CP multiplicity $N$ associated with the ends of the open strings is set to the value $N=32$ for consistency. Other values of $N$ lead to inconsistent equations of motion (or, equivalently, to a tadpole) for an \emph{unphysical} R-R field, thus spoiling conformal invariance~\cite{Marcus:1986cm,Sagnotti:1987tw,Polchinski:1987tu} (see \cref{sec:sigma_models} and the discussion at the beginning of \cref{chapter:tadpoles}).
Type I string theory affords a spacetime geometric realization in terms of a spacetime-filling ${\text O9}^-$ plane together with $N=32$ D9 branes, which are mutually BPS, preserve 16 supercharges, and build a supersymmetric Yang-Mills multiplet with the gauge algebra of SO$(32)$.

\section{Non-supersymmetric strings}
\label{sec:non_susy_strings}

Among the theories presented in the previous section, type IIA, type IIB, type I, and the two heterotic theories denoted by HO and HE are the only tachyon-free ones. They also share another feature: they are spacetime supersymmetric.
However, these two requirements are non-exclusive, and three ten-dimensional string theories without tachyons in the spectrum and without spacetime supersymmetry are known. These models are the central topic of this thesis, and in this section, I introduce them from the worldsheet perspective.

One such model is a closed-string heterotic theory known as the SO$(16)$$\times$SO$(16)$ string~\cite{Alvarez-Gaume:1986ghj,Dixon:1986iz}.
It follows from the classification of~\cite{Dixon:1986iz,Kawai:1986vd,BoyleSmith:2023xkd} that this is the only tachyon-free non-supersymmetric heterotic model in ten dimensions.
It can be obtained from the HE model by a $(-1)^F \delta$ orbifold in the bosonic formulation, where $\delta$ is an order-two shift in the internal lattice, or equivalently by a $(-1)^{F+F_1+F_2}$ orbifold in the fermionic formulation, where $F_1$ and $F_2$ are the fermion numbers of the two groups of 16 right-moving fermions. Note that $F_1$ and $F_2$ can be interpreted as $2\pi$ rotations in the two O$(16)$ factors of the O$(16)$$\times$O$(16)$ subgroup of $E_8\times E_8$. In the language of~\cite{BoyleSmith:2023xkd}, it combines the $\mathfrak{D}_8\times\mathfrak{D}_8$ fermionic CFT
\begin{eq}
    \bar{\mathcal{Z}}_{\text{NS}}^{\text{NS}}&=\bar{O}_{16}\bar{O}_{16}+\bar{S}_{16}\bar{S}_{16}+\bar{V}_{16}\bar{C}_{16}+\bar{C}_{16}\bar{V}_{16}\mcomma \\ \bar{\mathcal{Z}}_{\text{NS}}^{\text{R}}&=\bar{O}_{16}\bar{O}_{16}+\bar{S}_{16}\bar{S}_{16}-\bar{V}_{16}\bar{C}_{16}-\bar{C}_{16}\bar{V}_{16}\mcomma \\
    \bar{\mathcal{Z}}_{\text{R}}^{\text{NS}}&=\bar{O}_{16}\bar{S}_{16}+\bar{S}_{16}\bar{O}_{16}+\bar{V}_{16}\bar{V}_{16}+\bar{C}_{16}\bar{C}_{16}\mcomma \\
    \bar{\mathcal{Z}}_{\text{R}}^{\text{R}}&=\bar{O}_{16}\bar{S}_{16}+\bar{S}_{16}\bar{O}_{16}-\bar{V}_{16}\bar{V}_{16}-\bar{C}_{16}\bar{C}_{16}\mcomma
\end{eq}
with the $\mathcal{N}=(0,1)$ sigma model in ten-dimensional flat Minkowski
\begin{eq}
    \mathcal{Z}_{\text{NS}}^{\text{NS}}&=\frac{O_8+V_8}{2}\mcomma \qquad \mathcal{Z}_{\text{NS}}^{\text{R}}=-\frac{O_8-V_8}{2}\mcomma \\ \mathcal{Z}_{\text{R}}^{\text{NS}}&=-\frac{S_8+C_8}{2} \mcomma \qquad \mathcal{Z}_{\text{R}}^{\text{R}}=-\frac{S_8-C_8}{2}\mperiod
\end{eq}
All these equivalent interpretations lead to the one-loop vacuum amplitude
\begin{eq}\label{eq:so1616_vacuum_ampl}
    {\cal T}_{\text{SO$(16)$$\times$SO$(16)$}} & = \int_{\cal F} \frac{d^2\tau}{\tau_2^2}\frac{1}{\tau_2^4 \eta^8 \bar{\eta}^8}\Bigg[O_8 \left(\bar{V}_{16}\bar{C}_{16}+\bar{C}_{16}\bar{V}_{16}\right)+V_8 \left(\bar{O}_{16}\bar{O}_{16}+\bar{S}_{16}\bar{S}_{16}\right)\\
    & - S_8 \left(\bar{O}_{16}\bar{S}_{16}+\bar{S}_{16}\bar{O}_{16}\right)-C_8 \left(\bar{V}_{16}\bar{V}_{16}+\bar{C}_{16}\bar{C}_{16}\right)\Bigg] \mperiod
\end{eq}
The bosonic massless spectrum can be extracted from \cref{eq:so1616_vacuum_ampl} and contains the standard bosonic sector, with a metric, a two-form field $B$, and a dilaton, together with gauge fields in the adjoint of the SO$(16)$$\times$SO$(16)$ gauge algebra. 
The fermionic massless spectrum consists of left-handed Majorana-Weyl spin-$\frac{1}{2}$ fermions in the $(128,1)\oplus(1,128)$ representation of the gauge group, and right-handed Majorana-Weyl spin-$\frac{1}{2}$ fermions in the $(16,16)$ representation of the gauge group. 
The last feature of this model that is important in the following sections is that the numerical value of the torus amplitude in \cref{eq:so1616_vacuum_ampl} is non-vanishing and finite~\cite{Alvarez-Gaume:1986ghj}, and in particular, it is negative, or equivalently
\begin{eq}
    -{\cal T}_{\text{SO$(16)$$\times$SO$(16)$}}>0\mperiod
\end{eq}
From a quantum field theory perspective, this is as expected because there is no bosonic-fermionic degeneracy, and the sign is consistent with the net excess of massless fermions, with a positive one-loop vacuum energy from \cref{eq:1_loop_amplitude_string}:
\begin{eq}
    \Gamma=-\frac{{\cal T}_{\text{SO$(16)$$\times$SO$(16)$}}}{2 (4\pi^2\alpha')^{5}}  >0 \mperiod
\end{eq}
Note that the presence of more massless fermions than massless bosons is not sufficient for a negative torus amplitude because, in principle, the tower of string oscillator modes might flip the sign of the massless contribution only.

The other two non-supersymmetric models are theories of open and closed strings, orientifolds of parent closed-string theories.
From the arguments presented in the previous section, and in particular from the one-loop vacuum amplitudes of the closed superstring models of \cref{eq:torus_of_closed_superstrings}, three string theories are left-right symmetric, and thus allow for ten-dimensional orientifold projections: type IIB, type 0A, and type 0B.
In fact, type IIB has two types of tachyon-free orientifolds, which differ in their vacuum amplitudes by relative signs in the contributions of the various open-string sectors: the supersymmetric type I projection $\Omega$, and the non-supersymmetric USp$(32)$ Sugimoto model~\cite{Sugimoto:1999tx}, for which the orientifold projection is $\Omega (-1)^{f_L}$ in the open sector.
The one-loop vacuum amplitudes of the USp$(32)$ model are the same as those of type I, with the exception of the M\"obius amplitude, which differs by the sign of the vector character $V_8$:
\begin{eq}\label[pluralequation]{eq:vacuum_ampl_Sugimoto}
    {\cal T}_{\text{USp$(32)$}} & = {\cal T}_{\text{I}} \mcomma \qquad {\cal K}_{\text{USp$(32)$}}  = {\cal K}_{\text{I}}\mcomma \qquad
    {\cal A}_{\text{USp$(32)$}} = {\cal A}_{\text{I}} \mcomma \\
    {\cal M}_{\text{USp$(32)$}} & = -\frac{N}{2}\int_0^\infty \frac{d\tau_2}{\tau_2^2}\frac{-\hat{V}_8-\hat{S}_8}{\tau_2^4 \hat{\eta}^8}\left(\frac{1}{2}+i\frac{\tau_2}{2}\right)\mperiod
\end{eq}
Consistency selects the value $N=32$ similarly to the case of type I string theory.
The signs of ${\cal M}_{\text{USp$(32)$}}$ have dramatic consequences, which are the focus of \cref{sec:tadpole_div}.
The overall geometric interpretation is a sign change in the tensions and charges of branes and orientifolds; thus, the spacetime picture of the Sugimoto model is type IIB with an $\text{O9}^+$ plane and $N=32$ $\overline{{\text{D9}}}$ branes, which are not mutually BPS, break all the spacetime supercharges, and build the gauge algebra of USp$(32)$. In these terms, the Sugimoto model is the simplest instance of a more general strategy for breaking supersymmetry in string theory with localized sources: brane supersymmetry breaking~\cite{Antoniadis:1999xk,Angelantonj:1999jh,Aldazabal:1999jr,Angelantonj:1999ms}.
The massless spectrum of the USp$(32)$ model includes the gravity multiplet of ten-dimensional type I supergravity together with non-supersymmetric matter. In detail, the bosonic sector involves a metric, a two-form field, a dilaton, and gauge bosons in the adjoint of the USp$(32)$ algebra. The fermionic sector includes the left-handed spin-$\frac{3}{2}$ and the right-handed spin-$\frac{1}{2}$ fields of the gravity multiplet together with left-handed Majorana-Weyl spin-$\frac{1}{2}$ fields in the antisymmetric $(496)$ representation of the gauge algebra.
This spectrum has at least two notable features: it contains a massless gravitino even though the theory is non-supersymmetric, and the antisymmetric representation of the USp$(32)$ algebra is reducible and contains a singlet $(495+1)$. Both properties are interpreted as the presence of a non-linear realization of supersymmetry in this model~\cite{Dudas:2000nv,Pradisi:2001yv,Kitazawa:2018zys}. 

The remaining non-supersymmetric string of interest arises from the bosonic type 0 theories.
For these, several sign choices in \cref{eq:Klein} are available because the theories have nontrivial automorphisms that reflect different options for the orientifold projection.
Both type 0 theories are bosonic and therefore they can be orientifolded by $\Omega (-1)^{F_{R,L}}$, where $F_{R,L}$ are the right- and left-moving spacetime fermion numbers. Yet, neither $\Omega$ nor $\Omega (-1)^{F_{R,L}}$ remove the closed-string tachyon of the parent models. 
Type 0B comes with an additional option that is compatible with the fusion rules because the right-moving worldsheet fermion number $f_R$ (or, equivalently, the left-moving $f_L$) is an order-two automorphism, and therefore one can perform the orientifold with $\Omega (-1)^{f_R}$.
The resulting open-string model, type 0'B string theory~\cite{Sagnotti:1995ga,Sagnotti:1996qj}, is tachyon-free, and is the last of the non-supersymmetric strings of interest in this thesis. 
The one-loop vacuum amplitudes are given by
\begin{eq} \label[pluralequation]{eq:vacuum_ampl_0primeB}
    {\cal T}_{\text{0'B}} & = \frac{1}{2}{\cal T}_{\text{0B}} \mcomma \\
    {\cal K}_{\text{0'B}} & = \frac{1}{2}\int_0^\infty \frac{d\tau_2}{\tau_2^2}\frac{-O_8+V_8+S_8-C_8}{\tau_2^4 \eta^8}(2i\tau_2)\mcomma \\
    {\cal A}_{\text{0'B}} & = \int_0^\infty \frac{d\tau_2}{\tau_2^2}\frac{N\bar{N} \, V_8 -\frac{1}{2}\left(N^2 + \bar{N}^2\right) C_8}{\tau_2^4 \eta^8}\left(i\frac{\tau_2}{2}\right)\mcomma \\
    {\cal M}_{\text{0'B}} & = \frac{N+\bar{N}}{2}\int_0^\infty \frac{d\tau_2}{\tau_2^2}\frac{\hat{C}_8}{\tau_2^4 \hat{\eta}^8}\left(\frac{1}{2}+i\frac{\tau_2}{2}\right)\mcomma
\end{eq}
where the notation of $N$ and $\bar{N}$ comes from the interpretation of the CP labels as complex charges in the (anti)fundamental representation of unitary gauge algebras. Consistency and absence of open-string tachyons set $N=\bar{N}=32$.
The massless bosonic spectrum that can be extracted from \cref{eq:vacuum_ampl_0primeB} contains a graviton and a dilaton from the NS-NS sector, a scalar, a two-form potential, and a self-dual four-form potential from the R-R sector, together with U$(32)$ gauge bosons.\footnote{At low energies, only the SU$(32)$ subalgebra survives because the remaining U$(1)$ gauge field eats the R-R scalar and becomes massive~\cite{Sagnotti:1996qj,Schwarz:2001sf}.}
On the other hand, the fermionic spectrum contains left-handed Majorana-Weyl spin-$\frac{1}{2}$ fields in the $496\oplus\overline{496}$ of the gauge algebra.
Type 0'B string theory also has a spacetime interpretation in terms of orientifold planes and D9 branes, which is more involved than in the Sugimoto case because the parent type 0B theory has two copies of O9s and D9s. The action of $f_R$ exchanges these two copies, and the picture that emerges is in terms of tensionless orientifold planes and 32 pairs of D9 branes. Spacetime fermions are generated in the open sector of strings stretched between branes. 

At first glance, these three strings are examples of consistent theories of gravity from string theory without spacetime supersymmetry, thus contradicting the provocative remark at the beginning of the thesis. As such, they must provide finite results and avoid UV divergences. 
Upon closer inspection, the one-loop vacuum amplitudes of the orientifolds already bring about an unwelcome surprise.

\subsection{Tadpole divergence}
\label{sec:tadpole_div}

Consider the USp$(32)$ vacuum amplitudes of \cref{eq:vacuum_ampl_Sugimoto}. Their sum, which would be the vacuum energy density in a quantum field theory interpretation, becomes
\begin{eq}\label{eq:Sugimoto_divergence}
    {\cal T}_{\text{USp$(32)$}}+{\cal K}_{\text{USp$(32)$}}+{\cal A}_{\text{USp$(32)$}}+{\cal M}_{\text{USp$(32)$}}=32\int_0^\infty \frac{d\tau_2}{\tau_2^6}\frac{\hat{V}_8}{\hat{\eta}^8}\left(\frac{1}{2}+i\frac{\tau_2}{2}\right)\mperiod
\end{eq}
This expression is divergent, and in particular, the integral diverges at $\tau_2=0$, which represents the UV in terms of the Schwinger parameter $\tau_2$. 
It should be striking to find a divergent result in a theory that is supposed to be fully consistent: does this mean that the Sugimoto model is inconsistent?
Remarkably, the divergence is not a UV problem, and even more importantly, it contains the physics of non-supersymmetric strings.

It is convenient to use open-closed duality and translate \cref{eq:Sugimoto_divergence} into the transverse channel using a $P=T^{\frac{1}{2}} S T^2 S T^{\frac{1}{2}}$ transformation~\cite{Angelantonj:2002ct}, so that the only non-vanishing contribution is 
\begin{eq}
    \widetilde{{\cal M}}_{\text{USp$(32)$}}=64\int_0^\infty d\ell \frac{\hat{V}_8}{\hat{\eta}^8}\left(\frac{1}{2}+i\ell\right)\mperiod
\end{eq}
This allows to interpret the divergence as an IR one, because $\tau_2\to0$ corresponds to $\ell \to\infty$. Indeed, the expansion of the $V_8$ character for small $q$,
\begin{eq}
    V_8\sim 8 q^{\frac{1}{3}}\left(1+8 q + \mathcal{O}(q^2)\right)\mcomma
\end{eq}
leads to
\begin{eq}\label{eq:tadpole_extraction}
    \tilde{{\cal M}}_{\text{USp$(32)$}}\propto \int_0^\infty d\ell \mperiod
\end{eq}
This expression conveys the crucial physical message: it explicitly states that the divergence, as $\ell\to\infty$, is in the IR from the point of view of the bulk closed-string modes. The identity
\begin{eq}
    \int_0^\infty d\ell = \int_0^\infty d\ell e^{-2\pi \ell m^2}\Big|_{m^2=0}=\frac{1}{2\pi m^2}\Big|_{m^2=0} \mcomma
\end{eq}
allows to interpret the divergence of \cref{eq:tadpole_extraction} as coming from massless states propagating for an infinite amount of Schwinger time and then annihilating into the vacuum, which requires non-vanishing tadpoles for these states.
Because the dilaton is the only Lorentz-scalar physical field,\footnote{There may be tadpoles for non-physical fields, and indeed, the dilaton tadpole is usually accompanied by a tadpole for the non-physical trace part of the metric. I shall not delve into these technical subtleties, which have been carefully addressed in~\cite{Polchinski:1988jq}.} it must be the massless field propagating in \cref{eq:Sugimoto_divergence}, and thus it must have non-zero tadpoles at half-loop.
The tadpole coefficient can be extracted by noting that if the spacetime effective action---that I shall introduce in \cref{chapter:tadpoles}---includes a tadpole term of the form
\begin{eq}
    \delta S  \sim -\int \gamma T \phi \mcomma
\end{eq}
the diagram that describes the $\phi$ propagation of \cref{eq:Sugimoto_divergence} is
\begin{eq}
    \sim  \frac{(\gamma T)^2}{p^2}\Big|_{p^2=0}\mcomma
\end{eq}
so that $T\gamma$ is obtained by comparing this expression with \cref{eq:tadpole_extraction}.

Similar considerations hold for type 0'B, whereas in the heterotic model one would find a divergence in the two-loop vacuum amplitude.
The aim of the next chapter is to gain a better understanding of the consequences of these tadpoles in string theory.
In fact, tadpoles in quantum field theories are usually taken into account by turning on background fields such that the currents of the background fields and the tadpoles give equal and opposite contributions, which then cancel. 
Interestingly, the outcome of string theory is similar to that of quantum field theory, but the requirement of two-dimensional conformal invariance involves a mechanism that solves this problem in a spectacular way.


%% file: chapters/chapter1.tex

\chapter{Dilaton tadpoles and vacuum solutions}
\label{chapter:tadpoles}

In this chapter, I address the main point of the thesis: the spacetime backreaction of the absence of supersymmetry, focusing on the non-supersymmetric strings introduced in \cref{sec:non_susy_strings}.
First, I review the mechanism that explains the divergences of \cref{sec:tadpole_div} and then use the results to understand the vacuum solutions of the models of interest.

The language that I need involves the quantum effective actions from string theory, which are the natural continuation of the discussion in \cref{sec:sigma_models}, linking the worldsheet to the spacetime physics.
Already in~\cite{Callan:1985ia}, the authors note that the lowest-order beta functional equations in the sigma model approach can be obtained from a target-space action.
However, there is more: the action reproduces the scattering amplitudes from the S matrix formulation of string theory, at the appropriate order of $\alpha'$.
In fact, it represents a quantum effective action for the massless modes of the string~\cite{Callan:1985ia,Sen:1985eb,Sen:1985qt,Fradkin:1985ys,Tseytlin:1986ti} and can be obtained from the renormalized partition function
\begin{eq}\label{eq:effective_action}
    e^{-S}=\mathcal{Z}=\sum_n \mathcal{Z}_n\mperiod
\end{eq}
The equations of motion for this quantum effective action are the embodiment of conformal invariance, and therefore of consistency. They effectively set the physics on shell, and the tree-level S matrix extracted from $S$ reconstructs the all-loop string scattering amplitudes for the massless modes. 
In this way, the renormalization procedure provides the $\alpha'$ corrections and the string loop expansion provides the $g_s$ corrections to the equations of motion.

Note that $S$ is a quantum effective action for the \emph{massless modes} only, so that it represents an all-loop sum over diagrams that are 1-particle irreducible with respect to the massless states. Massive states are allowed to propagate along internal lines, but the effective action does not capture processes in which they appear as external legs. 
The restriction to massless modes is not a concern for at least two reasons. First, at low spacetime energies, only massless states are relevant; and second, massive states that are not BPS decay into massless states at finite $g_s$; thus, massless states control the ultimate physics description.

In curved backgrounds, which are the only options for non-supersymmetric strings, the effective action approach is appropriate because it is slightly more than an on-shell formulation: it embodies several possible on-shell descriptions, which are obtained by 
\begin{enumerate}
    \item solving the classical equations of motion to obtain a vacuum;
    \item expanding the classical fields around their vacuum expectation values to compute the full scattering amplitudes from the tree-level S matrix defined by $S$.
\end{enumerate}
This is the true meaning of background independence that I mentioned in \cref{sec:sigma_models}, but in a full quantum version.

\section{IR divergences and the Fischler-Susskind mechanism}
\label{sec:FS_mechanism}

The aim of this section is to provide an idea of the mechanism that takes massless tadpoles into account. For simplicity, the discussion is confined to the case of closed strings, for which a single Riemann surface contributes to each order of string perturbation theory. For open strings, several technical subtleties arise because of the different topologies at the same order, but the basic mechanisms are the same; thus, it is sufficient to review the closed-string case and then extrapolate to open strings.
\Cref{sec:string_divergences} is a summary of the possible IR divergences that can occur, \cref{sec:susy_protections} explains why these divergences are absent when spacetime supersymmetry is present, and \cref{sec:FS} contains the string theory mechanism that takes tadpoles into account. Then, in \cref{sec:open_string_tadpoles} I briefly discuss the consequences for the Sugimoto model by solving the issues raised in \cref{sec:tadpole_div}, and \cref{sec:action_non_susy_strings} is the starting point for the analysis of the later sections.

\subsection{String divergences}
\label{sec:string_divergences}

\begin{figure}[!ht]
\centering
\begin{subfigure}{.5\textwidth}
  \centering
  \includegraphics[width=.5\linewidth]{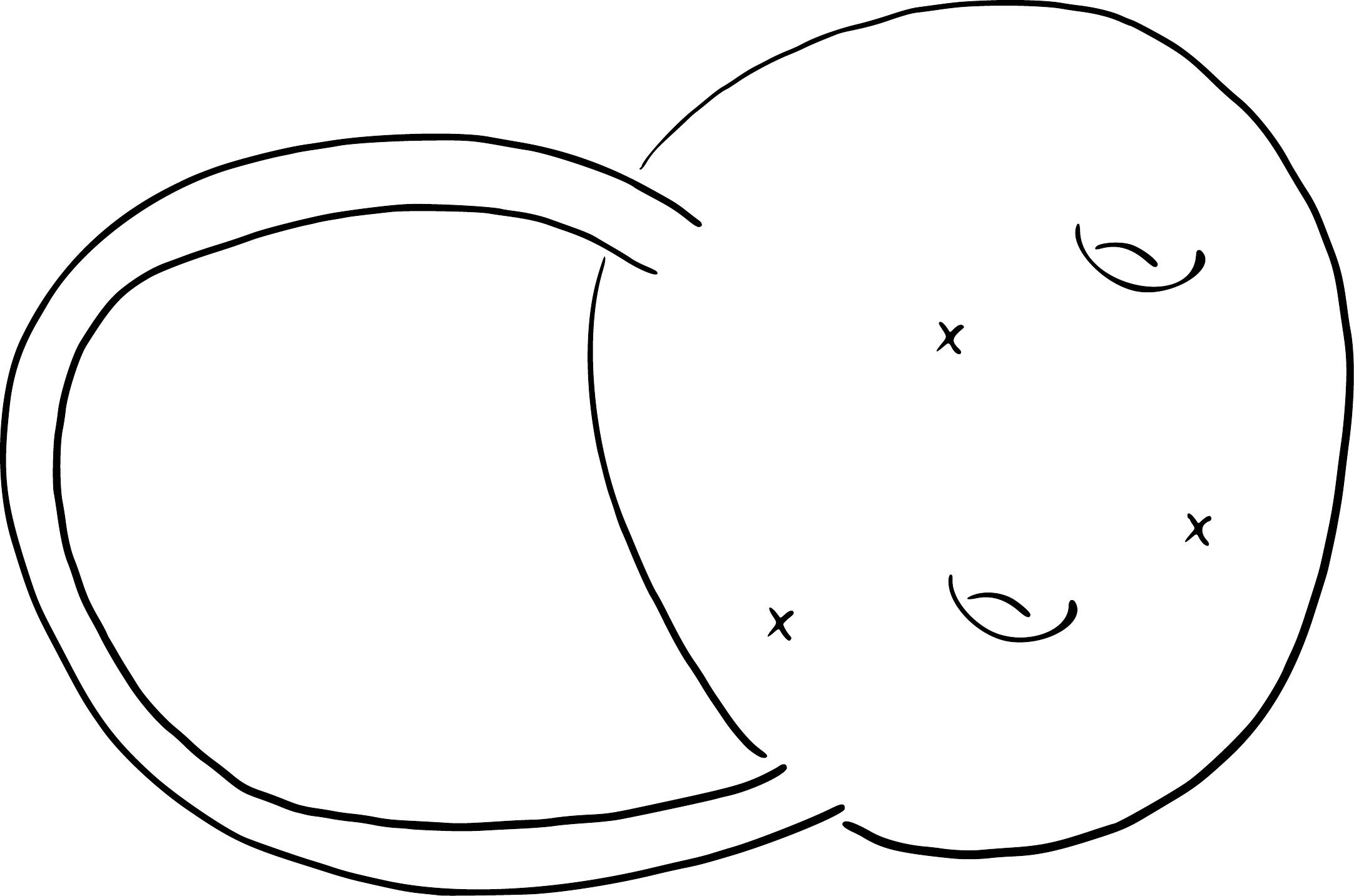}
  \caption{Infinitely long handle.}
  \label{fig:handle}
\end{subfigure}%
\begin{subfigure}{.5\textwidth}
  \centering
  \includegraphics[width=.9\linewidth]{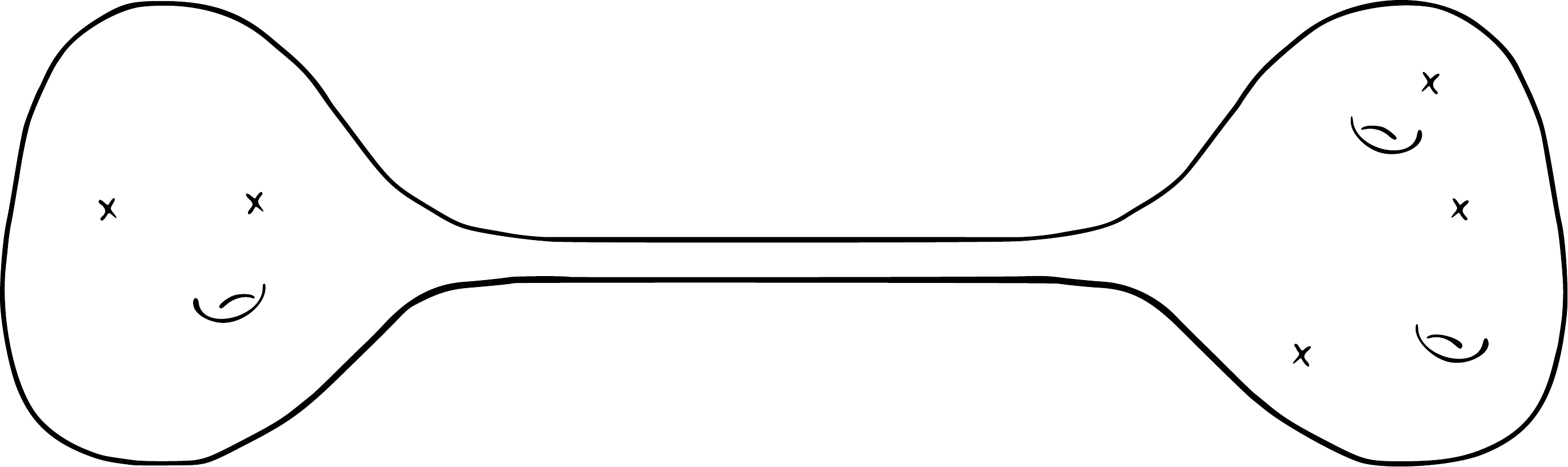}
  \caption{Infinitely long cylinder.}
  \label{fig:cylinder}
\end{subfigure}
\caption{Potentially dangerous degeneration limits of Riemann surfaces.}
\label{fig:degeneration_limits}
\end{figure}

IR divergences can develop at the boundaries of the moduli spaces of Riemann surfaces when some topological features shrink to zero size and when massless or tachyonic states are involved in the shrinking.
Focusing on closed-string settings, two limits are potentially relevant:
\begin{itemize}
    \item[--] A handle that shrinks. Choosing a constant-curvature metric on the surface, this is equivalent to an infinitely long handle, and the contribution can be approximated by a particle loop as in \cref{eq:1_loop_amplitude_particle} (see \cref{fig:handle}). 
    Explicitly, summing over the states that can run in the handle, one is effectively considering the large-$t$ behavior of the integrand
    \begin{eq}
        \sim  t^{-1-\frac{D}{2}} \text{ Str}\left(  e^{-t M^2} \right)\mcomma
    \end{eq}
    so that only tachyonic states yield divergences. For non-supersymmetric strings, this degeneration is not dangerous.
    \item[--] A cycle connecting two sub-diagrams that shrinks. One can again use a conformally-equivalent description that translates this degeneration into an infinitely long cylinder, which becomes a particle propagator (see \cref{fig:cylinder}). Then, using \cref{eq:propagator}, the contribution of each state takes the form
    \begin{eq}
        \sim \int^\infty e^{- t M^2} \mperiod
    \end{eq}
    Hence, tachyons and massless fields are the origins of infinities. 
\end{itemize}
For the vacuum amplitudes of the tachyon-free non-supersymmetric strings, the relevant degeneration is that of the second type with massless scalar propagators. Assuming a non-vanishing tadpole $\expval{V}_g$ for a massless state on a genus-$g$ surface and letting $q=e^{-2\pi t}$, the divergent part of the $\ell$-loop vacuum amplitude can then be written as
\begin{eq}\label{eq:tadpole_divergence}        
    \sim \int_0^1 \frac{dq}{q} \sum_{g=0}^{\ell} \expval{V}_g \expval{V}_{\ell-g} \mperiod
\end{eq}
Non-vanishing tadpoles become logarithmic divergences in the $q$-integral.
For closed strings, SL(2,$\C$) invariance forbids tadpoles on the sphere, $\expval{V}_0=0$, and therefore the one-loop vacuum amplitude $\mathcal{Z}_1$ is always finite. Analogous arguments starting from $\expval{V V}_0=0$ imply that genus-one tadpoles are never divergent. When these are non-vanishing, IR divergences are bound to appear: from \cref{eq:tadpole_divergence}, a non-vanishing $\expval{V}_1$ leads to a logarithmic divergence in the two-loop vacuum amplitude:
\begin{eq}\label{eq:2loop_div}
    \mathcal{Z}_2 \sim {\expval{V}_1}^2 \log\epsilon \mcomma 
\end{eq}
with $\epsilon\to 0$.
For the non-supersymmetric heterotic SO$(16)$$\times$SO$(16)$ string, such a two-loop divergence is expected.

\subsection{Supersymmetry protections}
\label{sec:susy_protections}

The divergence in \cref{eq:2loop_div} requires non-vanishing tadpoles $\expval{V}_1$.
Perhaps the most important consequence of spacetime supersymmetry in string theory is that these tadpoles vanish, and no divergence is present.
This is part of non-renormalization theorems, which protect, for instance, $0$-, $1$-, $2$-, and $3$-point functions of massless modes in a flat Minkowski background from string-loop corrections when all supercharges are present.
The argument uses only the existence of supersymmetry currents~\cite{Martinec:1986wa}, and for $0$-point functions, namely vacuum amplitudes, it is simple enough that I can mention it here. 

The idea is to start from a handle in a genus-$g$ vacuum amplitude, cut it, and then sum over a complete set of states to glue it again. The genus-$g$ vacuum amplitude is then replaced by a supertrace of genus-$(g-1)$ processes. 
States are related by supersymmetry, and therefore one can replace a fermionic $\ket{F}$ with its expression in terms of the bosonic partner, $\ket{F}=Q\ket{B}$, which is equivalent to adding a contour integral of the supersymmetry current around the cut.
The contour can be freely deformed because vacuum amplitudes have no insertions of vertex operators. In particular, it can reach the other side of the cut by sweeping over the entire surface.
Therefore, the fermionic contribution to the supertrace is equal in magnitude and opposite in sign to the bosonic contribution, and genus-$g$ vacuum amplitudes vanish.
Translating the text into formulas,
\begin{eq}
    \mathcal{Z}_g & = \sum_B \bra{B}(g-1)\text{-loop}\ket{B}-\sum_F\bra{F}(g-1)\text{-loop}\ket{F} \\
    &\overset{\scriptscriptstyle{\ket{F}=Q\ket{B}}}{=} \sum_B \bra{B}(g-1)\text{-loop}\ket{B}-\sum_B\bra{B}(g-1)\text{-loop}\ket{B}=0 \mcomma
\end{eq}
where ``$(g-1)\text{-loop}$'' is a genus-$(g-1)$ process.
Note that the vanishing of vacuum amplitudes only requires states to appear in supermultiplets.

\begin{figure}[!ht]
    \centering
    \includegraphics[scale=0.15]{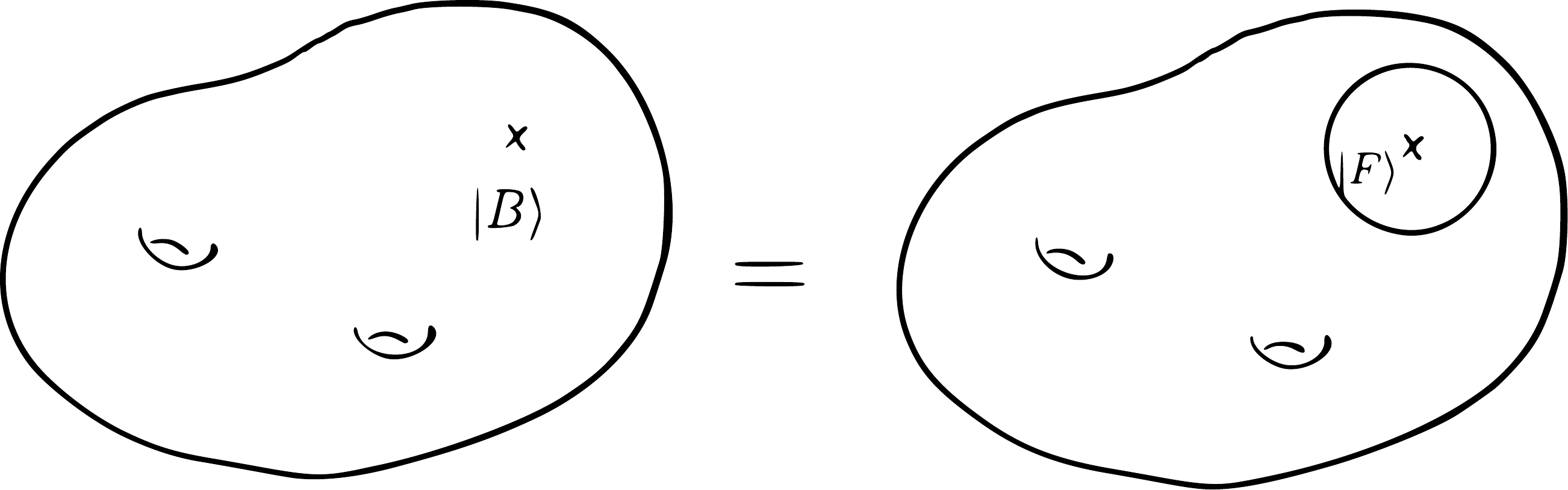}
    \caption{A bosonic tadpole can be traded for a fermionic one, together with a contour integral of the supersymmetry current, which can shrink to zero size.}
    \label{fig:susy_protection}
\end{figure}

A similar statement holds for $1$-point functions, namely the tadpoles that are of interest in this section. First, fermionic tadpoles are forbidden by Lorentz invariance. For bosonic tadpoles, a bosonic vertex operator can be replaced with a fermionic one surrounded by a contour integral of the supersymmetry current, see \cref{fig:susy_protection}. Since the contour can be deformed until it shrinks, Lorentz invariance makes all tadpoles vanish also in this case. A complete exposition of the technical details can be found in~\cite{Friedan:1985ge}. 

These protections hold for massless and BPS states, while amplitudes for generic massive states of the string are loop-corrected. In fact, a non-vanishing imaginary part in the mass corrections is physically needed to signal the instability that causes massive modes to decay to stable massless (or BPS) ones~\cite{Marcus:1988vs,Okada:1989sd}.
In non-supersymmetric strings, the absence of supersymmetry leads to the appearance of tadpoles and IR divergences.

\subsection{All-loop Weyl invariance}
\label{sec:FS}

The presence of IR divergences from massless tadpoles is problematic from the perspective of two-dimensional conformal invariance. From \cref{eq:effective_action} at a given loop order, it appears that one is not solving the string beta functional equations because the effective potential has non-vanishing first derivatives, which means that one is not on-shell. Then, the conformal mode of \cref{eq:conformal_mode} does not decouple and it makes no sense to consider genus-$g$ partition functions
\begin{eq}
    \mathcal{Z}_g= \int_{\mathcal{M}_g}d\tau_g\mathcal{D}[\phi]e^{-S[\phi,\tau_g]} \mcomma
\end{eq}
where $\mathcal{M}_g$ is the genus-$g$ moduli space, because the theory is not quantum conformally invariant.
This is clearly inconsistent with the worldsheet definition of string theory.

\begin{figure}
    \centering
    \includegraphics[scale=0.13]{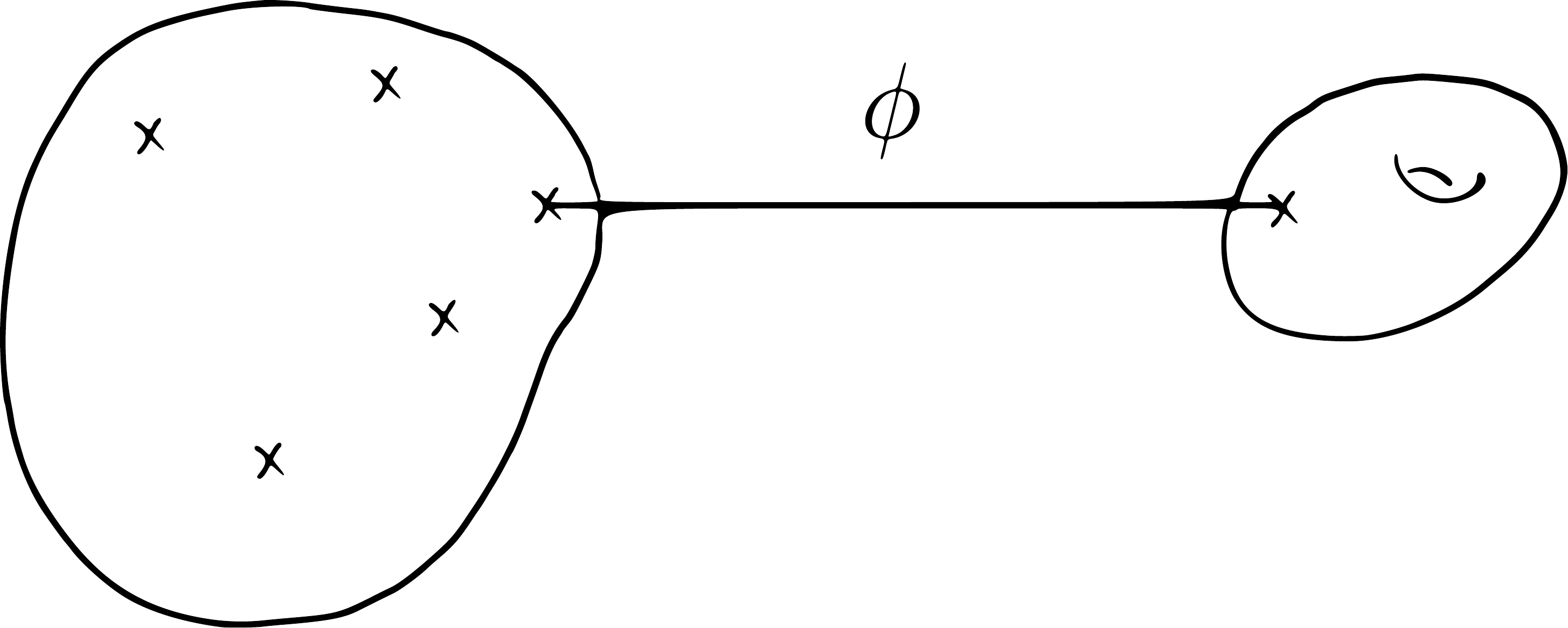}
    \caption{Zero-momentum dilaton, propagating for an infinite amount of Schwinger time and annihilating on a genus-one tadpole.}
    \label{fig:tadpole}
\end{figure}

The crucial observation is that Weyl invariance may be lost genus-by-genus but recovered in the complete sum over Riemann surfaces, so that consistency is regained in a quantum fashion and 
\begin{eq}
    \mathcal{Z}= \sum_{g=0}^\infty \int_{\mathcal{M}_g}d\tau_g\mathcal{D}[\phi]e^{-S[\phi,\tau_g]} 
\end{eq}
is well-defined.
The mechanism that implements such a procedure is known as the Fischler-Susskind mechanism~\cite{Fischler:1986ci,Fischler:1986tb}.
The idea is to regularize the tadpole IR divergences by introducing the same short-distance cutoff $\epsilon$ for both the local divergences of the sigma model and the degeneration points of the moduli spaces of \cref{sec:string_divergences}.
In the original language of~\cite{Fischler:1986ci,Fischler:1986tb}, the fields are regularized perturbatively in the string coupling. The bare dilaton $\phi^{(0)}$, for instance, would be 
\begin{eq}
    \phi^{(0)} \sim \phi-g_s^2 T_1 \log\epsilon \mcomma 
\end{eq}
where $T_1$ denotes the one-loop dilaton tadpole.
This corresponds to the result that the infinite part of a genus-one amplitude is the same amplitude at genus zero to which one adds a zero-momentum dilaton, which propagates for an infinite amount of Schwinger time and closes with a genus-one tadpole (see \cref{fig:tadpole}). 
Note that tadpoles are present not only for the dilaton, but also for the non-physical trace part of the metric, and counterterms must be introduced for both, although I shall not be concerned with this technical issue.

In the above formulation of the Fischler-Susskind mechanism, infinities in one-loop amplitudes cancel against counterterms inserted into tree-level amplitudes.
Pictorially, this is an \emph{all-genera renormalization}~\cite{Fischler:1986ci,Fischler:1986tb,Lovelace:1986kr,Callan:1986bc,Das:1986dy,Metsaev:1987ju,Callan:1988st,Callan:1988wz,Tseytlin:1988mw,Russo:1989nlt,Tseytlin:1990mv}, and consequently, the equations of motion contain simultaneous contributions from different worldsheet topologies.
Indeed, string theory is not a quantum field theory of sigma models, and the genus-by-genus renormalization is a bonus conferred by unbroken supersymmetry.
In generic setups, it is reasonable to expect consistency only in the complete theory, after the sum over Riemann surfaces.
After this generalized renormalization procedure, the renormalized partition function gives the effective action as in \cref{eq:effective_action}, whose equations of motion set all tadpoles to zero in the true vacuum. In practice, tadpoles are subtracted at the cost of shifting the background, which requires a complete change of perspective on the worldsheet formulation: the sum over surfaces is not only a perturbative expansion but is critical to achieve consistency. 

The ingredients that allow to perform the complete Fischler-Susskind tadpole subtraction are the genus-$g$ dilaton tadpoles $\expval{V_\phi}_g$.
While the computation of $\expval{V_\phi}_1$ is carried out with no subtleties, at two loops, one must be careful with the divergent sub-diagrams. In practice, $\expval{V_\phi}_2$ contains a diverging part originating from $\expval{V_\phi}_1$ and a finite part, $T_2$, which is the sought-after tadpole coefficient:
\begin{eq}\label{eq:two-loop_tadpole}
    \expval{V_\phi}_2 \sim \expval{V_\phi}_1 \expval{V_\phi V_\phi}_1 \left(\frac{1}{p^2}\right)_{p^2\to 0}+T_2\mperiod
\end{eq}
With these, one can build the effective action $S$. In particular, $S$ contains a scalar potential for the dilaton that follows from the tadpole subtraction: the \emph{tadpole potential} in the string frame:
\begin{eq}\label{eq:tadpole_potential_generic}
    V(\phi)=\sum_{g=1}^\infty T_g e^{2(g-1)\phi}\sim T_1 + T_2 e^{2\phi} + \dots \mperiod
\end{eq}

This is the manifestation of the absence of spacetime supersymmetry in string perturbation theory: the tadpole potential replaces the quantum cosmological constant problem in quantum field theory.
In fact, the scalar potential can be interpreted as loop corrections to the string-tree-level cosmological constant, but since the latter vanishes in all critical string models, the leading term in $V(\phi)$ must be taken into account in the tree-level equations of motion.
The physical consequences are dramatic: tadpole potentials from perturbative string theory are runaways for the dilaton, and therefore flat space is not a vacuum solution of non-supersymmetric strings.

Before turning to this problem, the next section discusses the open-string counterpart of the tadpole subtraction.

\subsection{Open string tadpoles}
\label{sec:open_string_tadpoles}

The combined results of \cref{sec:string_divergences} and \cref{sec:FS} can now explain the physical meaning of the divergence in the one-loop vacuum amplitude of \cref{sec:tadpole_div}. In the following discussion, I focus on the Sugimoto model; similar considerations apply to type 0'B string theory.

For open strings, dilaton tadpoles at the half-genus level are generically present, which means that the insertion of a dilaton vertex operator on a disk or a crosscap is non-vanishing and, using a variation of the arguments in \cref{sec:string_divergences}, finite.
The potentially divergent diagrams are those for which a dilaton is created from a disk/crosscap tadpole, propagates for an infinite amount of Schwinger time, and annihilates again through a disk/crosscap tadpole.

\begin{figure}[!ht]
    \centering
    \includegraphics[scale=0.2]{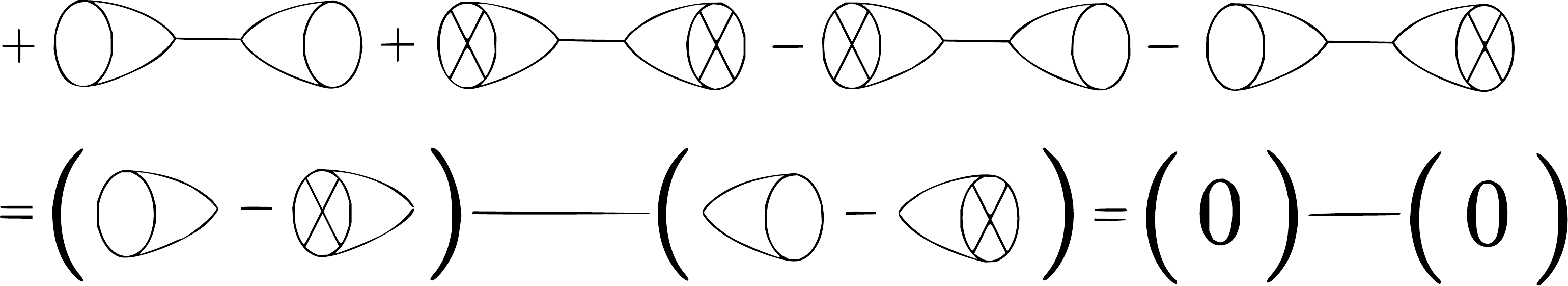}
    \caption{Degeneration of the type of \cref{fig:cylinder} for type I string theory: the 4 contributions cancel because of the relative signs of the disk/crosscap tadpoles.}
    \label{fig:type_I_tadpole}
\end{figure}

For type I and for the Sugimoto model, non-vanishing tadpoles are present for both the disk and crosscap geometries, and the two are equal, but the supersymmetric projection that leads to type I implements in an elegant way the supersymmetry protections of \cref{sec:susy_protections}. Indeed, reading the annulus and M\"obius strip amplitudes from \cref{eq:vacuum_ampl_type_I} in the transverse channel, one realizes that the signs and coefficients of $V_8$ are such that the contributions of the dilaton tadpoles to the dangerous degenerations are as in \cref{fig:type_I_tadpole}, and the net result vanishes because of the difference between the two tadpoles.

\begin{figure}[!ht]
    \centering
    \includegraphics[scale=0.2]{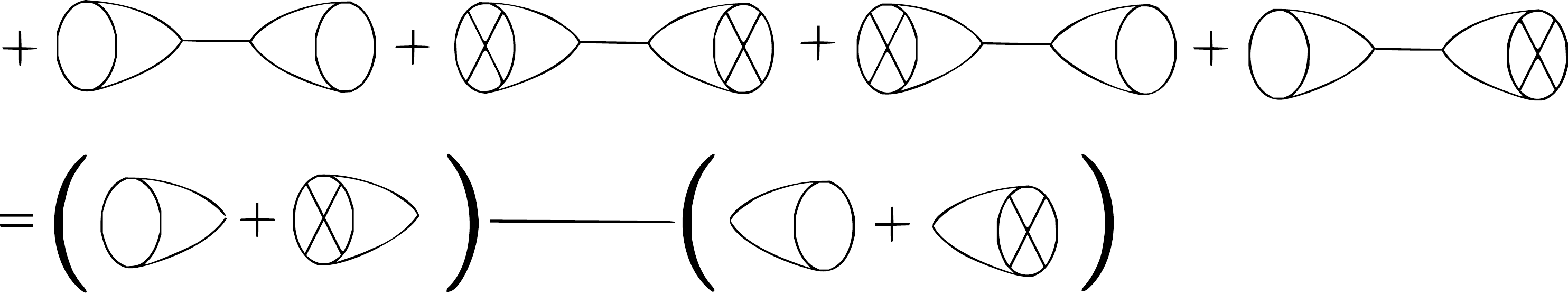}
    \caption{Degeneration of the type of \cref{fig:cylinder} for the Sugimoto string: the 4 contributions do not cancel and lead to a divergent one-loop vacuum amplitude.}
    \label{fig:sugimoto_tadpole}
\end{figure}

On the other hand, for the Sugimoto model, the amplitudes of \cref{eq:vacuum_ampl_Sugimoto} with the sign change in the M\"obius amplitude correspond to the GSO projection as in \cref{fig:sugimoto_tadpole}. The relative signs lead to the effective addition of tadpoles, thus generating the IR divergence of \cref{sec:tadpole_div} in the one-loop vacuum amplitude.
Then, the Fischler-Susskind mechanism for open string theories removes these half-loop tadpoles by changing the tree-level equations of motion in the same way as for closed strings.

From \cref{eq:Sugimoto_divergence}, one can extract the half-loop tadpole $T_{\frac{1}{2}}$, and the exponentiation required from \cref{eq:effective_action} leads to the open-string version of the tadpole potential of \cref{eq:tadpole_potential_generic}:
\begin{eq}\label{eq:tadpole_potential_open}
    V(\phi)=\sum_{2g=1}^\infty T_g e^{2(g-1)\phi}\sim T_{\frac{1}{2}}e^{-\phi} + T_1 + T_{\frac{3}{2}}e^{\phi} + \dots \mperiod
\end{eq}
In principle, the tadpole coefficient $T_1$ should be visible in the one-loop vacuum amplitude; however, the Fischler-Susskind mechanism does not specify any prescription for subtracting $T_{\frac{1}{2}}$ to obtain a finite $T_1$.
Naively, one can consider \cref{eq:Sugimoto_divergence}, which can be conveniently expressed in terms of $\eta$ functions as
\begin{eq}\label{eq:usp_32_1_loop}
    \mathcal{Z}_{1,\text{USp}(32)}=\int_0^\infty dt \frac{2^8}{t^6} 
    \frac{\left[\eta\left(i \frac{t}{2}\right) \eta\left(2 i t \right)\right]^{16}}{\left[\eta\left(i t \right)\right]^{40}}\mcomma
\end{eq}
and subtract the contribution of the massless genus-$\frac{1}{2}$ tadpole. Removing the $\frac{1}{t^2}$ term from the integrand of \cref{eq:usp_32_1_loop}, which corresponds to the transverse-channel logarithmic $q$-divergence of \cref{fig:sugimoto_tadpole}, yields a finite negative result:
\begin{eq}
    \mathcal{Z}_{1,\text{USp}(32)}^{\text{finite}}\sim -268 \mperiod
\end{eq}
Nevertheless, it is unclear whether this can be read as the true $T_1$ because different subtraction schemes lead to different results, and the current understanding of the Fischler-Susskind mechanism is not sufficient to determine the appropriate subtraction procedure. 
If this were the finite one-loop tadpole contribution to the effective action, $T_{\frac{1}{2}}$ and $T_1$ would be positive and of the same order ($T_1\sim \frac{1}{3} T_{\frac{1}{2}}$); therefore, the one-loop tadpole would not change the qualitative features of the scalar potential.

Note that, in general, closed-string contact interactions on the D9 branes also contribute to the one-loop vacuum amplitude, but the arguments of~\cite{Dudas:2004nd,Kitazawa:2008hv} suggest that they vanish for the Sugimoto string.

\subsection{Effective actions of non-supersymmetric strings}
\label{sec:action_non_susy_strings}

The analysis in the previous sections leads to the low-energy effective actions of the three non-supersymmetric strings of \cref{sec:non_susy_strings}. 
In general, without protecting principles such as supersymmetry, one expects a double expansion in $g_s$ and $\alpha'$, which, at the two-derivative level and at the lowest order in $\alpha'$, takes the form
\begin{eq} \label{eq:general_action_nosusy}
    S  = \frac{2\pi}{(2\pi\sqrt{\alpha'})^8}\int d^{10}x \, \sqrt{g} \sum_{i=0}^\infty \Bigg\{ e^{(-2+i)\phi} \left[ c_{R,i}R+4 c_{\phi,i}(\d\phi)^2 \right] - e^{(\gamma_s+i)\phi}c_{T,i}& \\
     -\sum_p \frac{c_{F_{p+2},i}}{2}e^{(-2\beta_{p,s}+i)\phi}\frac{F_{p+2}^2}{(p+2)!}-\frac{c_{g,i}}{2}e^{(-2\beta_{g,s}+i)\phi} \frac{\tr F_{2}^2}{2}\Bigg\} \mcomma &
\end{eq}
in the string frame, where $c_{R,i}$, $c_{\phi,i}$, $c_{T,i}$, $c_{F_{p+2},i}$ and $c_{g,i}$ come from the string-loop expansion, and where $\beta_{p,s}$ and $\beta_{g,s}$ are the lowest-order dilaton dressings of the form fields and the gauge fields. The dilaton tadpole potential, whose exponents start with $\gamma_s$, is built with the coefficients $c_{T,i}$, which correspond to the $T_g$ of \cref{eq:tadpole_potential_generic}.
Solving the equations of motion from \cref{eq:general_action_nosusy}, even with explicit expressions for the coefficients of the loop corrections, would require considerable effort. 
In the following sections, I focus on the leading orders in all the string-loop expansions, and in particular, on the leading term of the tadpole potential. 
The relevant string-frame action becomes
\begin{eq} \label{eq:action_nosusy}
    S  = \frac{2\pi}{(2\pi\sqrt{\alpha'})^8}\int d^{10}x \, \sqrt{g}\Bigg\{ e^{-2\phi} \left[ R+4(\d\phi)^2 \right] -e^{-2\beta_{g,s} \phi} \frac{\tr F_{2}^2}{4}-  T e^{\gamma_s\phi} & \\
     -\sum_p e^{-2\beta_{p,s} \phi}\frac{F_{p+2}^2}{2(p+2)!}\Bigg\} \mperiod &
\end{eq}

For the heterotic SO$(16)$$\times$SO$(16)$ string, the tadpole exponent is
\begin{eq}
    \gamma_s=0\mcomma 
\end{eq}
from its one-loop origin.
In a democratic-like formulation, the dilaton dressings of $F_3$ and $F_7$ are $\beta_{1,s}=1$ and $\beta_{5,s}=-1$ , while the gauge field strength has $\beta_{g,s}=1$.
On the other hand, for the two orientifolds, the tadpole exponent is
\begin{eq}
    \gamma_s=-1\mcomma
\end{eq}
from its disk/crosscap origin. Adopting a similar democratic-like formulation, the Sugimoto model has form fields with $p=1,5$, whereas type 0'B has $p=-1,1,3,5,7$, and in all cases $\beta_{p,s}=0$ since they are R-R fields. The gauge field strength has $\beta_{g,s}=\frac{1}{2}$ from its open-string origin.

The Einstein-frame version of the action in \cref{eq:action_nosusy} is
\begin{eq} \label{eq:action_nosusy_einstein_frame}
    S  = \frac{2\pi}{(2\pi\sqrt{\alpha'})^8}\int d^{10}x \, \sqrt{g}\Bigg[   R-\frac{1}{2}(\d\phi)^2  -e^{-2\beta_{g} \phi} \frac{\tr F_{2}^2}{4}-  T e^{\gamma \phi} & \\
     -\sum_p e^{-2\beta_{p} \phi}\frac{F_{p+2}^2}{2(p+2)!}\Bigg] \mcomma &
\end{eq}
where, for the heterotic model,
\begin{eq}\label{eq:heterotic_beta_p}
    \gamma=\frac{5}{2}\mcomma \qquad \beta_g=\frac{1}{4}\mcomma \qquad \beta_p=\frac{3-p}{4}\mcomma
\end{eq}
for the relevant values of $p$, while for the two orientifolds
\begin{eq}\label{eq:orientifold_beta_p}
    \gamma=\frac{3}{2}\mcomma \qquad \beta_g=-\frac{1}{4}\mcomma \qquad \beta_p=\frac{p-3}{4}\mcomma
\end{eq}
for the relevant values of $p$.
Note the Einstein-frame convention, which differs from the usual convention but is physically equivalent to it, for which the dilaton zero mode is included in the conformal rescaling of the metric. 
This choice is more natural in the presence of tadpole potentials because the dilaton zero mode loses its role as the string coupling at infinity.

In fact, the runaway nature of the tadpole potentials accounts for much of the effort required to find vacua in these non-supersymmetric models. The impossibility of having flat Minkowski regions with constant dilaton---the usual vacuum in supersymmetric setups---immediately poses a fundamental question: what is the gravity description of empty spacetimes? 
This involves finding vacuum solutions for the three non-supersymmetric strings, thus making the Fischler-Susskind mechanism concrete, leading to the vanishing of all tadpoles in the shifted vacuum. This is the topic to which I now turn: the low-energy description of empty spacetimes.

\section{Vacuum solutions}
\label{sec:vacuum_solutions}

The minimal set of ingredients that can solve the equations of motion, turning off all fluxes, is given by the metric and the dilaton. After removing the common $\alpha'$ factor, the relevant string-frame action is
\begin{eq}
    S = \int d^{10}x \, \sqrt{g}\left[e^{-2\phi}\left(R+4(\d\phi)^2\right)-T e^{\gamma_s\phi}\right] \mcomma
\end{eq}
with $\gamma_s=-1$ or $0$. In the Einstein frame, this is a dilaton-gravity system with an exponential potential for the dilaton,
\begin{eq} \label{eq:DM_action}
    S = \int d^{10}x \, \sqrt{g}\left[R-\frac{1}{2}(\d\phi)^2-T e^{\gamma\phi}\right] \mcomma
\end{eq}
with $\gamma=\frac{3}{2}$ and $\frac{5}{2}$ for the two cases of interest; therefore it is not possible to find solutions with unbroken ten-dimensional isometries or with constant dilaton.
Guided by the principle that vacuum solutions must be the most symmetric ones, the authors of~\cite{Dudas:2000ff} proposed to use codimension-one solutions, which are backgrounds of the type
\begin{eq} \label[pluralequation]{eq:codim1_ansatz}
    ds^2 & = e^{2A(y)}\eta_{\mu\nu}dx^\mu dx^\nu + e^{2B(y)}dy^2 \mcomma \\
    \phi & = \phi(y)\mperiod
\end{eq}
The resulting gravity profiles for non-supersymmetric strings are known as the Dudas-Mourad vacua, and they will be the main focus of this section, which is based on~\cite{Raucci:2022jgw,Basile:2022ypo}.

\subsection{Free codimension-one solutions}
\label{sec:free_codim_1}

It is instructive to begin with the action without tadpole potential to gain familiarity with the codimension-one backgrounds, and because the resulting solutions will play a role later. The Einstein-frame action
\begin{eq}
    S = \int d^{10}x \, \sqrt{g}\left[R-\frac{1}{2}(\d\phi)^2 \right] \mcomma 
\end{eq}
with the codimension-one ansatz in \cref{eq:codim1_ansatz}, has both the trivial flat space solution and two classes of non-trivial singular solutions, which in the gauge $B=0$ read
\begin{eq} \label[pluralequation]{eq:free_codim_1}
    ds^2 = \left(y_0\pm 9y\right)^{\frac{2}{9}}\eta_{\mu\nu}dx^\mu dx^\nu+ dy^2 \mcomma \qquad \phi=\phi_0\pm\frac{4}{3}\log\left(y_0\pm 9y\right)\mperiod
\end{eq}
The double sign that accompanies the variable $y$ means that both choices are viable, and the additional sign in $\phi$ is an independent sign ambiguity that describes two different behaviors for the backgrounds. These are singular solutions: for instance, by using the freedom to shift and reflect $y$ so that $y_0=0$ and $y>0$, the scalar curvature diverges as the inverse of the squared distance from the origin. As $y\to\infty$, the scalar curvature decreases, and the two sign choices for $\phi$ lead to two different behaviors: one with a diverging $e^\phi$ and the other with $e^\phi\to 0$.

The spacelike nature of the $y$ coordinate, or equivalently the timelike nature of the singularity, is consistent with the interpretation of these backgrounds as domain walls. 
To investigate this possibility, one can either add the Gibbons-Hawking-York boundary term and compute the junction conditions or, equivalently, introduce a localized term in the action that represents the 8-brane domain wall. The presence of the dilaton allows for a coupling $s(\phi)$ on this localized source; thus, in the Einstein frame the 8-brane action can be written as
\begin{eq}
    -\int d^9 x \, \sqrt{\gamma} s(\phi) \mcomma
\end{eq}
which in the string frame would be $s(\phi)e^{-\frac{9}{4}\phi}$. Inspired by string theory, I consider only exponential couplings to the dilaton in the following.
The presence of this localized term makes the first derivatives of the metric and the dilaton discontinuous, and the equations of motion, or equivalently the matching conditions, with a source at $y=0$ are compatible with two different types of sources. In the Einstein frame, these are
\begin{eq}
    s(\phi)\propto e^{\mp\frac{3}{4}\phi} \mcomma
\end{eq}
which translate into the string-frame couplings
\begin{eq}
    e^{-3\phi} \mcomma \qquad e^{-\frac{3}{2}\phi}\mperiod
\end{eq}
This result is not promising for a string theory interpretation because no source in perturbative string theory can lead to such couplings. The full solution with the source at $y=0$ is then~\cite{Raucci:2022jgw}
\begin{eq} \label[pluralequation]{eq:free_codim_1_DW}
    A=\frac{1}{9}\log(y_0\pm9|y|) \mcomma \quad \phi=\phi_0\pm\frac{4}{3}\log(y_0\pm9|y|) \mcomma \quad s(\phi)=\mp32 e^{\mp\frac{3}{4}(\phi-\phi_0)}\mcomma
\end{eq}
where $y\in\R$ and where there are two independent sign choices: one that enters the argument of the logarithm and the sign of the tension and the other that controls the sign ambiguity in the dilaton and the exponential coupling of the source.
Note the peculiar $\phi_0$ dependence. As I stressed after \cref{eq:orientifold_beta_p}, the present convention for the Einstein frame is a Weyl rescaling of the complete dilaton; therefore, the source $s(\phi)$ in \cref{eq:free_codim_1_DW} implies that the tension depends on $\phi_0$, which is not an expected feature in string theory. 
This is not necessarily an issue, precisely because $\phi_0$ is not the asymptotic value of the dilaton, but it remains unclear how $s(\phi)$ should be interpreted physically. This puzzling feature will reappear with non-supersymmetric strings.

Other related codimension-one backgrounds exist, one of which solves the Euclidean equations of motion with a curved internal manifold~\cite{Raucci:2022jgw}, and another, which is codimension-one in time, providing a singular time-dependent background. This last case is the double analytic continuation of \cref{eq:free_codim_1} and it has a counterpart in non-supersymmetric strings. 

\subsection{Dudas-Mourad solutions}
\label{sec:DM}

The inclusion of tadpole potentials as in \cref{eq:DM_action} leads to similar codimension-one backgrounds, which were introduced in~\cite{Dudas:2000ff} in the context of non-supersymmetric strings, to which I now turn.
Consider the background ansatz of \cref{eq:codim1_ansatz} in the gauge 
\begin{eq}
    B=-\frac{1}{2}\gamma\phi \mperiod
\end{eq}
The equations can be expressed in terms of a single function
\begin{eq}
    f(y)=\log\left(\sqrt{1+\frac{72 (A')^2}{T}}+6\sqrt{\frac{2}{T}}A'\right)\mcomma
\end{eq}
which determines $A$ and $\phi$ from
\begin{eq}
    A'=\frac{1}{6}\sqrt{\frac{T}{2}}\sinh f \mcomma \qquad \phi'=\pm \sqrt{2T}\cosh f \mperiod
\end{eq}
The remaining equation for $f$ depends on the exponent $\gamma$ of the tadpole potential.
\begin{itemize}
    \item[-] For the orientifold models with $\gamma=\frac{3}{2}$, the equation for $f$ is given by
    \begin{eq}
        4f'+6 \sqrt{\frac{T}{2}}\cosh f \pm 6\sqrt{\frac{T}{2}}\sinh f = 0\mcomma
    \end{eq}
    and its solution, up to an additive constant in $y$, is
    \begin{eq}
        f=\mp\log\left(\pm\frac{3}{2}\sqrt{\frac{T}{2}}y\right)\mperiod
    \end{eq}
    The complete background is then, selecting the range $y>0$,
    \begin{eq} \label[pluralequation]{eq:DM_orientifold}
        ds^2 & = \left(\sqrt{\frac{T}{2}}y\right)^{\frac{1}{9}}e^{-\frac{T}{16}y^2} \eta_{\mu\nu}dx^\mu dx^\nu + e^{-\frac{3}{2}\phi_0}\left(\sqrt{\frac{T}{2}}y\right)^{-1}e^{-\frac{9T}{16}y^2} dy^2\mcomma \\
        e^\phi & = e^{\phi_0}\left(\sqrt{\frac{T}{2}}y\right)^{\frac{2}{3}}e^{\frac{3T}{8}y^2} \mperiod
    \end{eq}
    $y=0$ and $y\to\infty$, separated by a finite proper distance, host two timelike curvature singularities, while $e^\phi$ vanishes at one end and diverges at the other.
    \item[-] For the heterotic case with $\gamma=\frac{5}{2}$, the equation for $f$ is
    \begin{eq} \label{eq:f_eq_het}
    2f'+3\sqrt{\frac{T}{2}}\cosh f \pm 5\sqrt{\frac{T}{2}}\sinh f = 0\mperiod
    \end{eq}
    There are three possible solutions. The first is trivial, with constant $f$, and has the same gravitational profile as the bulk of the D8 brane solutions in massive type IIA. In fact, at the lowest order in $\alpha'$, and only for the dilaton-gravity system, the Romans' mass enters the equations of motion in the same way as a one-loop vacuum energy. Physically, since the heterotic model contains no D8-like sources, this solution is probably only an artifact of the lowest order terms in the effective action.
    The remaining two solutions of \cref{eq:f_eq_het} are given by
    \begin{eq}
        e^f = \pm 2^{\mp1}\frac{e^{\sqrt{\frac{T}{2}}y}+\epsilon e^{-\sqrt{\frac{T}{2}}y}}{e^{\sqrt{\frac{T}{2}}y}-\epsilon e^{-\sqrt{\frac{T}{2}}y}} \mcomma
    \end{eq}
    with $\epsilon$ a real constant. The sign ambiguity can be reabsorbed by an $\epsilon\to-\epsilon$ transformation, and therefore it suffices to consider the upper signs. In addition, bulk solutions are only affected by the sign of $\epsilon$, so that only two classes of bulk solutions are present. The first class, with $\epsilon=1$, leads to the background
    \begin{eq} \label[pluralequation]{eq:DM_het}
        ds^2 & = \left(\sinh\sqrt{\frac{T}{2}}y\right)^{\frac{1}{12}}\left(\cosh\sqrt{\frac{T}{2}}y\right)^{-\frac{1}{3}}\eta_{\mu\nu}dx^\mu dx^\nu \\
        &+ e^{-\frac{5}{2}\phi_0}\left(\sinh\sqrt{\frac{T}{2}}y\right)^{-\frac{5}{4}}\left(\cosh\sqrt{\frac{T}{2}}y\right)^{-5 } dy^2 \mcomma \\
        e^\phi & = e^{\phi_0} \left(\sinh\sqrt{\frac{T}{2}}y\right)^{\frac{1}{2}}\left(\cosh\sqrt{\frac{T}{2}}y\right)^{2} \mcomma
    \end{eq}
    with two curvature singularities at $y=0$ and $y\to\infty$, whose proper distance is finite.
    The other case with $\epsilon=-1$ corresponds to \cref{eq:DM_het} in which one exchanges $\cosh\leftrightarrow\sinh$. With this choice, the proper length of the $y$ direction and the resulting nine-dimensional Planck mass are infinite; therefore, this solution is less interesting than \cref{eq:DM_het} as a gravitational background. 
\end{itemize}
The two codimension-one profiles of \cref{eq:DM_orientifold} and \cref{eq:DM_het} are known as the Dudas-Mourad solutions and are used throughout this thesis. 
They are singular solutions, and they can be part of string theory only in specific regimes, in which the $\alpha'$ and the string loop corrections of \cref{eq:general_action_nosusy} are subleading.
Their ultimate fate in string theory is unclear, and in particular, there is no worldsheet understanding of these types of spacetimes that can resolve the singularities.
The approach that I adopt here is to trust the results of the effective action guided by the considerations of \cref{sec:FS_mechanism}, deferring a more profound understanding of these solutions to future investigations and only noting a peculiarity of the backgrounds. 
This involves the asymptotic behavior~\cite{Mourad:2023wjg}: both endpoints of the orientifold solution, as $y\to0$ and $y\to\infty$, and the weakly coupled endpoint of the heterotic solution as $y\to 0$, reduce to the tadpole-free codimension-one solution in \cref{eq:free_codim_1}.
On the other hand, the strongly coupled heterotic endpoint approaches a different background, schematically, of the form
\begin{eq} \label[pluralequation]{eq:DM_other_asympt}
    ds^2 \sim z^{\frac{2}{25}}\eta_{\mu\nu} dx^\mu dx^\nu + dz^2 \mcomma \qquad e^\phi \sim z^{-\frac{4}{5}} \mcomma
\end{eq}
which is the same limiting behavior that emerges in type IIA string theory in the presence of D8 branes with non-vanishing local R-R tadpole, near the singularity that closes spacetime at a finite proper distance. These comments will be relevant in \cref{sec:brane_isometries}.

One can understand why this is happening by writing the equations of motion in a different gauge. Consider \cref{eq:codim1_ansatz} with $B=9A$, and let
\begin{eq}
    W=9 A +\frac{\gamma}{2}\phi \mperiod
\end{eq}
The field $W$ characterizes the contribution of the tadpole potential in the equations of motion: $T$ always appears together with the exponential $e^{2W}$.
The equation of motion for $W$ is 
\begin{eq} \label{eq:DM_W_eq}
    W''=\frac{T}{2}\left(\gamma^2-\frac{9}{4}\right) e^{2W} \mperiod
\end{eq}
This differential equation appears several times in \cref{sec:brane_isometries}, and the reader can consult one of the appendices of~\cite{Mourad:2021qwf,Mourad:2021roa} for a detailed discussion of its solutions. The only information that I need here is that only two asymptotic behaviors for $W$ are possible: it can become a linear function of $y$ as $y\to\pm\infty$, or have a logarithmic divergence at some finite value of $y$, which can be turned into $W\sim - \log y$ as $y\to 0$ by a $y$-translation.
From \cref{eq:DM_W_eq}, for the orientifold $\gamma=\frac{3}{2}$ case, $W$ is a globally linear function; therefore, its asymptotics are necessarily of the linear type. However, linear asymptotics are also solutions of the complete tadpole-free equations of motion, and therefore when $\gamma=\frac{3}{2}$ the solution is forced to end in the same way as in \cref{eq:free_codim_1}.
On the other hand, for the one-loop $\gamma=\frac{5}{2}$ tadpole potential, it is convenient to use $W$ and the other combination
\begin{eq}
    K=8\gamma A+\phi \mperiod
\end{eq}
The equations of motion are then \cref{eq:DM_W_eq}, which is equivalent to 
\begin{eq}
    (W')^2 = E + 2 T e^{2W}\mcomma 
\end{eq}
where $E$ is a positive energy-like quantity, and 
\begin{eq}
    (K')^2 = \frac{16}{9}E \mperiod
\end{eq}
In this case, the logarithmic divergence of $W$ is an allowed possibility, which physically corresponds to a tadpole-dominated collapse. In fact, given
\begin{eq}
    W\sim - \log\left(\sqrt{2T} y\right) \mcomma
\end{eq}
one obtains the background
\begin{eq}
    ds^2 \sim \left(\sqrt{2T} y\right)^{\frac{1}{8}} \eta_{\mu\nu}dx^\mu dx^\nu +\left(\sqrt{2T} y\right)^{\frac{9}{8}} dy^2 \mcomma \qquad e^\phi \sim  \left(\sqrt{2T} y\right)^{-\frac{5}{4}} \mcomma
\end{eq}
with $y\to0$, which is \cref{eq:DM_other_asympt} in different coordinates.

These results are related to the topics of \cref{sec:brane_isometries}, where the notation in terms of $W$ and $K$ is clarified and these asymptotics become elements of wider classes of solutions. 
For the Dudas-Mourad case, the conclusion is that the tadpole drives the background to the collapse, but the endpoints of the collapse itself are, in all but one instance, described by tadpole-free solutions.
This has a series of interesting consequences. 
If one trusts the two-derivative action until the collapse, all cases with the asymptotic behavior as in \cref{eq:free_codim_1} close at finite proper distance in the same way, which means that if the free codimension-one solution describes a domain wall, or an end-of-the-world brane, the same domain wall is expected in the Dudas-Mourad case. This is related to a recent line of research known as dynamical cobordism~\cite{McNamara:2019rup,Antonelli:2019nar,Buratti:2021yia,Buratti:2021fiv,Angius:2022aeq,Blumenhagen:2022mqw,Angius:2022mgh,Blumenhagen:2023abk,Angius:2023xtu,Huertas:2023syg,Angius:2023uqk,Angius:2024zjv}.
Perhaps even more interestingly, connecting with the initial issues of having a worldsheet description of the codimension-one profiles, one can argue that if the solutions of \cref{eq:free_codim_1} and of \cref{eq:DM_other_asympt} uplift to complete backgrounds in the UV, then one can believe in the Dudas-Mourad solutions as gravitational backgrounds for non-supersymmetric strings.
In practice, this involves gluing the interior of the Dudas-Mourad backgrounds, which can be made arbitrarily weakly coupled and weakly curved, to the UV complete asymptotic behaviors that replace \cref{eq:free_codim_1} and \cref{eq:DM_other_asympt} once $\alpha'$ and $g_s$ corrections are included.
Nevertheless, even from a purely classical perspective, the existence of only two types of collapses is an interesting remark.

Two straightforward generalizations are possible. One can consider an arbitrary number of dimensions, which is physically relevant for non-critical strings or for compactifications of non-supersymmetric strings, and allow for a generic tadpole exponent $\gamma$. These cases are not qualitatively different from the two ten-dimensional cases that I reviewed, and the reader can find all the details in~\cite{Basile:2022ypo}. Another natural generalization is to consider a nine-dimensional Ricci-flat manifold instead of nine-dimensional Minkowski. From the point of view of the equations of motion, any Ricci-flat manifold solves the leading order equations, which is of primary importance in \cref{sec:branes}. However, Ricci-flat manifolds are generically equipped with massless moduli that mix with the dilaton and develop runaway potentials akin to the ten-dimensional ones, see for instance~\cite{Dudas:2010gi}. This makes the story more intricate and requires moduli stabilization mechanisms.

Notably, solutions where curved Einstein manifolds replace Ricci-flat ones are unknown, except for a single case with time dependence~\cite{Raucci:2022jgw}, which is a special case of a class of scaling solutions. This cosmological background balances the tadpole with internal curvature, is of the codimension-one type, and takes the form 
\begin{eq}
    ds^2 = - e^{2B(t)} dt^2 + e^{-\gamma\phi(t)} ds_{(9)}^2 \mcomma \qquad \phi=\phi(t) \mcomma
\end{eq}
with a nine-dimensional Einstein manifold of negative constant curvature,
\begin{eq}
    R_{mn} = - \left(\gamma^2 -\frac{1}{4}\right) \frac{T}{2}  g_{mn}\mperiod
\end{eq}
The remaining equation of motion,
\begin{eq}
    (\phi')^2 = \frac{T}{4} e^{\gamma\phi+2B} \mcomma
\end{eq}
allows the metric to be expressed in terms of the dilaton as
\begin{eq}
    ds^2 = e^{-\gamma\phi}\left[-\frac{4}{T}(\phi')^2 dt^2 + ds_{(9)}^2\right] \mperiod
\end{eq}
In the gauge $B=0$, this time-dependent background becomes
\begin{eq}
    ds^2 = -dt^2 + \left(t_0\pm\frac{\gamma \sqrt{T}}{4} t\right)^2 ds_{(9)}^2 \mcomma \qquad e^\phi = \left(t_0\pm\frac{\gamma \sqrt{T}}{4} t\right)^{-\frac{2}{\gamma}} \mcomma
\end{eq}
where an arbitrary integration constant can be added to the time variable, as in the spatial codimension-one case.

Before concluding this section, I want to emphasize the importance of the codimension-one solutions from the perspective of realizing the Fischler-Susskind mechanism. From \cref{sec:FS}, perturbative string theory requires solving the modified equations of motion; therefore, the outcomes must be considered seriously. I can think of four possible interpretations for the codimension-one Dudas-Mourad backgrounds and their generalizations.
\begin{enumerate}
    \item They could be non-physical, which means that they are not part of complete spacetime backgrounds of non-supersymmetric strings. In practice, this would manifest as the absence of a string theory completion of the asymptotics. As a result, one should look for alternative options to take the tadpole into account and possibly add more ingredients. 
    \item If the solutions are physical, they could represent domain walls, as originally argued in~\cite{Antonelli:2019nar}. As I briefly mentioned, this possibility raises several open questions because of the unusual 8-brane source that is required by the matching conditions. In~\cite{Raucci:2022jgw}, I listed some of the possibilities, but in all cases, one has either a $\phi_0$-dependent tension, where $\phi_0$ is the integration constant that enters the dilaton profile, or a non-stringy dilaton coupling on the brane worldvolume. This indicates the absence of a perturbative worldsheet understanding of the domain walls and therefore of the full gravity solution, which resonates with the exotic branes of~\cite{Bergshoeff:2011zk,Bergshoeff:2011qk,Bergshoeff:2012ex}. In principle, the dynamics of the domain walls may alter the stability of the system.
    \item Alternatively, the Dudas-Mourad backgrounds may represent fully static vacuum solutions, indicating that the ten-dimensional spacetime dynamically compactifies to nine dimensions. This was the original motivation of~\cite{Dudas:2000ff}, which was further strengthened by~\cite{Basile:2018irz} where the authors proved that the codimension-one spacelike solution is perturbatively stable, making the Dudas-Mourad solutions natural candidates to realize the Fischler-Susskind mechanism. This interpretation is not inconsistent with the previous one, provided that the domain walls are stable and that the dimensional reduction takes their presence into account (see the discussion on singularities in \cref{sec:fake_susy} and \cref{sec:branes}).
    \item Perhaps the most interesting possibility is to view these backgrounds as time-dependent vacua. This interpretation matches the physical expectation that the absence of spacetime supersymmetry leads to dynamics. Supersymmetry breaking would provide a natural explanation for time evolution, which is of utmost importance when engineering cosmological models in string theory. Rephrasing my provocative introductory statement, this would teach us that apples can also fall in string theory.
\end{enumerate}

\subsection{The spontaneous compactification}
\label{sec:spont_cpt}

I now interpret these backgrounds as static vacuum solutions with or without (stable) domain walls to better understand the physical implications of the absence of spacetime supersymmetry. The following analysis is a summary of the content of~\cite{Basile:2022ypo}, in which we generalized the results to arbitrary dimensions and tadpole exponents.

The Fischler-Susskind mechanism does not directly compute the spectrum in the shifted vacuum, where some of the tree-level massless modes may become gapped, for example, because of boundary conditions.
As I stressed at the beginning of this chapter, one must follow the effective action because it encodes the all-string-loop physics of the massless modes, although only in regimes where string perturbation theory can be trusted.
Therefore, in the Dudas-Mourad static vacua, one must study the effective nine-dimensional theory that remains after the spontaneous compactification, and extract which massless fields are dynamical in the nine-dimensional EFT.
In the following analysis I mostly focus on zero modes in the internal direction. This is the natural expectation for the massless nine-dimensional sector, although it is not the most general possibility because non-trivial wavefunctions in the internal Dudas-Mourad interval can combine with the warped metric to yield massless nine-dimensional modes. 
Allowing for generic internal dependence leads to more involved equations; here, I only comment on the dilaton case, referring to~\cite{Mourad:2023wjg} for a careful treatment of the boundary conditions. I shall return to the topic in \cref{sec:branes}, where I discuss the cases of form fields in detail.

Gravity remains dynamical in nine-dimensions for both Dudas-Mourad solutions, \cref{eq:DM_orientifold} and \cref{eq:DM_het}, as one can verify by computing the gravitational constant,
\begin{eq}\label{eq:DM_kappa9}
    \frac{1}{2 \kappa_9^2}= \frac{2\pi}{(2\pi\sqrt{\alpha'})^8}\int dy \,  e^{7A+B} \mcomma
\end{eq}
which is finite. Note that the integral scales with $\phi_0$ as
\begin{eq}
    \frac{1}{\kappa_9^2}\propto e^{-\frac{\gamma}{2}\phi_0} \mcomma
\end{eq}
the same $\phi_0$ dependence as the proper length of the compactification interval. As a result, $\phi_0$ can be traded for the nine-dimensional gravitational constant in all nine-dimensional expressions. As I have already stressed, this analysis ignores potential issues with boundary conditions, which were understood in~\cite{Mourad:2023wjg} in terms of self-adjoint Schr\"odinger systems. The final result is that there is a unique boundary condition for the graviton that is compatible with stability, with a massless graviton as the zero mode, thus substantiating the above analysis.

A similar investigation for the dilaton yields a surprising result: the nine-dimensional dilaton is gapped. A hint comes from the inconsistency of naively promoting the constant mode $\phi_0$ to a nine-dimensional field, because of the $\mu y$ metric equation~\cite{Basile:2022ypo}. 
In general, to prove that there is no massless nine-dimensional dilaton, one must consider all scalar perturbations to the background because they are generically mixed. Therefore, it is convenient to use the language of~\cite{Basile:2018irz}, where the authors studied the perturbative stability of the Dudas-Mourad vacua, and of~\cite{Basile:2022ypo}, where we generalized the results to arbitrary dimensions and tadpole exponents.
Scalar perturbations of the metric and the dilaton take the following form:
\begin{eq} \label[pluralequation]{eq:DM_conformal_gauge}
    ds^2 &= e^{2\Omega(z)}\left(e^{2a(z,x)}\eta_{\mu\nu}dx^\mu dx^\nu + e^{2d(z,x)}dz^2\right) \mcomma \\
    \phi &= \phi(z)+\varphi(z,x) \mcomma
\end{eq}
where the background is written in the conformal gauge, in which $\Omega(z)$ and $\phi(z)$ satisfy
\begin{eq}
\Omega_{zz} + 8 \Omega_z^2 + \frac{T}{8} e^{2\Omega+\gamma\phi} &= 0 \mcomma \\
72\Omega_z^2- \frac{1}{2}\phi_z^2 + T   e^{2\Omega+\gamma\phi} &= 0 \mcomma \\
\phi_{zz} + 8\Omega_z\phi_z -  T\gamma  e^{2\Omega+\gamma\phi} &= 0 \mcomma
\end{eq}
and where $[ \ ]_z$ denotes a $z$-derivative.
The equations of motion imply that
\begin{eq}
    d & = - 7 a \mcomma\\
    \varphi & = -\frac{16}{\phi_z}\left(a_z+7\Omega_z a\right)\mcomma
\end{eq}
and one is left with a single equation for scalar perturbations
\begin{eq} \label{eq:DM_scalar_pert}
    \Box a +\left(8\Omega_z-2\frac{\phi_{zz}}{\phi_z}\right)a_z - \frac{7}{4}\left(1+8\gamma \frac{\Omega_z}{\phi_z}\right) T e^{2\Omega+\gamma\phi} a = 0 \mperiod
\end{eq}
Expanding $a$ in nine-dimensional momentum modes, with $p^2 = - m^2$, and letting
\begin{eq}
    a = \phi_z e^{- 4\Omega} \psi \mcomma
\end{eq}
\cref{eq:DM_scalar_pert} can be turned into the Schr\"odinger form
\begin{eq}
    \left({\cal A}^\dagger {\cal A}+ b\right)\psi = m^2 \psi \mcomma
\end{eq}
with 
\begin{eq}
    {\cal A} &= -\frac{d}{dz}+4\Omega_z-\frac{\phi_{zz}}{\phi_z} \mcomma \\
    b & = \frac{7}{4}\left(1+8\gamma \frac{\Omega_z}{\phi_z}\right) T e^{2\Omega+\gamma\phi}\mcomma
\end{eq}
which is formally positive for all relevant cases, with $\gamma=\frac{3}{2}$ and $\frac{5}{2}$. 
In fact, to prove that the Schr\"odinger operator is positive, one must consider its self-adjoint extension, which is required to work with a complete set of modes, and explore the existence of stable self–adjoint boundary conditions. This has been done in~\cite{Mourad:2023wjg}, where it was found that, consistently with the results of~\cite{Basile:2022ypo}, no massless scalars remain in the nine-dimensional EFT. 
Both~\cite{Basile:2022ypo} and~\cite{Mourad:2023wjg} estimated the mass of the first massive dilaton mode in nine-dimensions, in two different ways, leading to similar results: the mass is inversely proportional to the length of the Dudas-Mourad interval, as expected by a KK-like compactification.

The final type of tree-level massless fields are ten-dimensional form fields, including the gauge sector. 
In the orientifold models, assuming Einstein-frame dilaton dressings as in \cref{eq:action_nosusy_einstein_frame}, the nine-dimensional gauge couplings are given by
\begin{eq}
\frac{1}{g_{{\text{9D}},p}^2} = \frac{2\pi}{(2\pi\sqrt{\alpha'})^8} \int dy \, e^{(5-2p)A+B-2\beta_p\phi}\propto e^{-\left(\frac{\gamma}{2}+2\beta_p\right)\phi_0}\mperiod
\end{eq}
A necessary condition for the zero mode of a form field to be dynamical after the spontaneous compactification is that this integral must be finite, which translates into the following condition:
\begin{eq} \label{DM_9D_orient_ineq}
    -\frac{7-p}{12}<\beta_p < \frac{7-p}{12}\mperiod
\end{eq}
For the R-R fields of both orientifold models, $\beta_p$ is given in \cref{eq:orientifold_beta_p} and the inequality of \cref{DM_9D_orient_ineq} holds for $p<5$ and $p>1$, leaving $p=3$ as the only available option. 
Note that this case is subtle because the field strength is self-dual and the action in \cref{eq:action_nosusy_einstein_frame} is not enough to capture its physics. I shall comment on this in \cref{sec:branes}, also considering the existence of appropriate self-adjoint boundary conditions. 
For the gauge fields of both orientifold models, $\beta_g$ is again given in \cref{eq:orientifold_beta_p} and satisfies \cref{DM_9D_orient_ineq} with $p=0$; thus, zero modes from this sector are also allowed. 
In the heterotic SO$(16)$$\times$SO$(16)$ string, similar considerations lead to the condition
\begin{eq} \label{DM_9D_het_ineq}
    -\frac{15-p}{20}<\beta_p < \frac{7-p}{12}\mperiod
\end{eq}
For the Kalb-Ramond field and its dual, $\beta_p$ is given in \cref{eq:heterotic_beta_p} and the inequality of \cref{DM_9D_het_ineq} holds for $p<5$ and $p>1$, thus excluding both fields with $p=1,5$. For the gauge fields, $\beta_g$ is again given in \cref{eq:heterotic_beta_p} and satisfies \cref{DM_9D_het_ineq} with $p=0$; thus, the nine-dimensional EFT only contains the gauge sector.

To summarize the content of this section, for all non-supersymmetric strings, the lower-dimensional EFT after the spontaneous compactification of the Dudas-Mourad solution contains a massless graviton and massless gauge fields. For the type 0'B model, a five-form field strength can be present, although its self-duality nature raises subtleties. Similar results hold in different dimensions, and the reader can consult~\cite{Basile:2022ypo} for a complete presentation. In \cref{sec:branes}, I shall return to the topic from a complementary perspective and address cases with internal dependence.

\section{Fake supersymmetry with tadpole potentials}
\label{sec:fake_susy}

\Cref{sec:vacuum_solutions} rests on a metric ansatz of the codimension-one type, whose physical motivation is that empty spacetime must be the most symmetric background, preserving as many spacetime isometries as possible. 
Nevertheless, one expects several families of vacuum solutions, analogous to the Calabi-Yau compactifications of supersymmetric strings, providing lower-dimensional purely gravitational solutions. The issue here is that they are more difficult to engineer in the presence of runaway tadpole potentials.
A simple option would be to embed nine- or lower-dimensional Ricci-flat spaces in the Dudas-Mourad vacua, for instance, by adding an internal Calabi-Yau space. However, these are useful only as Ricci-flat Riemannian manifolds that solve the two-derivative equations of motion, thereby losing their attractive properties as complex manifolds and exact sigma models. 
The truly interesting question is whether there are non-trivial ten-dimensional solutions resembling the Calabi-Yau compactifications of supersymmetric strings, for which one can have a higher degree of control.
There is a precise reason why this question cannot be easily answered: unlike supersymmetric setups, we lack appropriate technology to classify all solutions. In the absence of complete worldsheet descriptions, one must follow the effective action without the assistance of any protecting principle, which means solving the complete second-order coupled differential equations of motion.  

In this section, based on~\cite{Raucci:2023xgx}, I outline a possible strategy for solving the equations of motion in a way that mimics supersymmetry as a solution-generating technique. 
In fact, supersymmetric vacua can be found by solving first-order differential equations that impose unbroken spacetime supersymmetry on the background ansatz under consideration. It is sufficient to solve these equations, supplemented with the Bianchi identities, to obtain a solution to the complete second-order equations of motion~\cite{Lust:2004ig,Gauntlett:2005ww,Lust:2008zd}, up to technical subtleties related to the annihilator of ten-dimensional Majorana-Weyl spinors. 
When only the metric field and the dilaton are present, the supersymmetry variations of the gravitino and the dilatino, which can be schematically written as
\begin{eq}
    \delta  \psi_M & = D_{M}^{\mbox{\scriptsize susy}}  \varepsilon=0\mcomma\\
    \delta  \lambda & =\mathcal{O}^{\mbox{\scriptsize susy}}  \varepsilon= 0\mcomma
\end{eq}
reduce to 
\begin{eq}
    \delta  \psi_M &  =\nabla_M  \varepsilon=0\mcomma\\
    \delta  \lambda & =  d\phi  \varepsilon=0\mperiod
\end{eq}
For non-supersymmetric strings, the strategy of~\cite{Raucci:2023xgx} is to find two conditions of the type
\begin{eq}\label[pluralequation]{eq:fake_susy_general}
    D_{M}  \varepsilon & =0 \mcomma \\
    \mathcal{O} \varepsilon & = 0 \mcomma
\end{eq}
with two appropriate operators $D_{M}$ and $\mathcal{O}$, such that \cref{eq:fake_susy_general} imply the ten-dimensional equations of motion. The expectation is that only particular types of solutions to the complete equations of motion can be recovered in terms of \cref{eq:fake_susy_general}, but for these special classes of solutions, first-order equations can be employed to engineer vacua systematically, as in the supersymmetric cases.
I focus on settings with only gravity and the dilaton, postponing some comments on the inclusion of form fields to \cref{sec:fluxes}. 

The string-frame equations that arise from \cref{eq:action_nosusy} are
\begin{eq} \label[pluralequation]{eq:eom_gravity_dilaton}
    R_{MN}+2 \nabla_M \nabla_N\phi + \frac{T}{2}\left(1+\frac{\gamma_s}{2}\right)e^{(2+\gamma_s)\phi}g_{MN} &=0 \mcomma \\
    R+4\nabla^2 \phi - 4 (\d\phi)^2 + \frac{T\gamma_s}{2}e^{(2+\gamma_s)\phi} & = 0\mperiod
\end{eq}
The most general form of the operators $D_{M}$ and $\mathcal{O}$, considering only one spinor $\varepsilon$, is
\begin{eq}
    \label[pluralequation]{eq:fake_susy_eq_variations}
    D_{M} \varepsilon & =\left(\nabla_M + \mathcal{W}(\phi) \Gamma_M\right)\varepsilon\mcomma \\
    \mathcal{O} \varepsilon&= \left(d\phi+g(\phi)\right)\varepsilon\mperiod
\end{eq}
Equations of this type are known in the literature as fake supersymmetry~\cite{Boucher:1984yx,Townsend:1984iu,Skenderis:1999mm,Freedman:2003ax,Townsend:2007aw,Trigiante:2012eb,Giri:2021eob}, and the function $\mathcal{W}(\phi)$ as the fake superpotential.
It is important to stress here that these equations are only a formal tool to obtain solutions to the equations of motion and that they carry no physics: they are not required to be variations of any field under any physical transformation, and in fact, the formalism applies regardless of the presence of gravitinos and dilatinos in the spectrum.

Following the procedure reviewed in~\cite{Giri:2021eob,Tomasiello:2022dwe}, the equations of motion of \cref{eq:eom_gravity_dilaton} can be reconstructed from combinations of the two operators $D_M$ and $\mathcal{O}$:
\begin{eq} \label{eq:string_gravity_eom}
    0&=2\bigg[\Gamma^M\comm{D_M}{D_N}+\comm{D_N}{\mathcal{O}}+(\mathcal{W}'-2\mathcal{W})\Gamma_N\mathcal{O}\bigg]\varepsilon\\
    &=\bigg[\Gamma^M (R_{MN}+2\nabla_M\nabla_N\phi)+(36\mathcal{W}^2+2 \mathcal{W}'g - 4 \mathcal{W} g)\Gamma_N + \\ & +(2g'-4\mathcal{W}-16\mathcal{W}')\nabla_N\phi\bigg]\varepsilon\mcomma
\end{eq}
where $\mathcal{W}'$ denotes the $\phi$-derivative of $\mathcal{W}$, and
\begin{eq} \label{eq:string_dilaton_eom}
    0&=\bigg[(D-\mathcal{O})^2  - (\nabla_M-2\d_M\phi)D^M+(g'-2g-9\mathcal{W}'+18\mathcal{W})\mathcal{O} \\
    & - (19\mathcal{W}-2g) D \bigg]\varepsilon \\
    & = \left[-\frac{1}{4}\left(R+4\nabla^2\phi-4(\d\phi)^2\right)+(-9\mathcal{W}'g + g g'+18 \mathcal{W} g - g^2 -90 \mathcal{W}^2)\right]\varepsilon\mcomma
\end{eq}
with $D=\Gamma^M D_M$.
\Cref{eq:eom_gravity_dilaton} are then recovered from \cref{eq:string_gravity_eom} and \cref{eq:string_dilaton_eom} provided
\begin{eq}
    \label[pluralequation]{eq:the_three_conditions}
    36\mathcal{W}^2+2\mathcal{W}'g-4\mathcal{W} g & = \frac{T}{2}\left(1+\frac{\gamma_s}{2}\right)e^{(2+\gamma_s)\phi}\mcomma \\
    g'-2\mathcal{W}-8\mathcal{W}' & = 0\mcomma\\
    -9\mathcal{W}'g+ g'g +18\mathcal{W} g -g^2-90\mathcal{W}^2 & = -\frac{T\gamma_s}{8}e^{(2+\gamma_s)\phi}\mperiod
\end{eq}
Letting
\begin{eq}
    h(\phi)\equiv g(\phi)-6\mathcal{W}(\phi)\mcomma \qquad k(\phi)=g(\phi)-12 \mathcal{W}(\phi)\mcomma
\end{eq}
the three conditions in \cref{eq:the_three_conditions} reduce to two independent equations
\begin{eq}\label[pluralequation]{eq:hk_system}
    hk&=-\frac{T}{4}e^{(2+\gamma_s)\phi} \mcomma\\
    h-k & =  2h'+k' \mperiod
\end{eq}
A sign ambiguity is present in \cref{eq:hk_system}: $(h,k)\to(-h,-k)$. For this reason, it is convenient to define two combinations, $A(\phi)$ and $B(\phi)$, given by
\begin{eq}
    A & = -hk \mcomma \\
    B & = -\frac{k}{2h} \mcomma
\end{eq}
which are positive for exponential scalar potentials and therefore can be solved with no sign ambiguity. Explicitly, \cref{eq:hk_system} now read
\begin{eq}\label[pluralequation]{eq:AB_system}
    A & = \frac{T}{4}e^{(2+\gamma_s)\phi} \mcomma \\
    \frac{B'}{B}+B' & = (1+\gamma_s)-(4+\gamma_s)B \mcomma
\end{eq}
thus reducing to a single differential equation for $B$. 

Ten-dimensional non-supersymmetric strings have $\gamma_s=-1$ or $\gamma_s=0$, while \cref{eq:AB_system} is more general and can also be applied to other scalar potentials. For instance, an Einsten-frame cosmological constant with $\gamma_s=-\frac{5}{2}$, which is not an available tadpole potential in perturbative string theory, leads to $B=B_0 e^{-\frac{3}{2}\phi}$ and a fake superpotential that is the sum of two exponentials. Another physically interesting value is $\gamma_s=-2$, which describes non-critical strings and corresponds to $B^{-1}=\sqrt{1+2 e^{2(\phi-c)}}-1$. The reader can find a complete analysis in~\cite{Raucci:2023xgx}, here I only focus on the two cases of interest.
\begin{itemize}
    \item[-] For the two orientifold theories, with $\gamma_s=-1$, one can change variables in the second of \cref{eq:AB_system}, writing it in terms of $e^{3\phi}$. The resulting differential equation corresponds to the one that defines (the inverse of) the Lambert $W$ function, which is implicitly defined as the $W(x)$ that solves $W(x)e^{W(x)}=x$. The final result is 
    \begin{eq}
        B=\left[W\left(e^{3(\phi-c)}\right)\right]^{-1}\mcomma
    \end{eq}
    where $c$ is an integration constant. In terms of $\mathcal{W}$ and $g$, this corresponds to 
    \begin{eq}\label[pluralequation]{eq:string_frame_fake_superpotential_gravity_scalar}
    \mathcal{W}(\phi) & = \frac{1}{6}\sqrt{\frac{T}{4}}e^{\frac{1}{2}\phi}\left[\sqrt{\frac{W(e^{3(\phi-c)})}{2}}+\sqrt{\frac{2}{W(e^{3(\phi-c)})}}\right]\mcomma \\
    g(\phi)& = \sqrt{\frac{T}{2}}e^{\frac{1}{2}\phi}\left[\sqrt{W(e^{3(\phi-c)})}+\sqrt{\frac{1}{W(e^{3(\phi-c)})}}\right]\mcomma        
    \end{eq}
    up to overall sign ambiguities.
    \item[-] In the heterotic case, with $\gamma_s=0$, a constant solution is present:
    \begin{eq}
        B=\frac{1}{4}\mcomma
    \end{eq}
    which recovers the supersymmetry equations of massive type IIA supergravity,
    \begin{eq}
    \mathcal{W}(\phi)&=\frac{1}{2}\sqrt{\frac{T}{8}}e^\phi \mcomma \\
    g(\phi)&=5\sqrt{\frac{T}{8}}e^\phi \mperiod
    \end{eq}
    This is expected because the Romans' mass has the same dilaton potential as a one-loop cosmological constant, as stressed in the previous sections. There is another non-trivial solution that can be defined implicitly as follows:
    \begin{eq} \label{eq:B_eq_implicit_sol}
        \log{B}-\frac{5}{4}\log\left|B-\frac{1}{4}\right|=\phi - c\mcomma
    \end{eq}
    where $c$ is an integration constant. Two possible fake superpotentials arise from this case: the first is a $\mathcal{W}$ with a single minimum, and the second increases monotonically from $0$ to infinity.
\end{itemize}

Similar results can be found in the Einstein frame. Using the same notation as in \cref{eq:fake_susy_eq_variations}, the metric and dilaton equation follow from the fake supersymmetry equations if 
\begin{eq}
    g=-32 \mathcal{W}'\mcomma
\end{eq}
and a single differential equation for $\mathcal{W}$ holds. Letting
\begin{eq}
    \mathcal{W} (\phi)=\frac{1}{12}\sqrt{\frac{T}{2}}e^{\frac{\gamma}{2}\phi}\sinh f(\phi)\mcomma 
\end{eq}
the remaining differential equation for $\mathcal{W}$ reads
\begin{eq}\label{eq:Einstein_frame_fake_W_diffeq}
    16 (\cosh{f})^2 (f')^2+16\gamma  f' \sinh{f} \cosh{f} + (4\gamma^2-9)(\sinh{f})^2=9 \mperiod
\end{eq}
\begin{itemize}
    \item[-] For the orientifold tadpole potential, $\gamma=\frac{3}{2}$, \cref{eq:Einstein_frame_fake_W_diffeq} becomes
    \begin{eq}
        \label{eq:lambda_eq}
    \left[\frac{1}{3}(1+e^{-2f})2f'+1\right]\left[\frac{1}{3}(1+e^{2f})2f'-1\right]=0\mperiod
    \end{eq}
    Up to an overall sign ambiguity, this reduces to only one of the two factors, and therefore to
    \begin{eq}
        \label{eq:lambda_sol}
        2f-e^{-2f}=-3\phi-3 c \Rightarrow 2f=W(e^{3(\phi-c)})-3(\phi-c)\mperiod
    \end{eq}
    Using the properties of the $W$ function, the final expression for the fake superpotential in the Einstein frame becomes
    \begin{eq}\label{eq:einstein_frame_fake_W}
    \mathcal{W}(\phi) = \frac{1}{24}\sqrt{\frac{T}{2}}e^{\frac{3}{4}\phi}\left[\left(W(e^{3(\phi-c)})\right)^{-\frac{1}{2}} - \left(W(e^{3(\phi-c)})\right)^{\frac{1}{2}} \right]\mcomma      
    \end{eq}
    where $c$ is an integration constant. This $\mathcal{W}$ is a monotonically decreasing function.
    \item[-] For the one-loop heterotic tadpole potential, $\gamma=\frac{5}{2}$, \cref{eq:Einstein_frame_fake_W_diffeq} has a constant solution corresponding to 
    \begin{eq}
        \mathcal{W}\propto e^{\frac{5}{4}\phi}\mcomma
    \end{eq}
    which is again the massive IIA case, together with a non-trivial solution that can be defined implicitly as
    \begin{eq}\label{eq:f_for_generic_beta}
        -3f+5\log\left|5 \sinh{f}+3\cosh{f}\right|=-4(\phi-c)\mperiod        
    \end{eq}
    This is the analog of \cref{eq:B_eq_implicit_sol}.
\end{itemize}

These results demonstrate that it is possible to find fake supersymmetry equations for non-supersymmetric strings, at least at the two-derivative level, by only turning on the metric field and the dilaton.
As I stressed, a natural use of this formalism would be to engineer vacuum solutions that replace the Calabi-Yau compactifications. This is the topic of \cref{sec:vacuum_sol_fake_susy}, but before that, it is interesting to consider other massless fields in the spectrum: form fields.

\subsection{Fluxes}
\label{sec:fluxes}

The inclusion of form fields is a subtle issue because the simplest possibilities do not work, as explained in detail in~\cite{Raucci:2023xgx}. In this section, I only outline the argument.

The type I supergravity equations for the variations of the gravitino and the dilatino in the democratic formalism are
\begin{eq}\label[pluralequation]{eq:susy_type_I}
    \delta\psi_M&=D_{M}^{\mbox{\scriptsize susy}}  \varepsilon=\left(\nabla_M+\frac{1}{16}e^\phi F \Gamma_M\right)\varepsilon\mcomma \\
    \delta\lambda & =\mathcal{O}^{\mbox{\scriptsize susy}}  \varepsilon= \left(d\phi+\frac{1}{16}e^\phi\Gamma^M F\Gamma_M \right)\varepsilon\mcomma    
\end{eq}
in the conventions of~\cite{Tomasiello:2022dwe}, and where $F=F_3+F_7$. The equations of motion,
\begin{eq}\label[pluralequation]{eq:equations_of_motion_type_I}
    R_{MN}+2\nabla_M\nabla_N\phi-\frac{1}{4}e^{2\phi}(F^2)_{MN} & = 0\mcomma \\
    R+4\nabla^2\phi-4(\d\phi)^2&=0\mcomma    
\end{eq}
follow from the combinations
\begin{eq} \label[pluralequation]{eq:type_I_susy_equations}
    0& = \left[\Gamma^M\comm{D^{\mbox{\scriptsize susy}}_M}{D^{\mbox{\scriptsize susy}}_N}+\comm{\nabla_N}{\mathcal{O}^{\mbox{\scriptsize susy}}}-\frac{1}{16}e^\phi \mathcal{O}^{\mbox{\scriptsize susy}}F\Gamma_N\right]\varepsilon \\
    & = \left[\frac{1}{2}(\mbox{Gravity EoM})_{MN}\Gamma^M +\frac{1}{16}e^\phi (dF) \Gamma_N \right]\varepsilon\mcomma
\end{eq}
and
\begin{eq}\label{eq:dilaton_eom_susy}
    0&=\left[(D^{\mbox{\scriptsize susy}}-\mathcal{O}^{\mbox{\scriptsize susy}}+\frac{1}{16}e^\phi F)(D^{\mbox{\scriptsize susy}}-\mathcal{O}^{\mbox{\scriptsize susy}})-(\nabla^M -2 \d^M\phi)D^{\mbox{\scriptsize susy}}_M\right]\varepsilon \\
    &=\left[-\frac{1}{4}(\mbox{Dilaton EoM})+\frac{1}{8}e^\phi (d F)\right]\varepsilon\mperiod
\end{eq}

For non-supersymmetric strings, and in general when a scalar potential $V(\phi)$ is present in the string frame, the equations of motion become
\begin{eq}\label[pluralequation]{eq:equations_of_motion_tadpole_and_fluxes}
    R_{MN}+2\nabla_M\nabla_N\phi+\frac{1}{2}e^{2\phi}\left(V+\frac{1}{2}V'\right) g_{MN}-\frac{1}{4}e^{2\phi}(F^2)_{MN} & = 0\mcomma \\
    R+4\nabla^2\phi-4(\d\phi)^2+\frac{1}{2}e^{2\phi}V'&=0\mperiod
\end{eq}
Two crucial features of \cref{eq:type_I_susy_equations} and \cref{eq:dilaton_eom_susy} seem problematic to accommodate with a scalar potential. First, the R-R quadratic terms in \cref{eq:type_I_susy_equations} emerge from certain combinations, $\Gamma^M \comm{D^{\mbox{\scriptsize susy}}_M}{D^{\mbox{\scriptsize susy}}_N}$ and $e^\phi {O}^{\mbox{\scriptsize susy}}F\Gamma_N$, and no other first-order susy-like system reproduces these quadratic terms. In other words, there are no known fake supersymmetry equations for type I supergravity, other than those of type I supergravity itself.
The second crucial feature is that obtaining the dilaton equation from \cref{eq:dilaton_eom_susy} rests on the presence of the same dilaton dressing in $D$ and $\mathcal{O}$. Unless the first-order equations are drastically changed, the only natural ansatz for the two fake supersymmetry equations is
\begin{eq}\label[pluralequation]{eq:proposal_fake_susy_forms}
    D_{M}\varepsilon & =\left(\nabla_M + \mathcal{W}(\phi)\Gamma_M+\frac{1}{16}e^\phi F \Gamma_M\right)\varepsilon\mcomma \\
    \mathcal{O}\varepsilon&= \left(d\phi+g(\phi)+\frac{1}{16}e^\phi \Gamma^M F \Gamma_M\right)\varepsilon\mperiod
\end{eq}
However, these do not imply the equations of motion in \cref{eq:equations_of_motion_tadpole_and_fluxes} unless the scalar potential vanishes, $V=0$, together with the fake superpotential $\mathcal{W}$ and $g$.

Some simple modifications are discussed in~\cite{Raucci:2023xgx}, where I show that a similar argument also applies to the heterotic case. To date, no fake supersymmetry equations that include form fields are known for non-supersymmetric strings.

\subsection{Vacuum solutions}
\label{sec:vacuum_sol_fake_susy}

Returning to the main motivation of \cref{sec:fake_susy}, a natural application of the fake supersymmetry equations with gravity and the dilaton is to find vacuum solutions. Retracing the steps of the supersymmetric case~\cite{Candelas:1985en}, one starts from a metric ansatz of the type
\begin{eq}
    ds^2=e^{2A}ds_4^2+ ds_6^2\mcomma
\end{eq}
where the warping $A$ and the dilaton depend only on internal coordinates. Splitting ten-dimensional spinors in the usual way, as a product of four-dimensional and six-dimensional spinors, one obtains, schematically~\cite{Tomasiello:2022dwe},
\begin{eq} \label{eq:susy_spinor_ansatz}
    \varepsilon=\zeta_+ \otimes \eta_+ + \zeta_- \otimes \eta_-\mcomma
\end{eq}
where $\zeta$ and $\eta$ are chiral spinors in the four-dimensional and six-dimensional spaces.
The isometries force $\zeta$ to live in the space of four-dimensional Killing spinors with
\begin{eq}
    \nabla_\mu\zeta_{\pm}=\frac{\mu}{2}\gamma_\mu \zeta_{\mp}\mcomma
\end{eq}
where $\gamma_\mu$ are four-dimensional Gamma matrices.
Unfortunately, this natural splitting yields no solutions because $\eta_+$ and $\gamma^i \eta_-$ are independent spinors. Indeed, the terms in \cref{eq:fake_susy_general} that generate the scalar potential carry a number of Gamma matrices that has opposite parity with respect to the supersymmetric parts of $D$ and $\mathcal{O}$, so that the fake supersymmetry equations reduce to 
\begin{eq}
    \mu e^{-A}\eta_- + dA \eta_+ +2\mathcal{W} \eta_+&=0\mcomma \\
    \nabla_i \eta_+ + \mathcal{W} \gamma_i \eta_+ & = 0\mcomma \\
    (d\phi+g)\eta_+ & = 0\mcomma
\end{eq}
whose solution is $d\phi=0$, $\mu=0$ and $A=0$; thus, they are only consistent with a vanishing scalar potential.
Modifications of the fake supersymmetry equations may enlarge the space of solutions in this formalism, and this issue deserves further investigation.

One case that can avoid these problems is given by codimension-one vacua, for which the independence of the spinors does not hold. This is the case to which I now turn.

\subsection{Codimension-one}

Consider the ansatz of \cref{eq:codim1_ansatz} in the gauge $B=0$ and a ten-dimensional spinor $\varepsilon(y)$. The fake supersymmetry equations in the Einstein frame imply that
\begin{eq}\label[pluralequation]{eq:codim_1_fake_W}
    \phi'(y) & = \pm g(\phi)\mcomma\\
    \d_y \varepsilon & = \pm \mathcal{W} \varepsilon\mcomma \\
    A'(y) & = \pm 2\mathcal{W} \mcomma
\end{eq}
and therefore 
\begin{eq} \label{eq:codim_1_spinor}
    \varepsilon=e^{\frac{1}{2}A(y)}\varepsilon_0\mcomma
\end{eq}
where $\varepsilon_0$ is a constant spinor. Then, the Einstein-frame differential equations for the fake superpotential,
\begin{eq}\label[pluralequation]{eq:fake_susy_einst_frame_conditions}
    g+32\mathcal{W}' & = 0\mcomma \\
    2^9 (\mathcal{W}')^2 - 3^2 2^5 \mathcal{W}^2 & = V\mcomma \\
    2^{10} \mathcal{W}' \mathcal{W}'' - 3^2 2^6 \mathcal{W}\mathcal{W}' & = V'\mcomma
\end{eq}
become exactly the Dudas-Mourad equations for $A$ and $\phi$ in the $B=0$ gauge. It is not surprising, then, that the Dudas-Mourad vacua of \cref{sec:DM} are solutions to the fake supersymmetry equations.

Explicitly, for the orientifold case in the string frame, the integration constant $c$ in \cref{eq:einstein_frame_fake_W} is related to $\phi_0$ by
\begin{eq}
    c=\phi_0+\frac{2}{3}\log\frac{2}{3}\mcomma
\end{eq}
and the two functions $\mathcal{W}$ and $g$ become, in terms of $\phi_0$,
\begin{eq} \label[pluralequation]{eq:DM_string_frame_fake_superpotential}
    \mathcal{W}(\phi) & = \pm\frac{1}{12}\sqrt{T}e^{\frac{1}{2}\phi}\left[\sqrt{\frac{W(\frac{9}{4}e^{3(\phi-\phi_0)})}{2}}+\sqrt{\frac{2}{W(\frac{9}{4}e^{3(\phi-\phi_0)})}}\right]\mcomma \\
    g(\phi)& = \pm\sqrt{T}e^{\frac{1}{2}\phi}\left[\sqrt{\frac{W(\frac{9}{4}e^{3(\phi-\phi_0)})}{2}}+\sqrt{\frac{1}{2 W(\frac{9}{4}e^{3(\phi-\phi_0)})}}\right]\mperiod
\end{eq}
The ten-dimensional string-frame spinor for the Dudas-Mourad solution is 
\begin{eq}\label[pluralequation]{eq:DM_String_frame_spinor}
    \varepsilon_S&=e^{\frac{T}{32}y^2}\left(\sqrt{\frac{T}{2}}y\right)^{\frac{1}{9}}\varepsilon_0\mcomma\\
    \gamma_y\varepsilon_0&=\mp\varepsilon_0\mcomma
\end{eq}
where the sign choice matches the overall sign ambiguity in $\mathcal{W}$ and $\gamma_y$ is the Gamma matrix in flat space. In the Einstein frame, the relevant spinor is
\begin{eq}\label[pluralequation]{eq:DM_Einstein_frame_spinor}
    \varepsilon_E&=e^{-\frac{T}{64}y^2}\left(\sqrt{\frac{T}{2}}y\right)^{\frac{1}{36}}\varepsilon_0\mcomma\\
    \gamma_y\varepsilon_0&=\mp\varepsilon_0\mperiod
\end{eq}
Similar considerations apply to the heterotic versions of the Dudas-Mourad solutions, which match the implicit fake superpotentials with $\gamma=\frac{5}{2}$.

One can show that all the codimension-one solutions studied in~\cite{Basile:2022ypo} can be captured by this formalism, and that \cref{eq:codim_1_spinor} holds in all cases. Even when no dilaton potential is present, the non-vanishing $\mathcal{W}$ leads to the free codimension-one solutions of \cref{sec:free_codim_1}. This is not a coincidence: in~\cite{Townsend:2007aw, Trigiante:2012eb} the authors proved that for particular classes of solutions that depend on a single variable, as in the case of codimension-one vacua, the second-order equations of motion can be traded for the same number of first-order equations, together with a non-linear differential equation that links the scalar potential to the fake superpotential. The condition that grants this feature is the vanishing of a conserved charge in the Hamilton-Jacobi formalism of the associated one-dimensional classical mechanics system. The first-order equations need not descend from expressions of the type of \cref{eq:fake_susy_general}, so that the cases that I have discussed are special instances of the ones in~\cite{Townsend:2007aw, Trigiante:2012eb}, for which the more constraining susy-like form of the equations is required.

\subsection{A positive-definite quantity}

It is interesting to determine the extent to which the analogy with the supersymmetry equations can be pushed. As I have already stressed, there is no hidden physics in \cref{eq:fake_susy_eq_variations}, which only replace supersymmetry as a solution-generating technique. 
However, it is possible to define a positive-definite quantity starting from \cref{eq:fake_susy_eq_variations}, which in supersymmetric setups would be interpreted as the Witten-Nester energy~\cite{Witten:1981mf,Nester:1981bjx}. The Witten-Nester procedure relies on the inequality
\begin{eq}
    \expval{\acomm{Q}{Q}}>0 \mcomma
\end{eq}
where $Q$ denotes a preserved real supercharge in a supersymmetric theory, and the gravitational energy is given by
\begin{eq}
    I(\varepsilon)\sim \expval{\acomm{Q}{Q}} \mcomma
\end{eq}
which is motivated by the presence of minima whenever supersymmetric states saturate the bound.

The formalism of~\cite{Witten:1981mf} relies on the existence of spinors and on the first-order supersymmetry equations. It was already noted in~\cite{Boucher:1984yx} that the spinors of~\cite{Witten:1981mf} need not carry any physics: they may well be mere auxiliary objects.
Hence, it is interesting to define an analogous quantity for non-supersymmetric strings.
In the Einstein frame, let
\begin{eq}\label{eq:NW_2_form}
   E^{MN}=-\bar{\varepsilon}\Gamma^{MNP}D_P \varepsilon\mcomma    
\end{eq}
where $D_P$ denotes the improved derivative operator from \cref{eq:fake_susy_eq_variations}. This is usually called an energy two-form. It enters Witten's definition of energy as
\begin{eq} \label{eq:energy_codim_1}
    I(\varepsilon)=\int_\Sigma d\star E_2 \mcomma
\end{eq}
where $\varepsilon$ is a commuting Majorana spinor, constant at infinity up to terms that decay sufficiently fast, and $\Sigma$ is a codimension-one surface acting as an initial-value surface. Typically, one uses Stokes' theorem and defines the energy as an integral over the boundary of $\Sigma$, as
\begin{eq}
    I(\varepsilon)=\int_{\d\Sigma} \star E_2 \mcomma
\end{eq}
while in the present setup, it is more convenient to use the form in \cref{eq:energy_codim_1}, for reasons that will become clear later.
Using the energy two-form in \cref{eq:NW_2_form}, and with Witten's definition of energy as an integral over $\Sigma$,
\begin{eq}\label{eq:fake_energy}
    I(\varepsilon)=\int_\Sigma \nabla_N E^{MN} d\Sigma_M\mcomma
\end{eq}
the computation reduces to finding $\nabla_N E^{MN}$. \Cref{eq:NW_2_form} and the derivative operator of \cref{eq:fake_susy_eq_variations} lead to
\begin{eq}\label{eq:divergence_two_form}
    \nabla_M E^{MN}  = \overbar{D_M\varepsilon}\Gamma^{MPN}D_P\varepsilon+\frac{1}{2}\bar{\varepsilon}\left({\mbox{Gravity EoM}}\right)^{MN}\Gamma_M\varepsilon-\frac{1}{8}\overbar{\mathcal{O}\varepsilon}\Gamma^N \mathcal{O}\varepsilon\mcomma
\end{eq}
and choosing a frame in which the codimension-one surface $\Sigma$ is spacelike, which is a natural setting from a physical perspective, the on-shell energy becomes
\begin{eq}
\label{eq:energy_from_NW}
    I(\varepsilon) = \int_\Sigma -\overbar{D_m\varepsilon}\Gamma^{0mp} D_p\varepsilon+\frac{1}{8}\overbar{\mathcal{O}\varepsilon}\Gamma^0 \mathcal{O}\varepsilon\geq \int_\Sigma -\left(D_m\varepsilon\right)^\dagger \Gamma^{m}\Gamma^{p}\left(D_p\varepsilon\right)\mperiod
\end{eq}
If the generalized Witten condition,
\begin{eq}\label{eq:Witten_condition}
    \Gamma^m D_m\varepsilon=0\mcomma
\end{eq}
can be imposed with the modified operator $D_M$, one is finally left with a positive-definite $I(\varepsilon)$, which vanishes if and only if the fake supersymmetry equations hold.

For non-supersymmetric strings, this occurs for the Dudas-Mourad solution with the spinor in \cref{eq:DM_Einstein_frame_spinor} for the orientifold cases.
It is tempting to speculate about the possible physical implications of this result, particularly regarding the stability of the Dudas-Mourad vacua.
However, there are subtleties, even assuming that \cref{eq:fake_energy} is a proper definition of energy. In fact, these are the same subtleties that I mentioned when I chose not to use Stokes' theorem: the formalism heavily relies on fixed boundary conditions for the surface $\Sigma$. In the Dudas-Mourad case, $\Sigma$ contains the internal interval, and therefore $\d\Sigma$ contains the two singular endpoints. To make physical statements, one must first understand whether it is reasonable to view the two singularities as fixed boundary conditions. The reader can consult~\cite{Mourad:2023wjg} for related discussions.

Even without a convincing argument that proves whether \cref{eq:fake_energy} can be physically interpreted as an energy, note that interpreting the Dudas-Mourad vacua as spontaneous compactifications leads to the Witten-Nester energy in nine dimensions. In fact, the Einstein-frame spinor in \cref{eq:DM_Einstein_frame_spinor} is smooth at the two endpoints of the interval, and one can integrate over $y$ and promote $\varepsilon_0$ to an $x$-dependent spinor. The final result for the energy is the nine-dimensional Witten-Nester energy for pure gravity in nine dimensions, which resonates with the dilaton being gapped in the lower-dimensional theory.
This suggests, for instance, that no bubbles of nothing can exist for the Dudas-Mourad solution, unless they involve the compact interval.

%% file: chapters/chapter2.tex

\chapter{Charting the landscape of non-supersymmetric strings}
\label{chapter:landscape}

The codimension-one vacuum solutions in the previous chapter are all singular, which prompts the question of whether including other ingredients, such as massless form fields, can realize the Fischler-Susskind mechanism with more regular backgrounds. This is the topic of this chapter.

The absence of systematic procedures to engineer vacua, such as the attempts of \cref{sec:fake_susy} in the presence of form fields, implies that exploring the landscape of the three ten-dimensional tachyon-free non-supersymmetric string theories is not as straightforward as exploring supersymmetric cases. To obtain a vacuum, one must solve the full ten-dimensional equations of motion from \cref{eq:action_nosusy} or \cref{eq:action_nosusy_einstein_frame}, which means dealing with coupled second-order partial differential equations.
The general procedure can be summarized as a three-step process.
\begin{enumerate}
    \item The first step is to solve the ten-dimensional equations of motion.\footnote{An alternative starting point is to consider a string theory compactification, in which the internal part is captured by the 2D worldsheet description, and then solve the equations of motion for the resulting large dimensions. Additional care is required to properly take the tadpole potential into account because now other tree-level moduli are present. See~\cite{Ginsparg:1986wr,Baykara:2022cwj,Fraiman:2023cpa} for explicit examples of toroidal compactifications of the heterotic SO$(16)$$\times$SO$(16)$ string.} This is already non-trivial, and only a relatively simple background ansatz can lead to solvable equations. In addition, one must ensure that $g_s$ and $\alpha'$ corrections can be treated as small perturbations of the vacuum, which typically restricts the applicability of the first step. Note that using the equations of motion allows one to control only the Riemannian structure of the compactification manifolds, which does not capture the complete physics. 
    \item The next step is to investigate the perturbative stability of the vacuum. In a worldsheet perturbative description, one would look for tachyons; however, after the Fischler-Susskind mechanism, one is left with only the quantum effective action, or equivalently with the equations of motion. In this language, perturbative stability can be captured by the absence of growing linear perturbations in the chosen vacuum, which have the same physical meaning as string tachyons. A true vacuum must be perturbatively stable.
    Note that with non-supersymmetric strings, one often works with singular spaces, and the correct boundary conditions for perturbations in singular backgrounds are not clear. In \cref{sec:linearized_branes}, I partially address the issue in the context of brane solutions, following~\cite{Mourad:2023wjg}.
    \item The last step, after finding a stable vacuum, is to check its non-perturbative stability or to understand its metastability. 
    Typically, this is achieved through a combination of kinematical (topological) and dynamical arguments~\cite{GarciaEtxebarria:2020xsr}. Control over dynamics is accomplished by spacetime supersymmetry, and without a replacement principle, I shall not be able to comment on this further.
\end{enumerate}

Keeping these three steps in mind, I now proceed to study three types of backgrounds in the landscape of non-supersymmetric strings: Freund-Rubin solutions (\cref{sec:FR}, based on~\cite{Raucci:2022bjw}), vacua with brane-like isometries (\cref{sec:brane_isometries}, based on~\cite{Mourad:2024dur}), and brane solutions in Dudas-Mourad backgrounds (\cref{sec:branes}, based on~\cite{Mourad:2024mpg}).

\section{Freund-Rubin electro- and magneto-vacua}
\label{sec:FR}

The AdS$_5\times$S$^5$ solution of type IIB string theory and the AdS$_4\times$S$^7$ and AdS$_7\times$S$^4$ solutions of eleven-dimensional supergravity, together with their generalizations where Einstein manifolds replace the spheres, are examples of Freund-Rubin vacua. 
These are constant-dilaton vacua for which spacetime is the direct product of two Einstein spaces, one acting as the external spacetime and the other as the compact internal manifold, whose curvatures are stabilized by electric or magnetic form fluxes.
The characteristic feature of these solutions is that they arise from balancing curvatures and form fluxes in the equations of motion.

For non-supersymmetric strings, standard Freund-Rubin vacua do not exist because, in addition to curvature and fluxes, the equations of motion contain the tadpole potential. However, a generalization is possible, as suggested by the presence of the scalar potential itself: one can define generalizations of Freund-Rubin vacua as constant-dilaton backgrounds where curvatures, fluxes, and the tadpole potential balance each other.
Solutions of this type, involving the R-R and NS fluxes, were discussed in~\cite{Gubser:2001zr,Mourad:2016xbk}, and their stability was studied in~\cite{Basile:2018irz}.
In this section, based on~\cite{Raucci:2022bjw} and on a section of~\cite{Basile:2022ypo}, I present other families of generalized Freund-Rubin vacua for non-supersymmetric strings, which rest on the presence of gauge fields in all models of interest. The vacua are built by considering non-zero vacuum expectation values for U(1) abelian gauge fields in the Cartan of the full gauge algebra.\footnote{As I already mentioned, type 0'B string theory has an SU(32) gauge algebra, and not U(32), because there is an anomalous U(1)~\cite{Sagnotti:1996qj}. This U(1) cannot be used as the background U(1) of these solutions.}

\subsection{The orientifold electrovac}

I begin by considering the two orientifold models with the string-frame tadpole potential
\begin{eq}
    T e^{-\phi}\mcomma
\end{eq}
where the kinetic terms of the gauge fields are dressed by an open-string dilaton coupling
\begin{eq}
    \sim e^{-\phi} F^2\mperiod
\end{eq}
The Freund-Rubin metric ansatz is
\begin{eq}
    ds^2 =\lambda_{\mu\nu} \, dx^\mu dx^\nu + \gamma_{ij} \, dy^i dy^j\mcomma
\end{eq}
where the dilaton is stabilized at a constant value, and both Einstein spaces are maximally symmetric. The latter condition is needed to explicitly perform the analysis of linear perturbations later, but in principle, any Einsten manifold with the appropriate curvature would yield a solution.
The string-frame equations of motion from \cref{eq:action_nosusy} for gravity, dilaton, and gauge fields require
\begin{eq}\label{eq:maxwell_propto_metric}
    F_{MA}{F_N}^A \propto g_{MN} \mcomma
\end{eq}
and the simplest choice that is compatible with this condition is, as expected from the Freund-Rubin ansatz, constant flux $F_2$ for a U(1) field strength. With this choice, the complete equations for the orientifold models become
\begin{eq}\label[pluralequation]{eq:orientifold_eom_u(1)}
    & \frac{\alpha'}{8}F^2+3T = 0 \mcomma \qquad \nabla_M {F}^{MN} =0 \mcomma  \\
    & R_{MN}-\frac{1}{2}T  \, e^{\phi_0}g_{MN}-\frac{\alpha'}{4} \,  e^{\phi_0}F_{MA}{F_N}^A = 0\mcomma
\end{eq}
and require a two-dimensional external spacetime with a constant electric field:
\begin{eq} \label{eq:electric_background}
    F_{\mu\nu} = \epsilon_{\mu\nu}f \mperiod
\end{eq}
This is a constant electric background because, in terms of the vielbein, one has
\begin{eq}\label{eq:constant_electric_and_magnetic_field}
    F_{MN}=E_{i}(e_M^0 e_N^i-e_N^0 e_M^i)+H_{ij}(e_M^i e_N^j-e_N^i e_M^j)\mcomma 
\end{eq}
and \cref{eq:electric_background} translates into a vanishing $H_{ij}$ and a constant $E_i$ along the single spatial direction of the two-dimensional manifold.
A solution of this type with a constant electric field was considered in a different context in~\cite{Freedman:1983xa,Antoniadis:1989mn}, where it was termed \emph{electrovac}. For this reason, I call the solution to \cref{eq:orientifold_eom_u(1)} an electrovac solution of the Freund-Rubin type:
\begin{eq}\label[pluralequation]{eq:electric_vacuum_string_frame}
    R_{\mu\nu}  = - \frac{5}{2} T  \,  e^{\phi_0}\lambda_{\mu\nu} \mcomma \qquad R_{mn} = \frac{1}{2} T \,  e^{\phi_0} \gamma_{mn}\mcomma \qquad \alpha' f^2 =12T  \mcomma    
\end{eq}
which is an AdS$_2\times$S$^8$ with constant dilaton $\phi_0$.

\subsubsection{Higher derivative corrections}

In the first step of the procedure outlined at the beginning of \cref{chapter:landscape}, I stressed that for a vacuum to be reliable one needs higher $\alpha'$ and $g_s$ orders to produce small perturbations.
Rather than performing a complete analysis, which would involve frame choices to avoid redundancies from field redefinitions, here, simple considerations are used to understand whether the solution can be under control or if it may be dismantled by $\alpha'$ and $g_s$ corrections.

In the electrovac, the radii of the two spaces and the gauge field strength scale with $\alpha'$ and $e^{\phi_0}$ as
\begin{eq}
    \frac{1}{l^2_{\text{AdS}_2}}\sim \frac{1}{R^2_{\text{S}^8}}\sim e^{\phi_0} (\alpha')^{-1}\mcomma \qquad F^2\sim (\alpha ' )^{-2}\mperiod
\end{eq}
Metric curvature corrections are generically subleading in an $e^{\phi_0}$ expansion for dimensional reasons because they contribute to the action as
\begin{eq}
    (\alpha')^{n-1}R^n\mcomma
\end{eq}
with an additional dilaton dressing that comes from the worldsheet sphere topology.
On the other hand, higher derivative terms in the field strength, which can be obtained, for instance, from the DBI action of the D9 branes, can become the dominant contributions because they only scale with $\alpha'$, and therefore they can invalidate the classical solution.

This is a potential problem; however, as explained below, this is not the only problem associated with the electrovac solution.

\subsubsection{The issue of perturbative stability}

Recall once more the three-step procedure: after finding a solution, one must understand whether the solution is stable or tachyonic, starting from the linearized equations of motion. Hence, I move on to study linear perturbations to the electrovac, choosing to work in the Einstein frame because it leads to results with clearer physical interpretations. 
The background transformation from the string frame to the Einstein frame is straightforward because the dilaton is constant.
To begin with, note that the non-abelian perturbations do not mix with the others; therefore, I focus on linearized perturbations of the dilaton, the metric, and the U(1) gauge field with non-trivial profile, $\varphi$, $h_{MN}$, and $a_M/f$. Here, $f$ is the Einstein-frame electric-field parameter, which is expressed as follows:
\begin{eq}
    \alpha' f^2=12 T \, e^{\phi_0}  \mperiod
\end{eq}
I shall comment on non-abelian perturbations at the end of this section.

The linearized fields depend on the AdS$_2$ coordinates $x^\mu$ and on the S$^8$ coordinates $y^i$, and in the following, I consider their expansions in spherical harmonics.
The linearized metric equations are
\begin{eq}\label[pluralequation]{eq:linearization_electric_metric}
    \Box h_{MN} &=-\nabla_M \nabla_N h +2\nabla_{(M} \nabla^R h_{N)R}+2{R^B}_{(N} h_{M)B}-2{R^B}_{MAN}{h_B}^A-R_{MN}\varphi\\
        & +\frac{1}{2}T \,  e^{\frac{3}{2}\phi_0}{g}_{MN}\varphi-T \,  e^{\frac{3}{2}\phi_0}h_{MN}-\frac{\alpha'}{f}  \,  e^{\frac{1}{2}\phi_0}\bigg[{F_M}^A \, \nabla_{[N}a_{A]}+ {{F}_N}^A \, \nabla_{[M}a_{A]}\\
            & -\frac{f}{2}{F}_{MA}{F}_{NP}h^{PA}\bigg]+{g}_{MN}\bigg[\frac{\alpha'}{8f} \,  e^{\frac{1}{2}\phi_0}{F}_{AB} \, \nabla^{[A} a^{B]}+\frac{1}{8}T \,  e^{\frac{3}{2}\phi_0}h-\frac{3}{4}T \, e^{\frac{3}{2}\phi_0}\varphi \\
                & -\frac{1}{4}R_{AB}h^{AB}\bigg]\mcomma
\end{eq}
while the dilaton equation is
\begin{eq}\label{eq:linearization_electric_dilaton}
    \Box\varphi &=  \frac{3}{2}T  \,  e^{\frac{3}{2} \phi_0}\varphi + \frac{\alpha'}{4f} \, e^{\frac{1}{2}\phi_0} F_{AB} \nabla^Aa^B  -\frac{1}{2}R_{MN}h^{MN} + \frac{1}{4} T  \,  e^{\frac{3}{2}\phi_0} h  \mcomma
\end{eq}
and the U(1) gauge field equations are
\begin{eq}\label[pluralequation]{eq:linearization_electric_gauge}
    \Box\frac{a^N}{f} &=\nabla_M \nabla^N \frac{a^M}{f}+{{F}_P}^N \nabla_M h^{MP}+ {{F}^M}_P \nabla_M h^{NP}-\frac{1}{2}{F}^{MN}\nabla_M (h+\varphi)\mperiod
\end{eq}
In this subsection only, let
\begin{eq}\label[pluralequation]{eq:units_orientifold}
    \tau= T  \, e^{\frac{3}{2}\phi_0}\mcomma \qquad L= \frac{l(l+7)}{14}\mperiod
\end{eq}
$\tau$ sets the units, and $L$ is the scale of the eigenvalues of the internal Laplacian. In the following, I denote the external and internal Laplacians as $\Box_2$ and $\Box_8$.

Tensor perturbations in the spacetime AdS would come from \cref{eq:linearization_electric_metric}, and would consist of a massless graviton and a KK tower with
\begin{eq}\label{eq:electro_tensor_modes}
    \Box h_{\mu\nu}=-\frac{2}{l_{\text{AdS}}^2} h_{\mu\nu}\mperiod
\end{eq}
However, the electrovac AdS is a two-dimensional spacetime, and these modes are pure gauge. Similar considerations hold for the vector modes from \cref{eq:linearization_electric_gauge} with $N=\mu$, which satisfy Maxwell's equations in AdS
\begin{eq}\label{eq:maxwell_AdS}
    \Box a_{\mu}=\nabla^\nu\nabla_\mu a_\nu \mcomma
\end{eq}
and for the graviphoton $h_{\mu i}$ in \cref{eq:linearization_electric_metric}, which only provides $l\geq1$ modes.

AdS scalar perturbations are more involved. Transverse traceless modes $h_{ij}$ are stable because from \cref{eq:linearization_electric_metric} they solve
\begin{eq}\label{eq:orientifold_electric_tensor_pert}
   \Box_2 h_{ij}  =L\tau  h_{ij} \mperiod
\end{eq}
Internal vectors come from $h_{\mu i}$ and $a_i$, and generically mix. Indeed, letting
\begin{eq}\label[pluralequation]{eq:electro_vector_pert_gauge}
    a_i= \frac{\Omega_i}{\alpha'} e^{-\frac{1}{2}\phi_0}\mcomma \qquad  h_{\mu i}=\tau^{-1}  \epsilon_{\mu\nu}\nabla^\nu V_i\mcomma 
\end{eq}
their linearized equations read
\begin{eq}\label[pluralequation]{eq:electro_vector_mass_matrix}
    \Box_2 \begin{pmatrix}V_i \\ \Omega_i \end{pmatrix} = 
    \begin{pmatrix}L-\frac{4}{7} & \frac{1}{2} \\ 12L-\frac{48}{7} & L+\frac{45}{7} \end{pmatrix}\tau \begin{pmatrix}V_i \\ \Omega_i \end{pmatrix}\mcomma  
\end{eq}
and the mass matrix has a vanishing eigenvalue when $l=1$, together with strictly positive eigenvalues for all other cases. Therefore, no unstable modes arise in this sector.

The last type of AdS scalars are singlet scalar perturbations, which are captured by
\begin{eq}\label[pluralequation]{eq:orientifold_electric_scalar_pert_definitions}
    h_{\mu\nu} = A \lambda_{\mu\nu} \mcomma \qquad h_{ij} = C \gamma_{ij} \mcomma \qquad h_{\mu i}  = \nabla_\mu \nabla_i D\mcomma \qquad  a_\mu  = e^{-\frac{1}{2}\phi_0}  \epsilon_{\mu\nu}\frac{\nabla^\nu \Omega}{\alpha'} \mcomma
\end{eq}
after removing the remaining gauge redundancy of the background.
The different tensor structures in the $\mu\nu$ and $ij$ coordinates must vanish separately, and one obtains
\begin{eq}\label[pluralequation]{eq:orientifold_electric_scalar_pert}
    &\Box \Omega  = -12\tau \left[-\frac{1}{2}\varphi+\Box_8 D + A - 4 C\right]\mcomma \\
    &\Box\varphi  = \frac{3}{2} \tau \varphi+3\tau A +\frac{1}{4}\Box_2 \Omega \mcomma \\
    &\Box A  = \frac{9}{4} \tau \varphi-\frac{11}{2} \tau  A -\frac{7}{8}\Box_2 \Omega \mcomma \\
    &\Box_8 D  = 4C\mcomma \\
    &\Box C  = - \tau C+\frac{1}{8}\Box_2 \Omega +\frac{3}{2} \tau A-\frac{3}{4} \tau \varphi\mcomma\\
    &\Box_2 D  = A+3C \mcomma \\
    &0  =  A+7C+ \tau D+\frac{1}{2}\Omega \mperiod
\end{eq}
Perturbations with $l=0$ are the simplest because there is no $D$ mode; therefore, $C=0$ and $\Omega$ becomes pure gauge. The two remaining fields, $A$ and $\varphi$, lead to a mass matrix with positive eigenvalues and therefore no unstable modes.
The case $l=1$ must be dealt with separately. It implies
\begin{eq}
    L=\frac{4}{7}\mcomma \qquad \Omega=-2A\mcomma
\end{eq}
and the mass matrix contains two positive and one negative modes. The latter, $-\frac{3}{7}\tau$, is still compatible with the B-F bound, $-\frac{5}{8}\tau$, and therefore all these modes are stable.
In the remaining $l>1$ cases, $C$ and $D$ satisfy algebraic constraints, leaving
\begin{eq}\label[pluralequation]{eq:electro_scalar_instab}
    \Box_2 \begin{pmatrix}A \\ \Omega \\ \varphi \end{pmatrix} = 
    \begin{pmatrix}
    L+5 & -\frac{7}{8}L & -3 \\ 
    -12 & L & 6 \\
    0 & \frac{1}{4}L & L+3 \end{pmatrix}\tau \begin{pmatrix}A \\ \Omega \\ \varphi \end{pmatrix}\mperiod
\end{eq}
By numerically comparing the eigenvalues with the B-F bound, one finds three sets of unstable modes with $l=2$, $3$ and $4$.

The final result is that the electrovac solution is perturbatively unstable and does not survive the second step of the general strategy to find vacua. 
Note that stability can be rephrased as a property of the Laplacian eigenvalues. Different Einstein manifolds correspond to different values for $L$, and a Laplacian gap $L\gtrapprox4$ would uplift the scalar instability of \cref{eq:electro_scalar_instab}.

\subsection{The heterotic magnetovac}

A similar construction applies to the heterotic SO$(16)$$\times$SO$(16)$ string with the positive one-loop tadpole potential, for which the kinetic terms of the gauge fields have closed-string dressings,
\begin{eq}
    \sim e^{-2\phi } F^2\mperiod
\end{eq}
In this case, the equations of motion require a constant magnetic gauge field, $F^2>0$, confined to a two-dimensional internal space. The resulting Freund-Rubin solution is AdS$_8\times$S$^2$, where 
\begin{eq}
    F_{mn}=\epsilon_{mn} f\mcomma
\end{eq}
and
\begin{eq}
    R_{\mu\nu} = -\frac{1}{2}T \,  e^{2\phi_0} \, \lambda_{\mu\nu}\mcomma \qquad R_{mn} = \frac{9}{2}T  \, e^{2\phi_0} \,  \gamma_{mn}\mcomma \qquad  \alpha' f^2 = 20 T \, e^{2\phi_0} 
\end{eq}
in the string frame. The constant value of the dilaton, $\phi_0$, can be traded for the U(1) magnetic flux
\begin{eq}
    N_m\propto e^{-\phi_0}\mcomma
\end{eq}
but in the following I shall keep $\phi_0$ as the free parameter.

\subsubsection{Higher derivative corrections}

One can closely follow the analysis of the $\alpha'$ and $g_s$ corrections to the orientifold electrovac, and in this heterotic case, the two metric curvatures and the gauge field strength scale as
\begin{eq}
    \frac{1}{l^2_{\text{AdS}_8}}\sim \frac{1}{R^2_{\text{S}^2}}\sim e^{2\phi_0} (\alpha')^{-1}\mcomma \qquad F^2\sim e^{2\phi_0} (\alpha ' )^{-2}\mperiod
\end{eq}
To compare the background with the higher-derivative terms, note that the classical contribution is dressed with $e^{2\phi_0} (\alpha ')^{-1}$, while the higher derivative corrections are accompanied by $e^{k\phi_0}(\alpha ')^{-1}$, where $k>2$. 
Therefore, the heterotic magnetovac is a reliable string theory solution when $e^{\phi_0}\ll 1$, or equivalently for large fluxes, $N_m\gg1$.

\subsubsection{The issue of perturbative stability}

The stability analysis proceeds as in the orientifold case, linearizing the Einstein-frame equations of motion and turning on gauge perturbations only along the U(1) that supports the background magnetic field.
In terms of the Einstein-frame magnetic field parameter $f$, the linearized equations for the metric are
\begin{eq}\label[pluralequation]{eq:linearization_magnetic_metric}
    \Box h_{MN} &=-\nabla_M \nabla_N h +2\nabla_{(M} \nabla^R h_{N)R}+2{R^B}_{(N} h_{M)B}-2{R^B}_{MAN}{h_B}^A+R_{MN}\varphi\\
            &+\frac{1}{2}T \,  e^{\frac{5}{2} \phi_0}{g}_{MN}\varphi+T  \, e^{\frac{5}{2} \phi_0}h_{MN}-\frac{\alpha'}{f} e^{-\frac{1}{2}\phi_0}\bigg[{{F}_M}^A \nabla_{[N} a_{A]} + {{F}_{N}}^A \nabla_{[M} a_{A]} \\
                    & - \frac{f}{2}{F}_{MA}{F}_{NP}h^{PA}\bigg] +{g}_{MN}\bigg[\frac{\alpha'}{8f} \, e^{-\frac{1}{2}\phi_0}{F}_{AB} \, \nabla^{[A} a^{B]}-\frac{1}{8}T \,  e^{\frac{5}{2} \phi_0}h \\
                            & -\frac{5}{4}T  \, e^{\frac{5}{2} \phi_0}\varphi-\frac{1}{4}R_{AP}h^{AP}\bigg]\mcomma
\end{eq}
while the dilaton equation is
\begin{eq}\label{eq:linearization_magnetic_dilaton}
    \Box\varphi &= \frac{15}{2}T  \, e^{\frac{5}{2} \phi_0}\varphi - \frac{\alpha'}{4f} \, e^{-\frac{1}{2}\phi_0} F_{AB}\nabla^A a^B+ \frac{1}{2}R_{MN}h^{MN}+\frac{1}{4}T \,  e^{\frac{5}{2} \phi_0} h  \mcomma
\end{eq}
and the gauge field equations are
\begin{eq}\label[pluralequation]{eq:linearization_magnetic_vacua}
    \Box\frac{a^N}{f} &=\nabla_M \nabla^N \frac{a^M}{f}+{{F}_P}^N \nabla_M h^{MP} + {{F}^M}_P \nabla_M h^{NP}-\frac{1}{2}{F}^{MN}\nabla_M (h-\varphi)\mperiod
\end{eq}
Analogously to \cref{eq:units_orientifold}, one can simplify the notation by letting, in this subsection only,
\begin{eq}\label[pluralequation]{eq:units_heterotic}
    \tau=T  \,  e^{\frac{5}{2}\phi_0}\mcomma \qquad L= \frac{9l(l+1)}{2}\mperiod
\end{eq}

AdS$_8$ tensor modes from \cref{eq:linearization_magnetic_metric} again describe a massless graviton and a tower of KK excitations.
Vector modes arise from the mixing of the divergence-free $h_{\mu i}$ and from gauge field perturbations $a_\mu$. Letting
\begin{eq}
    a_{\mu}= \frac{\Omega_\mu}{\alpha'} e^{\frac{1}{2}\phi_0}  \mcomma \qquad h_{\mu i}=\tau^{-1} \epsilon_{ji}\nabla^j V_\mu\mcomma
\end{eq}
the mass matrix becomes
\begin{eq}\label[pluralequation]{eq:heterotic_vector_modes}
    \Box_8 \begin{pmatrix}V_\mu \\ \Omega_\mu \end{pmatrix} =
    \begin{pmatrix}L+\frac{1}{2} & -\frac{1}{2}  \\ -20 {L} & L-\frac{1}{2} \end{pmatrix}\tau \begin{pmatrix}V_\mu \\ \Omega_\mu \end{pmatrix}\mperiod
\end{eq}
The $l=0$ sector, in which only the $a_\mu$ modes are present, leads to the AdS Maxwell equations, as in \cref{eq:maxwell_AdS}. For $l=1$, \cref{eq:heterotic_vector_modes} describe a triplet of massless vectors, whereas in all other cases, the mass matrix only leads to massive modes.

Perturbations that are AdS scalars but tensors or vectors with respect to the internal rotation group are stable in a way that closely follows the orientifold case.
However, unlike the electrovac, singlet scalar perturbations,
\begin{eq}\label[pluralequation]{eq:heterotic_magnetic_scalar_pert_definitions}
        h_{\mu\nu} = A \lambda_{\mu\nu} \mcomma \qquad h_{ij}  = C \gamma_{ij} \mcomma \qquad h_{\mu i}  = \nabla_\mu \nabla_i D\mcomma \qquad a_i = e^{\frac{1}{2}\phi_0} \epsilon_{ij} \frac{\nabla^j\Omega}{\alpha'} \mcomma
\end{eq}
are stable for the heterotic magnetovac.
The two different tensor structures from \cref{eq:linearization_magnetic_metric} must be dealt with separately, and the complete set of equations is reduced to
\begin{eq}\label[pluralequation]{eq:heterotic_magnetic_scalar_pert}
    &\Box \Omega  = -20\tau\left[\frac{1}{2}\varphi+\Box_8 D + C - 4 A\right]\mcomma \\
    &\Box\varphi  = \frac{15}{2}\tau \varphi+5\tau  C +\frac{1}{4}\Box_2 \Omega \mcomma \\
    &\Box A  = \tau A -\frac{1}{8}\Box_2 \Omega -\frac{5}{2}\tau C-\frac{5}{4}\tau \varphi\mcomma\\
    &\Box_2 D  = 3A+C \mcomma \\
    &\Box C  = \frac{15}{4}\tau \varphi+\frac{17}{2}\tau  C +\frac{7}{8}\Box_2 \Omega \mcomma \\
    &\Box_8 D   = 4A\mcomma \\
    &0  = C+7A-\tau D-\frac{1}{2}\Omega\mperiod
\end{eq}
Modes with $l=0$ are the simplest: $\Omega$ and $D$ are absent, the fourth of \cref{eq:heterotic_magnetic_scalar_pert} becomes 
\begin{eq}
    C=-3A\mcomma
\end{eq}
and the remaining mass matrix for $A$ and $\varphi$ has positive eigenvalues.
For modes with $l>0$, two equations are algebraic and allow one to remove $A$ and $D$ in terms of the other fields.
The mass matrix is then reduced to
\begin{eq}\label[pluralequation]{eq:heterotic_scalar_mass_matrix}
    \Box_8 \begin{pmatrix}C \\ \Omega \\ \varphi \end{pmatrix} =
    \begin{pmatrix}
    L+\frac{17}{2} & -\frac{7}{8}L & \frac{15}{4}  \\
    -20 & L & -10 \\
    5 & -\frac{1}{4}L & L+\frac{15}{2}  \\
    \end{pmatrix} \tau \begin{pmatrix} C \\ \Omega \\ \varphi \end{pmatrix}\mcomma
\end{eq}
whose eigenvalues lie all above the B-F bound.

This analysis shows that the perturbations $\varphi$, $h_{MN}$, and $a_M/f$ are stable. However, this is not sufficient. In fact, at the beginning of the discussion, non-abelian perturbations were excluded because of the absence of mixing with the other linear modes.
A complete study of these modes is more involved, but a simple limit suggests that unstable modes are indeed present in the non-abelian gauge sector.
For large fluxes, or equivalently small string coupling, the solution is a nearly flat space with a constant magnetic field. In~\cite{Chang:1979tg,Sikivie:1979bq}, the authors showed that instabilities arise for non-abelian perturbations in the presence of strong magnetic fields, and this happens precisely when the magnetic field is constant throughout an extended region of space, such that the product of the magnetic field and the surface area is of order 1.
For the heterotic magnetovac, this instability condition can be expressed as
\begin{eq}
    H R_{S^2}^2 =\mathcal{O}(1)\mcomma
\end{eq}
and the left-hand side is proportional to $e^{-\phi_0}$, or, equivalently, to the inverse of the flux. Therefore, in the parameter space under control from the string theory perspective, $N_m\gg1$, unstable modes are present in the non-abelian gauge sector for the magnetovac. A more comprehensive investigation is required to settle the issue, which is an interesting problem in its own right.
Note that this is reminiscent of what happens for magnetic monopoles in non-abelian gauge theories, which decay through non-abelian radiation, and only one monopole in each topology class is stable, protected by kinematical arguments.

\subsection{Freund-Rubin after spontaneous compactification}

Similar solutions with constant electric or magnetic fluxes are present in the nine-dimensional effective field theory that survives after the Dudas-Mourad compactification. In fact, gauge fields remain dynamical from the results of \cref{sec:spont_cpt}, and by solving the nine-dimensional equations of motion, one finds AdS$_2\times$S$^7$ and AdS$_7\times$S$^2$ Freund-Rubin vacua. 
These are parametrically weakly coupled in the large flux regime, and linear perturbations of the dilaton, metric, and U(1) gauge field are stable~\cite{Basile:2022ypo}. 
Nevertheless, they are expected to be unstable for the same reason as the heterotic magnetovac is: non-abelian gauge perturbations are in the regime where the analysis of~\cite{Chang:1979tg,Sikivie:1979bq} applies and signals the presence of unstable modes. 
The reader can find a complete description of the vacua, including the dependence of the radii on the flux numbers, in~\cite{Basile:2022ypo}.

The outcome of this section is that the simplest types of vacua, Freund-Rubin compactifications, are possible solutions for non-supersymmetric strings, although unstable modes are present in all cases under computational control.

\section{Vacua with brane-like isometries}
\label{sec:brane_isometries}

The gravity solutions of the BPS $p$-branes of string theory are captured by the two-derivative action as black hole-like backgrounds with ISO$(1,p)\times$SO$(9-p)$ isometries, which generically interpolate between a singularity, the brane core, and the flat space vacuum at infinity. 
These solutions are, in a broad sense, part of the landscape of string theory, and in some cases, one can relax the assumption of asymptotic flatness and obtain solutions that can be interpreted as flux vacua. 
This section and the next one address the question of how brane-like solutions are deformed in non-supersymmetric strings in the presence of tadpole potentials. Related work on this topic can be found in~\cite{Basile:2021mkd,Basile:2022zee}.

I proceed in steps, first by adding information about isometries and then by adding the condition that the solutions reduce to a vacuum far from a singularity. In this way, the first step produces solutions that can be interpreted as flux vacua, thus effectively exploring a sector of the non-supersymmetric string landscape.

In this section, based on~\cite{Mourad:2024dur}, I focus on imposing ISO$(1,p)\times$SO$(9-p)$ isometries. I only address the first step of the procedure outlined at the beginning of \cref{chapter:landscape}, leaving the stability analysis for future work.
Explicitly, the background ansatz is 
\begin{eq} \label[pluralequation]{eq:ansatz}
ds^{2} &= e^{2A(r)} \eta_{\mu\nu} dx^\mu dx^\nu + e^{2B(r)}dr^2 +  e^{2C(r)}\alpha' d\Omega_{8-p}^2\mcomma  \\
\phi &= \phi(r) \mcomma \\
F_{p+2} &=  H_{p+2}\, e^{2\beta_p\phi + B +(p+1)A-(8-p)C}\ dx^0 \wedge \ldots \wedge dx^p \wedge dr \mcomma
\end{eq}
where the electric form-field profile automatically solves the Bianchi identities and the equations of motion. Magnetic counterparts can be obtained from electric-magnetic duality. 
As I have stressed, the physical interpretation is twofold: \cref{eq:ansatz} can describe $p$-branes, uncharged or electrically charged ones, for which $r$ is the radial distance from the source; however, they can also be used to engineer cohomogeneity-one (flux) compactifications, namely vacua in which one direction is singled out and the warping factors only depend on it. 
In the case of \cref{eq:ansatz}, the special variable is $r$, and the metric contains a $(p+1)$-dimensional spacetime and an internal $(8-p)$-dimensional sphere.
In order to distinguish between vacua and branes, one must add some physics; for instance, the condition that a solution becomes a vacuum for large values of $r$ can characterize what is typically called a brane. I shall return to this in \cref{sec:branes}.

Regardless of whether \cref{eq:ansatz} describe a brane or a warped compactification, the Einstein-frame equations of motion from \cref{eq:action_nosusy_einstein_frame} become
\begin{eq} \label[pluralequation]{eq:cohom_1_full_eq}
    A''+A'F'& = -\frac{T}{8}e^{2B+\gamma\phi}+\frac{7-p}{16}e^{2B + 2\beta_p\phi - 2(8-p)C} H_{p+2}^2 \mcomma \\
    C''+ C'F' & = -\frac{T}{8}e^{2B+\gamma\phi}+\frac{7-p}{\alpha'}e^{2(B-C)}-\frac{p+1}{16}e^{2B+2\beta_p\phi-2(8-p)C}H_{p+2}^2 \mcomma \\
    \phi''+\phi'F' & = T\gamma e^{2B+\gamma\phi}+\beta_p e^{2B+2\beta_p\phi-2(8-p)C}H_{p+2}^2 \mcomma
\end{eq}
where 
\begin{eq}\label{eq:F_combination}
    F = (p+1)A-B+(8-p)C\mcomma
\end{eq}
together with a first-order condition known as the \emph{Hamiltonian constraint} from the associated one-dimensional classical mechanics system,
\begin{eq}
    &(p+1)A'\left[p A'+ (8-p)C'\right]+ (8-p)C'\left[(7-p)C'+(p+1)A'\right] - \frac{1}{2}(\phi')^2 \\
    &+ T e^{2B+\gamma\phi}-\frac{(7-p)(8-p)}{\alpha'}e^{2(B-C)}+\frac{1}{2}e^{2B+2\beta_p\phi-2(8-p)C}H_{p+2}^2 = 0 \mperiod
\end{eq}
The residual gauge freedom of reparametrizing $r$ allows $B$ to be any desired function of $r$, and two physically relevant choices are 
\begin{eq}
    B=0\mcomma
\end{eq}
for which $r$ is the proper radial distance, and 
\begin{eq}
    B= C- \log{\frac{r}{\sqrt{\alpha'}}}\mcomma
\end{eq}
which defines the isotropic coordinates, where the orthogonal $(9-p)$-dimensional space is conformal to the flat metric.
In the presence of tadpole potentials, other coordinates may be more convenient, such as the gauge
\begin{eq}
    B=-\frac{\gamma}{2}\phi \mcomma
\end{eq}
of~\cite{Dudas:2000ff} that allows to find the nine-dimensional vacuum analytically. 
In this section, another gauge choice is useful: the \emph{harmonic gauge}
\begin{eq}
    B=(p+1)A+(8-p)C\mcomma
\end{eq}
for which the combination $F$ of \cref{eq:F_combination} vanishes and the first-derivative terms in \cref{eq:cohom_1_full_eq} are absent altogether. These coordinates are particularly useful for classifying all possible asymptotic behaviors, which are the ultimate focus of this section.
The advantage of the harmonic gauge is that it allows physically different contributions to be identified and separated in the equations of motion. Indeed, let
\begin{eq}\label[pluralequation]{eq:harmonic_XYW}
    X&=(p+1)A+(7-p)C\mcomma \\
    Y&=(p+1)A+\beta_p\phi\mcomma \\
    W&=(p+1)A+(8-p)C+\frac{\gamma}{2}\phi\mperiod
\end{eq}
In terms of these combinations, the original fields that define the background read
\begin{eq}
    A&=\frac{2(8-p)\beta_p}{\Theta}X + \frac{(7-p)\gamma}{\Theta}Y-\frac{2(7-p)\beta_p}{\Theta}W \mcomma \\
    B&=\frac{(p+1)(8-p)\gamma}{\Theta}X-\frac{(p+1)\gamma}{\Theta}Y+\frac{2(p+1)\beta_p}{\Theta}W \mcomma \\
    C&=\frac{(p+1)(\gamma-2\beta_p)}{\Theta}X-\frac{(p+1)\gamma}{\Theta}Y+\frac{2(p+1)\beta_p}{\Theta}W \mcomma \\
    \phi&=-\frac{2(p+1)(8-p)}{\Theta}X+\frac{2(p+1)}{\Theta}Y+\frac{2(p+1)(7-p)}{\Theta}W \mcomma
\end{eq}
where
\begin{eq}
\Theta = (p+1)\left[2\beta_p+ (7-p)\gamma\right]\mperiod    
\end{eq}
Note that $\Theta$ never vanishes in the cases of interest.
In terms of $X$, $Y$, and $W$, the equations of motion are
\begin{eq}\label[pluralequation]{eq:brane_isometry_eom}
    X''&=\frac{(7-p)^2}{\alpha'}e^{2X}- T e^{2W}\mcomma \\
    Y''&= H_{p+2}^2 e^{2Y}+\left(\beta_p\gamma-\frac{p+1}{8}\right)T e^{2W}\mcomma \\
    W''&=\frac{(8-p)(7-p)}{\alpha'}e^{2X}+\left(\beta_p\gamma-\frac{p+1}{8}\right)\frac{H_{p+2}^2}{2}e^{2Y}+\frac{\gamma^2-\gamma_c^2}{2}T e^{2W}\mcomma
\end{eq}
and the Hamiltonian constraint is given by
\begin{eq}
    0&=\frac{(7-p)(8-p)}{\alpha'}e^{2X}-T e^{2W}-\frac{H_{p+2}^2}{2}e^{2Y}+\frac{p+1}{8 \Theta^2}\Bigg\{-(8-p)\Bigg[ \frac{p+1}{16}\Xi -9\\
    &+(p+1)\left(\beta_p\gamma-\frac{p+1}{8}\right)\Bigg](X')^2+\frac{\Xi}{2} (Y')^2+8(7-p) (W')^2\\
    &-16(8-p) X'W'+8\left(\beta_p\gamma-\frac{p+1}{8}\right) Y'\left[(8-p)X'-(7-p)W'\right]\Bigg\}\mcomma
\end{eq}
where 
\begin{eq}\label[pluralequation]{eq:deltasigmagamma}
    \Xi= (p+1)+4(7-p)\gamma^2\mcomma \qquad \gamma_c=\frac{3}{2}\mperiod
\end{eq}
In these coordinates, it is transparent that each combination in \cref{eq:harmonic_XYW} corresponds to a different physical contribution: 
\begin{itemize}
    \item[-] $X$ accompanies the curvature of the orthogonal sphere.
    \item[-] $Y$ accompanies the contribution of form fluxes.
    \item[-] $W$ accompanies the contribution of the tadpole potential.
\end{itemize}

These expressions can be generalized to $D$-dimensions, which is the original approach of~\cite{Mourad:2024dur}, and I shall refer to that paper for the nine-dimensional results that will be needed in \cref{sec:branes}. Here, I begin by introducing some familiar solutions in the harmonic gauge and then move on to the relevant cases of non-supersymmetric strings.

\subsection{Tadpole-free solutions}

It is instructive to begin with solutions without tadpole potentials, both to gain familiarity with the unusual coordinates and because some of the results are relevant in \cref{sec:branes}.
Flat space in harmonic coordinates is
\begin{eq}\label{eq:flat_space}
    ds^2 = \eta_{\mu\nu}dx^\mu dx^\nu + \left[\frac{(7-p)r}{\sqrt{\alpha'}}\right]^{-\frac{2(8-p)}{7-p}}dr^2 + \left[\frac{(7-p)r}{\sqrt{\alpha'}}\right]^{-\frac{2}{7-p}}\alpha' d\Omega_{8-p}^2 \mperiod
\end{eq}
$r\to\infty$ is the origin of the spherical coordinates, and $r=0$ lies at an infinite proper distance from any other finite-$r$ surface.
Note how flat space solves the harmonic-gauge equations of motion: the sphere curvature is the only contribution that is present in \cref{eq:brane_isometry_eom}, and the equation of motion for $X$,
\begin{eq}\label{eq:X_eq}
    X''=\frac{(7-p)^2}{\alpha'}e^{2X}\mcomma
\end{eq}
is solved by flat space with a logarithmic profile in terms of $r$,
\begin{eq}
    X=-\log\frac{(7-p)r}{\sqrt{\alpha'}}\mperiod
\end{eq}
For flat space, the contributions from second-derivative terms, like $X''$, and the ones from the exponential $e^{2X}$ are equally relevant.

\subsubsection{Uncharged branes}
\label{sec:uncharged_branes}

Uncharged branes without tadpole potentials or fluxes are included in the background ansatz of \cref{eq:ansatz}. The condition that far from the singularity they must reduce to flat space implies that, in harmonic coordinates, the metric of \cref{eq:flat_space} must be present as a limiting behavior in $r$.
For uncharged branes, one can consider $A$, $\phi$ and $X$ as independent functions, which simplifies the analysis because the first two are simply linear functions of $r$, while $X$ is the most general solution of \cref{eq:X_eq} that is compatible with the Hamiltonian constraint:
\begin{eq}
    X=-\log\left[\frac{(7-p)\sigma}{\sqrt{\alpha'}}\sinh\left(\frac{r}{\sigma}\right)\right]\mcomma
\end{eq}
with positive $\sigma$. The reader can find all the details in~\cite{Mourad:2024dur}, here I only recall the final result,
\begin{eq}\label[pluralequation]{eq:uncharged_harmonic}
    ds^2 & = e^{-\frac{2r}{R}}\eta_{\mu\nu}dx^\mu dx^\nu + e^{\frac{2(p+1)r}{(7-p)R}}\left[\frac{(7-p)\sigma}{\sqrt{\alpha'}}\sinh\left(\frac{r}{\sigma}\right)\right]^{-\frac{2(8-p)}{7-p}}dr^2 \\
    & + e^{\frac{2(p+1)r}{(7-p)R}}\left[\frac{(7-p)\sigma}{\sqrt{\alpha'}}\sinh\left(\frac{r}{\sigma}\right)\right]^{-\frac{2}{7-p}}\alpha' d\Omega_{8-p}^2\mcomma \\
    \phi & = \phi_0 + \phi_1 r \mcomma
\end{eq}
where the three parameters $\sigma\geq0$, $R$, and $\phi_1$ are related by the Hamiltonian constraint as follows:
\begin{eq}
    \frac{8(p+1)}{7-p}\frac{1}{R^2}+\frac{1}{2}\phi_1^2 = \frac{8-p}{7-p}\frac{1}{\sigma^2}\mperiod
\end{eq}
This metric describes a source with tension
\begin{eq}
    {\cal T}_p=\frac{8}{7-p}\frac{\Omega_{8-p}}{\kappa_{10}^2}\frac{\left(\alpha'\right)^{\frac{8-p}{2}}}{R}\mcomma
\end{eq}
where $\Omega_{8-p}$ is the area of an $(8-p)$-sphere with unit radius and $\kappa_{10}$ is the Einstein-frame gravitational coupling constant in ten dimensions that can be extracted from \cref{eq:action_nosusy_einstein_frame}. This source is embedded in flat Minkowski with constant dilaton, which is present in the solution in the limit $r\to0$.

For later convenience, the isotropic coordinates correspond to the reparametrization of $r$ given by
\begin{eq}
    \frac{2(7-p)\sigma}{\sqrt{\alpha'}}\tanh\left(\frac{r}{2\sigma}\right) = \left(\frac{\sqrt{\alpha'}}{\rho}\right)^{7-p}\mperiod
\end{eq}
In terms of the isotropic radial coordinate $\rho $, the background reads
\begin{eq} \label[pluralequation]{eq:uncharged_branes_isotropic}
    ds^2 & = \left[\frac{1+v(\rho)}{1-v(\rho)}\right]^{-\frac{2\sigma}{R}}\eta_{\mu\nu}dx^\mu dx^\nu \\
    & +  \left[\frac{1+v(\rho)}{1-v(\rho)}\right]^{\frac{2(p+1)\sigma}{(7-p)R}}\left[1-v^2(\rho)\right]^{\frac{2}{7-p}}\left(d\rho^2 + \rho^2d\Omega_{8-p}^2\right)\mcomma \\
    \phi & = \phi_0 + \phi_1 \sigma \log\left[\frac{1+v(\rho)}{1-v(\rho)}\right]\mcomma
\end{eq}
where 
\begin{eq}
    v(\rho) = \frac{\sqrt{\alpha'}}{2(7-p)\sigma}\left(\frac{\sqrt{\alpha'}}{\rho}\right)^{7-p} \mperiod
\end{eq}

\subsubsection{Electrically charged branes}

Including a non-vanishing electric form field profile leads to (electrically) charged branes, for which the relevant terms in the equations without the tadpole potential are given by the $X$- and the $Y$-exponentials, which are the curvature and flux contributions. It is convenient to use, as a third combination,
\begin{eq}
    Z=2\beta_p A-\frac{7-p}{8}\phi\mcomma
\end{eq}
whose equation is
\begin{eq}
    Z''=0\mperiod
\end{eq}
The decoupling of $Z$, with a linear profile in the harmonic coordinates, enables one to solve the system of \cref{eq:brane_isometry_eom} analytically.
The most general solution, up to reflection and translation of the variable $r$, can be expressed in terms of five real parameters, $\phi_0$, $r_1$, $z_1$, ${\cal E}_x$ and ${\cal E}_y$, constrained by the condition
\begin{eq}
    z_1^2 = \frac{1}{4(p+1)}\left[ 2(8-p) {\cal E}_x - (7-p){\cal E}_y\right]\mcomma
\end{eq}
and reads
\begin{eq}\label{eq:deformation_background_general}
    ds^2 &= \left|\frac{{\cal F}\left({\cal E}_y,r+r_1\right)}{{\cal F}\left({\cal E}_y,r_1\right)}\right|^{-\frac{7-p}{8}}e^{\beta_p z_1 r} \eta_{\mu\nu}dx^\mu dx^\nu \\
        &+\left|\frac{{\cal F}\left({\cal E}_y,r+r_1\right)}{{\cal F}\left({\cal E}_y,r_1\right)}\right|^{\frac{p+1}{8}}e^{-\frac{(p+1)\beta_p}{(7-p)} z_1 r}\left[ \frac{dr^2}{\left|\frac{7-p}{\sqrt{\alpha'}}{\cal F}\left({\cal E}_x,r\right)\right|^{\frac{2(8-p)}{7-p}}}+\frac{\alpha' d\Omega_{8-p}^2}{\left|\frac{7-p}{\sqrt{\alpha'}}{\cal F}\left({\cal E}_x,r\right)\right|^{\frac{2}{7-p}}}\right]\mcomma \\
    e^\phi &=e^{\phi_0}e^{-\frac{p+1}{2}z_1 r}\left|\frac{{\cal F}\left({\cal E}_y,r+r_1\right)}{{\cal F}\left({\cal E}_y,r_1\right)}\right|^{-\beta_p}\mcomma \\
    F_{p+2}& = -\epsilon e^{\beta_p\phi_0}\left|{\cal F}\left({\cal E}_y,r_1\right)\right|^{-1}\left|\frac{{\cal F}\left({\cal E}_y,r+r_1\right)}{{\cal F}\left({\cal E}_y,r_1\right)}\right|^{-2} dx^0 \wedge \ldots\wedge dx^p\wedge dr \mperiod
\end{eq}
Here, $\epsilon$ is a sign that distinguishes branes from anti-branes and ${\cal F}({\cal E},r)$ is defined as
\begin{eq}
    {\cal F}\left({\cal E},r\right)=\begin{cases}
    \frac{1}{\sqrt{{\cal E}}}\sinh{\left(\sqrt{{\cal E}}  r\right)} \qquad & \text{if   } {\cal E}>0 \mcomma \\
    r & \text{if   } {\cal E}=0 \mcomma \label{cases_F} \\
    \frac{1}{\sqrt{|{\cal E}|}}\sin{\left(\sqrt{|{\cal E}|} r\right)} & \text{if   } {\cal E}<0 \mperiod
\end{cases}
\end{eq}
The range of $r$ is delimited by two zeros of ${\cal F}\left({\cal E}_y,r+r_1\right)$ or ${\cal F}\left({\cal E}_x,r\right)$ (note that the Hamiltonian constraint forbids the case with two zeros of ${\cal F}\left({\cal E}_x,r\right)$).
When the range contains a zero of ${\cal F}\left({\cal E}_x,r\right)$, the background has an asymptotically flat region, and it is possible to define the tension and charge of the brane,
\begin{eq}\label[pluralequation]{eq:charged_tension_charge}
    {\cal T}_p &= e^{-\beta_p \phi_0} \frac{\Omega_{8-p}}{2\kappa_{10}^2} \left(\alpha'\right)^{\frac{8-p}{2}}\left[\frac{{\cal F}'\left({\cal E}_y, r_1\right)}{{\cal F}\left({\cal E}_y, r_1\right)} - \frac{8}{7-p} \beta_p z_1\right] \mcomma  \\
    Q_p &= \epsilon  e^{-\beta_p \phi_0} \frac{\Omega_{8-p}}{2\kappa_{10}^2}\left(\alpha'\right)^{\frac{8-p}{2}}\left|{\cal F}\left({\cal E}_y, r_1\right)\right|^{-1} \ .
\end{eq}
When ${\cal E}_x\geq0$ and ${\cal E}_y\geq0$, the limit $r_1\to\infty$ recovers the uncharged brane solutions of \cref{eq:uncharged_harmonic}. 

A particular case of these backgrounds is given by the BPS branes of string theory, which correspond to ${\cal E}_x={\cal E}_y=z_1=0$, with tension and charge that are identical, up to a sign, and are determined by $r_1$. 
In~\cite{Mourad:2024dur}, the reader can find insightful discussions about the general backgrounds defined by \cref{eq:deformation_background_general}, together with the analysis of the special cases $\beta_p=0$ and $r_1=0$, which require separate treatment, and some clarifications on the role of self-duality.

\subsection{Uncharged branes and vacua with tadpole potentials}
\label{sec:tadpole_uncharged}

I now include the tadpole potential contribution in \cref{eq:brane_isometry_eom}, while turning off form fluxes. This setting can describe both vacuum solutions with brane-like isometries, ISO$(1,p)\times$SO$(9-p)$, and the gravitational backreaction of uncharged branes in the three non-supersymmetric string theories.

The relevant combinations of \cref{eq:harmonic_XYW} are then $X$ and $W$, and it is convenient to take 
\begin{eq}
    K=\phi+8\gamma A 
\end{eq}
as the third independent one.
The equation for $K$ is 
\begin{eq}
    K''=0\mcomma
\end{eq}
so that $K$ is a linear function in the harmonic $r$ coordinate, namely
\begin{eq}
    K=k_0+k_1 r\mperiod
\end{eq}
The remaining two equations become
\begin{eq}
    X''&=\frac{(7-p)^2}{\alpha'}e^{2X}-T e^{2W}\mcomma \\
    W''&=\frac{(8-p)(7-p)}{\alpha'}e^{2X}+\frac{\gamma^2-\gamma_c^2}{2} T e^{2W}\mcomma
\end{eq}
and the Hamiltonian constraint is reduced to
\begin{eq}
    0&=\frac{(8-p)(7-p)}{\alpha'}e^{2X}-T e^{2W}+\frac{1}{\Xi}\Bigg[ 8(7-p)(W')^2-16(8-p)X'W'\\
        &-4(8-p)\left(\gamma^2-\gamma_c^2\right)(X')^2+\frac{p+1}{2}k_1^2\Bigg] \mperiod
\end{eq}

In~\cite{Mourad:2024dur}, we found no analytic solutions for the relevant values of $\gamma$. In addition, we found that asymptotically $X$ and $W$ never differ by a constant. However, this result, combined with the convexity of $W-X$ and $W$,
\begin{eq}\label[pluralequation]{eq:convexity}
    (W-X)''=\frac{7-p}{\alpha'}e^{2X}+\frac{1}{2}\left(\gamma^2-\frac{1}{4}\right)T e^{2W}>0\mcomma \qquad W''>0\mcomma
\end{eq}
yields a complete classification of possible asymptotic behaviors in the $r$ direction, together with partial information on how asymptotics match.
Only a finite number of limiting $r$ behaviors are available, and they are classified by how $X$ or $W$ diverge, thus terminating the $r$ evolution of the system of differential equations. Up to shifts and reflections of $r$, it is sufficient to focus on the two limits $r\to0^+$ and $r\to +\infty$, for which the classification of asymptotics is as follows.

The first possibility is to have $e^X\gg e^W$ asymptotically, which can only happen as $r\to+\infty$, regardless of the value of $\gamma$, leading to two different limiting behaviors:
\begin{enumerate}
    \item[$\mathfrak{a}$. \ ] $r\to+\infty$ and 
    \begin{eq}
        X &\sim  -  \log \frac{r}{\sqrt{\alpha'}} \mcomma  \\
        W &\sim  -  \frac{8-p}{7-p}\log \frac{r}{\sqrt{\alpha'}} \mcomma   \\
        K &\sim 0 \mperiod
    \end{eq}
    This asymptotic solution describes a regular point at $r\to+\infty$, with a constant dilaton. This is not a proper asymptotic behavior, but it is still important in the construction of vacuum solutions, and I shall comment on this at the end of this section.
    \item[$\mathfrak{b}$. \ ] $r\to+\infty$ and 
    \begin{eq}
        X &\sim  -  \frac{r}{\rho} \mcomma   \\
        W &\sim  \left(w_1 - \frac{8-p}{\left(7-p\right)\rho}\right) r \mcomma  \\
        K &\sim k_1 r\mcomma
    \end{eq}
    where the Hamiltonian constraint links $\rho>0$, $w_1$, and $k_1$ by
    \begin{eq}
        \frac{1}{\rho^2} = \frac{ (7-p)(p+1)k_1^2 +  16 (7-p)^2 w_1^2}{2(8-p)\Xi}\mperiod
    \end{eq}
    In order to be in the regime $e^X\gg e^W$, the inequality 
    \begin{eq}
        (7-p)\rho w_1<1
    \end{eq}
    must hold. This asymptotic behavior is a Kasner-like solution that can be mapped to a near-singularity region of the uncharged branes of \cref{eq:uncharged_harmonic}. The reader can find the complete map in~\cite{Mourad:2024dur}.
\end{enumerate}

The other asymptotic possibility is $e^W\gg e^X$, where the limiting behaviors depend on the value of $\gamma$. Here, I consider only the two cases relevant to the ten-dimensional non-supersymmetric strings, while~\cite{Mourad:2024dur} contains the complete discussion for all values of $\gamma$. I begin with the $\gamma=\frac{5}{2}$ heterotic case, which results in two possible asymptotics:
\begin{enumerate}
    \item[$\mathfrak{c}$. \ ] $r\to0^+$ and 
    \begin{eq}
        X &\sim  \frac{1}{2}\log\frac{r}{\sqrt{\alpha'}} \mcomma  \\
        W &\sim   - \ \log\frac{r}{\sqrt{\alpha'}}\mcomma  \\
        K &\sim 0 \mcomma
    \end{eq}
    which is the only asymptotic at a finite value of $r$. In~\cite{Mourad:2024dur}, we show how this asymptotic behavior can be mapped to a solution that resembles the codimension-one vacuum of~\cite{Dudas:2000ff}, albeit with an internal sphere.
    \item[$\mathfrak{d}$. \ ] $r\to+\infty$ and 
    \begin{eq}
        X &\sim  \left(x_1 \ + \ \frac{1}{2\rho}\right) r \mcomma  \\
        W &\sim   -  \frac{r}{\rho}\mcomma  \\
        K &\sim k_1 r \mcomma  
    \end{eq}
    where $x_1$ is given in terms of $\rho>0$ and $k_1$ by 
    \begin{eq}
        x_1=- \left[\frac{22-3p}{4(8-p)\rho^2}+  \frac{(p+1) k_1^2}{32(8-p)}\right]^{\frac{1}{2}} \mcomma
    \end{eq}
    and where $e^W\gg e^X$ demands
    \begin{eq}
        \rho^2 k_1^2 > \frac{16(25-3p)}{p+1} \mperiod
    \end{eq}
    This is again a Kasner-like solution that can be mapped to near-singularity limits of the uncharged branes of \cref{eq:uncharged_harmonic}. This concludes the available asymptotics for the heterotic one-loop tadpole potential. 
    \end{enumerate}
For the tadpole potential of the two orientifold models, with $\gamma=\frac{3}{2}$, the condition $e^W\gg e^X$ leads to two types of asymptotics\footnote{I use the same notation as in~\cite{Mourad:2024dur}, therefore labeling the next two cases by $\mathfrak{f}$ and $\mathfrak{g}$ and skipping $\mathfrak{e}$. In~\cite{Mourad:2024dur}, asymptotics of type $\mathfrak{e}$ are the ones for which $\gamma<\gamma_c$.}:
\begin{enumerate}
    \item[$\mathfrak{f}$. \ ] $r\to+\infty$ and
    \begin{eq}
        X &\sim  - \frac{T}{4 w_1^2} e^{2 w_1 r + 2 w_0} + \left[\frac{(7-p) w_1}{2(8-p)}+  \frac{(p+1)k_1^2}{32 (8-p) w_1}\right] r \mcomma  \\
        W &\sim   w_1 r \mcomma  \\
        K &\sim k_1 r \mperiod
    \end{eq}
    The sign of $w_1$ distinguishes two different cases. If $w_1>0$, the exponential in the $X$ profile dominates, and the full background approaches a near-singularity limit of uncharged branes that resembles the Dudas-Mourad solutions.
    On the other hand, when $w_1<0$ the exponential in $X$ is subleading, and the condition $e^W\gg e^X$ holds provided
    \begin{eq}
        w_1^2 < \frac{p+1}{16(9-p)}k_1^2\mperiod
    \end{eq}
    This case is also related to near-singularity regions of the uncharged branes of \cref{eq:uncharged_harmonic}. 
    \item[$\mathfrak{g}$. \ ] $r\to+\infty$ and
    \begin{eq}
        X &\sim - \frac{(p+1) k_1^2}{32(8-p)} r^2 \mcomma \\
        W &\sim  0 \mcomma \\
        K &\sim 0 \mcomma
    \end{eq}
    which is the $w_1\to 0$ limit of case $\mathfrak{f}$. This can be again mapped to the near-singularity limit of uncharged branes and to a codimension-one background with an internal sphere.
    \end{enumerate}

This classification provides all possible ways in which a solution with the isometries of a brane can end. 
In this flux-free case, the convexity considerations of \cref{eq:convexity} and the global linear profile of $K$ ($k_1$ remains the same at the two ends of the $r$ evolution) allow a more detailed analysis with partial matching of the asymptotics. Six groups of asymptotics can be identified:
\begin{itemize}
    \item[-] For the heterotic case, with $\gamma=\frac{5}{2}$,
    \begin{eq}
    k_1 = 0 \ \text{ and } \ \left\{  \quad \text{\underline{$\mathfrak{a}$}} \quad ; \quad \text{\underline{$\mathfrak{b}$}} \quad ; \quad \text{$\mathfrak{c}$} \quad \right\} \nonumber
    \end{eq}
    \begin{eq}
    k_1\neq0 \ \text{ and } \ \left\{ \quad \text{\underline{$\mathfrak{b}$}} \quad ; \quad \text{$\mathfrak{d}$}  \quad \right\} \nonumber
    \end{eq}
    \item[-] For the orientifold cases, with $\gamma=\frac{3}{2}$,
    \begin{eq}
     k_1 = 0 \ \text{ and } \ \left\{  \quad \text{\underline{$\mathfrak{a}$}} \quad ; \quad \text{\underline{$\mathfrak{b}$}} \quad ; \quad \text{$\mathfrak{f}$ \ with \ }  w_1 >0 \quad  \right\} \nonumber
    \end{eq}
    \begin{eq}
    k_1\neq0 \ \text{ and } \ \left\{ \quad \text{\underline{$\mathfrak{b}$}} \quad ; \quad \text{$\mathfrak{d}$}  \quad ; \quad \text{$\mathfrak{f}$} \quad ; \quad \text{$\mathfrak{g}$}  \quad\right\} \nonumber
    \end{eq}
\end{itemize}
Within each group, the combinations that are not excluded are obtained by combining two elements under the condition that the underlined ones are taken at most once. The two chosen asymptotics then represent the two limiting behaviors of a complete solution.
 
I conclude this section with a few remarks on these asymptotics, anticipating the content of the next section.

Case $\mathfrak{a}$ is necessarily part of a vacuum solution, akin to the Dudas-Mourad one, for which the broken isometry is radial-translation invariance rather than the $y$-translation of the codimension-one scenarios.

Cases $\mathfrak{b}$, $\mathfrak{d}$, $\mathfrak{f}$ and $\mathfrak{g}$ are Kasner-like solutions that can be matched to asymptotics of tadpole-free uncharged branes from \cref{sec:uncharged_branes}. Indeed, case $\mathfrak{b}$ lies in the regime where the tadpole is subleading; therefore, one recovers a solution of the free theory. However, there is more because $X\to-\infty$, implying that the curvature enters as a small perturbation in the equations of motion, which then read
\begin{eq}
    X''\sim 0 \mcomma \qquad W''\sim 0 \mcomma \qquad K''\sim 0\mcomma
\end{eq}
with linear solutions. 
In case $\mathfrak{d}$, the curvature contribution is subdominant with respect to the tadpole. Yet, the only solution has $W\to-\infty$, and therefore both curvature and tadpole are negligible, and one recovers a Kasner-like asymptotic behavior. The same holds for case $\mathfrak{f}$ with $w_1<0$. The two remaining cases, $\mathfrak{f}$ with $w_1>0$ and $\mathfrak{g}$, are more subtle because, although $e^W\gg e^X$, the profiles are such that $\left|X\right|\gg W$. The physical components of the metric and dilaton are linear combinations of $X$ and $W$; therefore, only $X$ determines the leading form of the background.
This is analogous to what occurs for the codimension-one Dudas-Mourad vacua, where both endpoints of the orientifold solution and the weakly coupled endpoint of the heterotic solution are the tadpole-free codimension-one solution of \cref{eq:free_codim_1}.

Case $\mathfrak{c}$ is different and corresponds to a tadpole-dominated collapse, which is only possible when $\gamma>\gamma_c$. The corresponding background is given by
\begin{eq}
    ds^2&\sim r^{\frac{1}{8}}\left(\eta_{\mu\nu}dx^\mu dx^\nu +\alpha'd\Omega_{8-p}^2\right)+ r^{\frac{9}{8}}dr^2 \mcomma \\
    e^\phi&\sim r^{-\frac{5}{4}}\mcomma
\end{eq}
and is analogous to the strongly coupled end of the heterotic Dudas-Mourad solution, \cref{eq:DM_other_asympt}. This limiting behavior also appears in the toroidal compactifications of section 3.2.2 of~\cite{Mourad:2021roa}, and in the unphysical singularities of massive type IIA in the presence of D8 branes.

\subsection{Charged branes and vacua with tadpole potentials}
\label{sec:harmonic_charged_tadpole_branes}

The complete set of \cref{eq:brane_isometry_eom}, with non-vanishing flux and tadpole contributions, can describe both flux vacua with brane-like isometries and the gravitational backreaction of charged branes in the three non-supersymmetric string theories.

In this most general case, one can again classify the possible asymptotics, but global convexity conditions are not sufficiently simple to allow partial matching of them. The reader can find more details in~\cite{Mourad:2024dur}, while here I present only the resulting catalog of asymptotics.

\subsubsection{Orientifold asymptotics}

The effective actions of the Sugimoto model and of type 0'B have the Einstein-frame tadpole exponent $\gamma=\frac{3}{2}$, and in the following, I focus on the R-R form fields with $\beta_p=\frac{p-3}{4}$, where the values of $p$ that are compatible with the background ansatz are $p=1,3,5$. There are $8$ types of asymptotics that can terminate the $r$ evolution of the differential system of \cref{eq:brane_isometry_eom}:
\begin{enumerate}
    \item The three combinations $X$, $Y$, and $W$ tend to $-\infty$ such that the leading contributions in \cref{eq:brane_isometry_eom} become the second derivative terms. This corresponds to linear profiles as $r\to\infty$, and to Kasner-like behaviors. 
    \item The three combinations $X$, $Y$, and $W$ are equally important and differ by constants.\footnote{In the classification of asymptotics, constants are irrelevant because they are generically subleading with respect to the limiting $r$ behavior. In fact, the numerical coefficients in \cref{eq:brane_isometry_eom} are irrelevant for the classification, and only the signs of the coefficients matter. In this type of asymptotic only, $X$, $Y$, and $W$ differ by constants as global solutions.} This is only possible for $p=1$ and corresponds to the globally defined AdS$_3\times$S$^7$ solution of~\cite{Mourad:2016xbk}, which in these settings is a fine-tuned asymptotic.
    \item $Y\sim W\gg X$, which contains only linear profiles in $r$, and therefore only Kasner-like solutions, for $p<5$.
    \item $X\sim W\gg Y$, which yields no solutions.
    \item $X\sim Y\gg W$, whose physical interpretation is that the tadpole can be included as a small perturbation. This contains both linear profiles in $r$, and therefore Kasner-like solutions, and two asymptotics from charged branes: the nonphysical side of the near-horizon BPS D1 solution as $r\to0$, and the physical near-horizon of the BPS D5 solution as $r\to\infty$. The latter case is a special subcase of \cref{eq:brane_isometry_eom}, because for the orientifold tadpole exponent, $p=5$, and $\beta_5=\frac{1}{2}$, $Y$ decouples from \cref{eq:brane_isometry_eom} and one can employ the classification with partial matching of \cref{sec:tadpole_uncharged}, up to a sign issue discussed in~\cite{Mourad:2024dur}.
    \item $W\gg X,Y$, which leads to asymptotics of type $\mathfrak{f}$ and $\mathfrak{g}$ from the classification of \cref{sec:tadpole_uncharged}.
    \item $Y\gg X,W$, which leads to both linear profiles in $r$, and therefore Kasner-like solutions, and dipole-like asymptotics for all values of $p$ as $r\to 0$, with
    \begin{eq}
        X\sim 0 \mcomma \qquad Y\sim -  \log\left(\left|H_{p+2}\right| r\right)\mcomma \qquad W\sim \frac{5-p}{8} \log\left(\left|H_{p+2}\right| r\right) \mperiod
    \end{eq}
    \item $X\gg Y,W$, which contains both linear profiles in $r$, and therefore Kasner-like solutions, and the regular point.
\end{enumerate}

All the orientifold asymptotics can be mapped to asymptotics of the tadpole-free equations or mild modifications of them, except the AdS$_3\times$S$^7$ solution. This result resonates with the codimension-one Dudas-Mourad solution, whose asymptotics are not controlled by the tadpole potential.

\subsubsection{Heterotic asymptotics}

The effective action of the SO$(16)$$\times$SO$(16)$ string has the Einstein-frame tadpole exponent $\gamma=\frac{5}{2}$, and in the following I focus on the Kalb-Ramond field and its dual with $\beta_p=\frac{3-p}{4}$, where the allowed values of $p$ are $p=1,5$. There are $8$ types of asymptotics that can terminate the $r$ evolution of the differential system of \cref{eq:brane_isometry_eom}:
\begin{enumerate}
    \item The three combinations $X$, $Y$, and $W$ tend to $-\infty$ such that the leading contributions in \cref{eq:brane_isometry_eom} become the second derivative terms. This corresponds to linear profiles as $r\to\infty$, and to Kasner-like behaviors. 
    \item The three combinations $X$, $Y$, and $W$ are equally important and differ by constants. This is only possible for $p=5$ and corresponds to the globally defined AdS$_7\times$S$^3$ solution of~\cite{Mourad:2016xbk}, which is a fine-tuned asymptotic in these settings.
    \item $Y\sim W\gg X$, which yields two interesting asymptotics. The first is for $p=1$ and corresponds to the $r\to 0$ region of  
    \begin{eq} \label[pluralequation]{eq:heterotic_new_asymptotic}
        X\sim \frac{1}{3}\log r \mcomma \qquad Y\sim- \log r \mcomma \qquad W\sim -\log r \mcomma
    \end{eq}
    which is part of an exact solution with an internal Ricci-flat space rather than the orthogonal sphere. The second asymptotic is for $p=5$ and corresponds to a case in which $X$ is the dominant negative contribution:
    \begin{eq}\label{eq:heterotic_another_new_asymptotic}
        X\sim -\frac{e^{2 y_1 r}}{4 y_1^2} \mcomma
    \end{eq}
    as $r\to\infty$, with $y_1>0$. This is again part of an exact solution with a Ricci-flat space instead of the sphere and is similar to the orientifold asymptotic of type $\mathfrak{f}$ from \cref{sec:tadpole_uncharged}.
    \item $X\sim W\gg Y$, which yields no solutions.
    \item $X\sim Y\gg W$, whose physical interpretation is that the tadpole can be included as a small perturbation. This contains both linear profiles in $r$, and therefore Kasner-like solutions, and two asymptotics from charged branes: the nonphysical side of the near-horizon BPS NS5 solution as $r\to0$, and the physical near-horizon of the BPS NS1 (fundamental string) solution as $r\to\infty$.
    \item $W\gg X,Y$, which leads to linear Kasner-like behaviors and to asymptotics of type $\mathfrak{c}$ from the classification of \cref{sec:tadpole_uncharged} for all relevant values of $p$.
    \item $Y\gg X,W$, which leads to both linear profiles in $r$, and therefore Kasner-like solutions, and dipole-like asymptotics for all values of $p$ as $r\to 0$, with
    \begin{eq}
        X\sim 0 \mcomma \qquad Y\sim -  \log\left(\left|H_{p+2}\right| r\right)\mcomma \qquad W\sim \frac{3p-7}{8} \log\left(\left|H_{p+2}\right| r\right) \mperiod
    \end{eq}
    \item $X\gg Y,W$, which contains both linear profiles in $r$, and therefore Kasner-like solutions, and the regular point.
\end{enumerate}

The results are again analogous to the codimension-one Dudas-Mourad solution because, while most of the asymptotics are dominated by tadpole-free regimes, some collapses crucially rely on the tadpole contribution $W$. These are the AdS$_7\times$S$^3$ solution, case $\mathfrak{c}$ from \cref{sec:tadpole_uncharged} and the limiting behavior of \cref{eq:heterotic_new_asymptotic} for $p=1$.

\section{Branes in Dudas-Mourad}
\label{sec:branes}

In this last section, I address the problem of identifying the gravitational backreaction of branes in the three non-supersymmetric strings. In addition to the isometries of \cref{sec:brane_isometries}, the gravitational solutions of branes must satisfy another condition: they must interpolate between the core of the brane and a vacuum solution far from it. 
The results of \cref{sec:vacuum_solutions} on vacuum solutions in non-supersymmetric strings leave one option: branes in the Dudas-Mourad vacua, which is the focus of this section. 

The worldsheet definitions of the three string theories predict a certain spectrum of branes for each model. In the SO$(16)$$\times$SO$(16)$ case, there must be charged NS1 (fundamental strings) and NS5 branes, and possibly also uncharged branes along the lines of~\cite{Kaidi:2023tqo}.
In the two orientifolds, the analysis of~\cite{Dudas:2001wd} shows that in the Sugimoto model, there must be charged D1 and D5 branes, together with stable K-theory charged (and therefore uncharged, from the point of view of \cref{eq:action_nosusy}) D3 and D4 branes, and unstable uncharged branes in all other cases.
On the other hand, type 0'B has the same brane spectrum as type IIB string theory: charged D$(-1)$, D1, D3, D5, and D7 branes, together with unstable uncharged branes in all other cases. Among the charged ones, I focus on the D1, D3, and D5 branes because the D$(-1)$ spacetime instanton and the D7\footnote{D7 branes in type 0'B string theory are unstable because tachyons are present in the D9-D7 open string spectrum with the background D9 branes~\cite{Dudas:2000sn,Dudas:2001wd}. These are the only examples of unstable $\Z$-charged branes that I am aware of, and presumably, they decay to non-trivial gauge configurations on the background D9 branes, as in~\cite{Loaiza-Brito:2001yer}.} would not be captured by the metric ansatz under consideration. 

These brane spectra are obtained from tree-level worldsheet arguments and do not account for the presence of tadpoles. Therefore, a complementary motivation for this section is to test whether the tree-level brane spectra can be matched, from the effective action point of view, to spacetime brane-like profiles in the shifted vacuum.
Note, however, that nothing guarantees that gravity solutions exist. In fact, the typical argument for the existence of spacetime backgrounds for branes is that despite their non-perturbative nature, with tension (for D$p$ branes)
\begin{eq} 
{\cal T}\sim g_s^{-1}\mcomma 
\end{eq}
the asymptotic value of the dilaton $\phi_0$ also enters the definition of the gravitational constant:
\begin{eq}
    \frac{1}{\kappa_{10}^2}\sim \frac{1}{G_N}\sim  g_s^{-2}\mperiod
\end{eq}
Therefore, the gravitational field generated by a brane scales as 
\begin{eq}
G_N {\cal T}\sim g_s\mcomma
\end{eq}
and at weak string coupling, one expects to be able to probe the gravitational backreaction of D$p$ branes in an asymptotically flat vacuum.
The issue with non-supersymmetric setups is that the asymptotic value of the dilaton is not a well-defined quantity; thus, this argument is not generally applicable. However, locally, in weakly coupled and weakly curved regions, the above argument suggests that partial gravitational descriptions of branes should be available.

The main question then becomes what backgrounds should be expected, given the absence of asymptotically flat vacua. This is the topic to which I now turn.

\subsection{The ansatz}

String theory branes, in their worldsheet characterization, have a flat $(p+1)$-dimensional portion along the directions of the worldvolume. 
The first issue is then how to include the appropriate isometries inside the Dudas-Mourad vacua. 
The following results are the content of~\cite{Mourad:2024mpg}.

It is convenient to consider the vacuum solution in the conformal gauge, as in \cref{eq:DM_conformal_gauge},
\begin{eq}
    ds^2 = e^{\Omega(z)}\left(\eta_{\mu\nu}dx^\mu dx^\nu + dz^2\right)\mcomma \qquad \phi=\phi(z)\mcomma
\end{eq}
denoting the finite range of the conformal variable $z$ as $z\in(0,z_m)$.
Close to the singular endpoints, the two profiles $\Omega(z)$ and $\phi(z)$ can be written explicitly:
\begin{eq}
    \Omega&\sim\frac{1}{8}\log\left(\frac{z}{z_m}\right) \mcomma \qquad \Omega\sim\frac{1}{8}\log\left(1-\frac{z}{z_m}\right) \mcomma \\
    \phi&\sim\frac{3}{2} \log\left(\frac{z}{z_m}\right) \mcomma \qquad \phi\sim - \frac{3}{2}\log\left(1-\frac{z}{z_m}\right) 
\end{eq}
for the orientifolds and 
\begin{eq}
    \Omega&\sim\frac{1}{24}\log\left(\frac{z}{z_m}\right) \mcomma \qquad \Omega\sim\frac{1}{8}\log\left(1-\frac{z}{z_m}\right) \mcomma \\
    \phi&\sim-\frac{5}{6} \log\left(\frac{z}{z_m}\right) \mcomma \qquad \phi\sim \frac{3}{2}\log\left(1-\frac{z}{z_m}\right)
\end{eq}
for the heterotic case.
The flat worldvolume requirement, together with the condition of keeping $z$ as a special direction, lead to the following ansatz for branes in Dudas-Mourad vacua:
\begin{eq} \label[pluralequation]{eq:branes_in_DM}
    ds^2&=e^{2A(z,r)}\eta_{\mu\nu}dx^\mu dx^\nu+e^{2B(z,r)}\left(dr^2 + r^2 d\Omega_{7-p}^2\right) + e^{2D(z,r)}dz^2 \mcomma \\
    \phi&=\phi(z,r)\mcomma \qquad F_{p+2}=F_{p+2}(z,r)\mcomma \qquad F_{p+3}=F_{p+3}(z,r)\mcomma
\end{eq}
where $\mu=0,\ldots,p$ and
\begin{eq}
    A(z,r),B(z,r),D(z,r)\overset{r\to\infty}{\to}\Omega(z)\mcomma \qquad  \phi(z,r)\overset{r\to\infty}{\to}\phi(z)\mcomma 
\end{eq}
with vanishing form fields as $r\to\infty$.
In this background, the radial direction transverse to the brane can be $r$ or a combination of $r$ and $z$. In particular, if the brane is smeared along $z$, the radial distance is given by $r$.

Denoting the local frame by
\begin{eq}
    \theta^\mu =e^A dx^\mu \mcomma \qquad \theta^r=e^B dr\mcomma \qquad \theta^z=e^D dz\mcomma \qquad \theta^a=r e^B \tilde{\theta}^a\mcomma
\end{eq}
where $\tilde{\theta}^a$ refers to the unit sphere, the two types of field strengths that are compatible with the isometries are
\begin{eq}\label{eq:brane_p+2_form}
    F_{p+2}=H(z,r) \theta^r\wedge\theta^0\wedge\ldots\wedge\theta^p + \tilde{H}(z,r)\theta^z\wedge\theta^0\wedge\ldots\wedge\theta^p\mcomma
\end{eq}
and 
\begin{eq}\label{eq:brane_p+3_form}
    F_{p+3}=h(z,r)\theta^0\wedge\ldots\theta^p\wedge\theta^r\wedge\theta^z\mperiod
\end{eq}

\subsection{Equations and exact solutions}

The equations for the form fields are the simplest. Introducing the potential
\begin{eq}
    {\cal B}_{p+1}= e^{-(p+1)A} {\cal B}(z,r)\theta^0\wedge\ldots\wedge\theta^p\mcomma
\end{eq}
the $(p+2)$-field strength of \cref{eq:brane_p+2_form} can be obtained from
\begin{eq}
    H=e^{-(p+1)A-B}{\cal B}_r\mcomma \qquad \tilde{H}=e^{-(p+1)A-D}{\cal B}_z\mcomma
\end{eq}
where $[ \ ]_{\{z,r\}}$ denotes a $\{z,r\}$-derivative.
The Bianchi identities are automatically satisfied, and the equations of motion are reduced to
\begin{eq}
    \left(e^{2G}{\cal B}_z\right)_z = - \left(e^{2(G-B+D)}{\cal B}_r\right)_r\mcomma
\end{eq}
where
\begin{eq}
    2G = -2\beta_p\phi - (p+1)A+(8-p)B+(7-p)\log r - D\mperiod
\end{eq}

The $(p+3)$-form field strength of \cref{eq:brane_p+3_form} affords an even simpler treatment because the Bianchi identities are satisfied and the equations of motion are solved by
\begin{eq}\label{eq:h_eq}
    h=\frac{h_0}{r^{7-p}}e^{2\beta_{p+1}\phi - (7-p)B}\mcomma
\end{eq}
with a constant $h_0$.

The dilaton equation of motion becomes
\begin{eq}
    &e^{-2B}\left(\phi_{rr}+\phi_r F^+_r\right)+e^{-2D}\left(\phi_{zz}+\phi_z F^-_z\right)= T\gamma e^{\gamma\phi} \\
    &+\beta_p\left(e^{-2B}{\cal B}_r^2+e^{-2D}{\cal B}_z^2\right)e^{-2\beta_p\phi-2(p+1)A}+\frac{h_0^2}{r^{2(7-p)}}\beta_{p+1}e^{2\beta_{p+1}\phi-2(7-p)B}\mcomma
\end{eq}
where 
\begin{eq}
    F^{\pm}=(p+1)A\mp B +(7-p)\left(B+\log r\right)\pm D\mperiod
\end{eq}

The metric equations of motion are more involved, and I refer to~\cite{Mourad:2024mpg} for the explicit expressions.

An exact brane-like solution can be extracted from these equations without much effort. The key observation is from \cref{sec:DM}: any nine-dimensional Ricci-flat manifold can replace the flat Minkowski sector in \cref{eq:DM_conformal_gauge}.
In particular, tadpole-free nine-dimensional uncharged branes without dilaton, which are the generalization of those in \cref{sec:uncharged_branes}, are Ricci-flat solutions and lead to backgrounds of the type covered by \cref{eq:branes_in_DM}. Therefore, they build exact solutions for uncharged branes in the Dudas-Mourad vacua,
\begin{eq}\label[pluralequation]{eq:DM_uncharged_branes}
    ds^2 & = e^{2\Omega(z)}\left[\frac{1+v(r)}{1-v(r)}\right]^{-2 \zeta}\eta^{(p+1)}_{\mu\nu}dx^\mu dx^\nu \\
    & +  e^{2\Omega(z)}\left[\frac{1+v(r)}{1-v(r)}\right]^{ \frac{2(p+1)}{6-p}\zeta}\left[1-v^2(r)\right]^{\frac{2}{6-p}}\left(dr^2 + r^2d\Omega_{8-p}^2\right)+e^{2\Omega(z)}dz^2\mcomma\\
    \phi&=\phi(z)\mcomma
\end{eq}
where $\Omega$ and $\phi$ are the Dudas-Mourad profiles and 
\begin{eq}
    \zeta = \pm \sqrt{\frac{7-p}{7(p+1)}} \mcomma \qquad v(r)=\left(\frac{r_0}{r}\right)^{6-p}\mperiod
\end{eq}
$r_0>0$ is a constant that determines the nine-dimensional tension of the source, 
\begin{eq}
    {\cal T}_p=\pm 14 \sqrt{\frac{7-p}{7(p+1)}} \frac{\Omega_{7-p}}{\kappa_9^2} r_0^{6-p}\mcomma
\end{eq}
where the two sign choices for $\zeta$ in \cref{eq:DM_uncharged_branes} correspond to the two signs in the tension and $\kappa_9$ is defined in \cref{eq:DM_kappa9}.
This solution describes ten-dimensional $(p+1)$-branes, which are smeared along the $z$ direction and become $p$-branes in the nine-dimensional compactification, in the same way as black string solutions become black holes after dimensional reduction.

Note that the same considerations with the time-dependent codimension-one vacua would yield Euclidean brane solutions, and this is part of the reason why here I focus on the spacelike vacuum. The case of branes in time-dependent backgrounds, or more generally in cosmological spacetimes, is an important problem that deserves further investigation.

\subsection{Linearized system}
\label{sec:linearized_branes}

In general, in the absence of other exact solutions, the strategy is to capture the backreaction of branes at large distances. To this end, one must linearize the equations of motion around the Dudas-Mourad background, letting
\begin{eq}
    A&=\Omega(z)+a(z,r)\mcomma \qquad B=\Omega(z)+b(z,r)\mcomma \qquad D=\Omega(z)+d(z,r)\mcomma \\
    \phi&=\phi(z)+\varphi(z,r)\mcomma \qquad {\cal B}={\cal B}(z,r)\mcomma
\end{eq}
with vanishing linear perturbations $a$, $b$, $d$, $\varphi$ and ${\cal B}$ as $r\to\infty$.

The form field equation decouples at the linear level, allowing one to study the dilaton-graviton system independently and then turn to the equations for ${\cal B}$. Note that this does not guarantee consistency at the level of higher-order perturbations. In other words, implications for full non-linear solutions are limited. 

\subsubsection{Dilaton-graviton sector}

The reader can find the linearized equations in~\cite{Mourad:2024mpg}. Here, I only present the results.
The $r z$ Einstein equation,
\begin{eq}
    -2\left[(p+1)a+(7-p)b\right]_{rz}+16\Omega_z d_r=\phi_z \varphi_r\mcomma
\end{eq}
can be integrated in $r$, taking into account that all perturbations must vanish as $r\to\infty$, and it determines $\varphi$ in terms of the other fields:
\begin{eq}
    \varphi=-\frac{2}{\phi_z}\left[(p+1)a_z+(7-p)b_z+\Omega_z d\right]\mperiod
\end{eq}
Analogously, combining the $\mu\nu$ and the $rr$ metric equations yields
\begin{eq}
    d=-(p+1)a-(6-p)b\mperiod
\end{eq}
Then, letting
\begin{eq}
    \rho=b-a\mcomma \qquad \chi=(p+1)a+(7-p)b\mcomma
\end{eq}
all metric equations are reduced to 
\begin{eq} \label[pluralequation]{eq:rho_chi_eq}
    &\rho_{rr}+\rho_{zz}+\frac{7-p}{r}\rho_r+8\Omega_z\rho_z=0\mcomma \\
    &\chi_{rr}+\chi_{zz}+\frac{7-p}{r}\chi_r +\frac{2}{\phi_z}\left(12\Omega_z\phi_z-\gamma T e^{2\Omega+\gamma\phi}\right)\chi_z=\\
    &\left(1+\frac{8\gamma\Omega_z}{\phi_z}\right)T e^{2\Omega+\gamma\phi}\left( \frac{7}{4}\chi-\frac{p+1}{4}\rho\right) \mperiod
\end{eq}
One can now approach \cref{eq:rho_chi_eq} by solving the first one for $\rho$, and then considering $\rho$ as a source for the second equation, which can be solved for $\chi$.

Consider the $\rho$ equation. Separating variables,
\begin{eq}
   \rho(z,r)=\rho_1(r)\rho_2(z) \mcomma
\end{eq}
the first of \cref{eq:rho_chi_eq} is equivalent to
\begin{eq}\label[pluralequation]{eq:rho12eq}
    (\rho_1)_{rr}+\frac{7-p}{r}(\rho_1)_r=m^2 \rho_1\mcomma \qquad (\rho_2)_{zz}+8\Omega_z(\rho_2)_z = - m^2 \rho_2\mcomma
\end{eq}
where $m^2$ is a real constant (possibly negative). Letting
\begin{eq}
    \rho_2=\psi e^{-4\Omega}\mcomma
\end{eq}
the second of \cref{eq:rho12eq} for $\rho_2$ takes the form of a Schr\"odinger equation,
\begin{eq}\label{eq:schrodinger_rho}
   \left(\frac{d}{dz}+4\Omega_z\right)\left(-\frac{d}{dz}+4\Omega_z\right)\psi = m^2\psi\mcomma
\end{eq}
which determines the allowed values of $m^2$. 
\begin{itemize}
    \item If $m^2=0$, then
    \begin{eq}
        \rho_1=\frac{\rho_0}{r^{6-p}}\mperiod
    \end{eq}
    \item If $m^2<0$, the mode is unstable and grows with time, which is not suitable for asymptotic behaviors of branes.
    \item If $m^2>0$, solutions that decay as $r\to\infty$ contain modified Bessel functions,
    \begin{eq}
        \rho_1=\rho_0 r^{-\frac{6-p}{2}}K_{\frac{6-p}{2}}(mr)\mperiod
    \end{eq}
\end{itemize}
\Cref{eq:schrodinger_rho} suggests that $m^2$ be positive or equal to zero. However, there is a potential subtlety with boundary conditions, as explained in~\cite{Mourad:2023wjg,Mourad:2023ppi}, because the potential of the Schr\"odinger equation is singular at $z=0$ and $z=z_m$.
Following~\cite{Mourad:2023wjg,Mourad:2023ppi}, singular potentials with double poles at $z=0$ and $z=z_m$,
\begin{eq}\label{eq:schrodinger_potential}
    V(z)\sim \frac{\mu^2 -\frac{1}{4}}{z^2}\mcomma \qquad V(z)\sim \frac{\tilde{\mu}^2 -\frac{1}{4}}{(z_m-z)^2}\mcomma 
\end{eq}
can be approximated using hypergeometric functions near the poles. The corresponding wavefunctions then behave as 
\begin{eq}\label[pluralequation]{eq:muneq0}
    \psi&\sim C_1\left(\frac{z}{z_m}\right)^{\frac{1}{2}+\mu}+C_2\left(\frac{z}{z_m}\right)^{\frac{1}{2}-\mu}\mcomma \\
    \psi&\sim C_3 \left(1-\frac{z}{z_m}\right)^{\frac{1}{2}+\tilde{\mu}}+ C_4\left(1-\frac{z}{z_m}\right)^{\frac{1}{2}-\tilde{\mu}}\mcomma
\end{eq}
when $\mu$ and $\tilde{\mu}$ are nonzero, and as 
\begin{eq}\label[pluralequation]{eq:mu=0}
    \psi&\sim C_1\left(\frac{z}{z_m}\right)^{\frac{1}{2}}+C_2\left(\frac{z}{z_m}\right)^{\frac{1}{2}}\log\left(\frac{z}{z_m}\right)\mcomma \\
    \psi&\sim C_3 \left(1-\frac{z}{z_m}\right)^{\frac{1}{2}}+ C_4\left(1-\frac{z}{z_m}\right)^{\frac{1}{2}}\log\left(1-\frac{z}{z_m}\right)\mcomma
\end{eq}
when $\mu=0$ and $\tilde{\mu}=0$.
If $\mu>1$ (or $\tilde{\mu}>1$), then $C_2$ (or $C_4$) must vanish for the modes to be normalizable, while if $0<(\mu,\tilde{\mu})<1$, the ratios
\begin{eq}
    \frac{C_2}{C_1} \quad \text{and} \quad \frac{C_4}{C_3}
\end{eq}
characterize the available self-adjoint boundary conditions. In general, it is not always true that a formally positive Hamiltonian has a positive spectrum.

Back to the $\rho_2$ equation in the Schr\"odinger form, \cref{eq:schrodinger_rho}, the singular potential for the two orientifold models is characterized by
\begin{eq}
    \mu=0 \mcomma \qquad \tilde{\mu}=0\mperiod
\end{eq}
Stable modes correspond to limiting behaviors as in \cref{eq:mu=0} with $C_2=C_4=0$, so that the logarithmic asymptotics are absent.
A normalizable zero mode is present, corresponding to 
\begin{eq}
    \psi =\psi_0 e^{4\Omega}\mcomma
\end{eq}
or equivalently to a constant $\rho_2$, and the complete zero-mode solution is then
\begin{eq} \label{eq:orientifold_rho_zeromode}
    \rho=\frac{\rho_0}{r^{6-p}}\mperiod
\end{eq}
On the other hand, massive solutions decay exponentially with the distance, as
\begin{eq}
    \rho \sim r^{-\frac{7-p}{2}}e^{-m r} e^{-4\Omega(z)}\psi_m(z)\mcomma
\end{eq}
with the eigenfunction $\psi_m$ still compatible with the above limiting behaviors.
At large distances, only the zero-mode solution in \cref{eq:orientifold_rho_zeromode} remains as the leading perturbation.

For the heterotic case, the singular potential is characterized by
\begin{eq}
    \mu=\frac{1}{3}\mcomma \qquad \tilde{\mu}=0\mcomma
\end{eq}
and the asymptotic behaviors near the endpoints of the Dudas-Mourad interval are as in the second of \cref{eq:muneq0} and the first of \cref{eq:mu=0}. In terms of $\rho_2$, one has
\begin{eq}\label[pluralequation]{eq:heterotic_rho_2_asymp}
    \rho_2&\sim C_1 \left(\frac{z}{z_m}\right)^{\frac{2}{3}}+C_2\mcomma \\
    \rho_2&\sim C_3+C_4\log\left(1-\frac{z}{z_m}\right)\mperiod
\end{eq}
In this case, a constant $\rho_2$ is not the only zero mode, because the Wronskian method yields another normalizable zero mode starting from it. The complete zero mode is then 
\begin{eq}
    \rho_2=A_1+A_2\int_{z_0}^z dz' e^{-8\Omega(z')}\mperiod
\end{eq}
As the reader can verify, close to the singular points this behaves as in \cref{eq:heterotic_rho_2_asymp}.
Finally, the complete zero-mode solution, which is the leading perturbation at large distances, is
\begin{eq}\label{eq:heterotic_zero_modes_rho}
    \rho=\frac{A_1}{r^{6-p}}+\frac{A_2}{r^{6-p}}\int_{z_0}^z dz' e^{-8\Omega(z')}\mperiod
\end{eq}

After having the $\rho$ perturbation, the second of \cref{eq:rho_chi_eq} can be solved for $\chi$ by adding one of its special solutions, $\chi_s(z,r)$, to the most general solution of the associated homogeneous differential equation, $\chi_0(z,r)$:
\begin{eq}
    \chi(z,r)=\chi_0(z,r) +\chi_s(z,r)\mperiod
\end{eq}

The special solution must have the same $r$-dependence as $\rho$, so that one can consider the separated form
\begin{eq}
    \chi_s(z,r)=\rho_1(r)\chi_{s,2}(z)\mperiod
\end{eq}
The equation for $\chi_{s,2}$ is similar to that for $\chi$ in \cref{eq:rho_chi_eq}, the only difference being that the source is $\rho_2$ instead of $\rho$.
The only explicit solution corresponds to the constant zero mode of $\rho_2$, allowed in all cases of interest, for which $\chi_s$ reads 
\begin{eq}\label{eq:chi_special_solution}
    \chi_s=\frac{p+1}{7}\frac{\rho_0}{r^{6-p}}\mperiod
\end{eq}
This special solution captures the large-distance behavior of the exact solution of \cref{eq:DM_uncharged_branes}, with 
\begin{eq}
    \rho_0=\pm\frac{2}{6-p}\sqrt{\frac{7(7-p)}{p+1}}r_0^{(6-p)}\mperiod
\end{eq}
In the heterotic case, there is the additional $z$-dependent zero mode, which gives another implicit $z$-dependent special solution. Similar considerations apply to the massive cases with $z$-dependence and exponential decay in $r$.

The homogeneous equation for $\chi_0$ becomes two ordinary differential equations after separating variables,
\begin{eq}
    \chi_0=\chi_1(r)\chi_2(z)\mperiod
\end{eq}
The equation for $\chi_1$ is analogous to the first of \cref{eq:rho12eq}, 
\begin{eq}
    (\chi_1)_{rr}+\frac{7-p}{r}(\chi_1)_r=M^2\chi_1\mcomma
\end{eq}
whose solution is a power for $M^2=0$ and a Bessel function for $M^2\neq0$. The equation for $\chi_2$, which determines the allowed values of $M^2$, can be turned into the Schr\"odinger form, letting
\begin{eq}
    \chi_2=\phi_z e^{-4\Omega }\Psi\mperiod
\end{eq}
Explicitly, it becomes
\begin{eq}
    \left[\left(\frac{d}{dz}+\alpha\right)\left(-\frac{d}{dz}+\alpha\right)+b_0\right]\Psi=M^2 \Psi\mcomma
\end{eq}
where 
\begin{eq}
    \alpha=4\Omega_z-\frac{\phi_{zz}}{\phi_z}\mcomma \qquad b_0=\frac{7}{4}\left(1+\frac{8\gamma\Omega_z}{\phi_z}\right)T e^{2\Omega+\gamma\phi}\mperiod
\end{eq}
The Hamiltonian is formally positive definite because for all ten-dimensional non-supersymmetric strings $b_0> 0$, and a careful analysis of the boundary conditions, as in the previous cases, leads to the conclusion that $M^2>0$. 
Therefore, the solution of the homogeneous equation is exponentially suppressed in $r$, and only the special solution remains at large distances.

To summarize, at large distances, the leading perturbations are given by the zero modes, as expected by analogy with KK reductions. For $\rho$, this corresponds to \cref{eq:orientifold_rho_zeromode} in both cases, together with an additional $z$-dependent mode for the heterotic string, \cref{eq:heterotic_zero_modes_rho}.
For $\chi$, only the special solution remains, corresponding to \cref{eq:chi_special_solution} in both cases, together with another implicit $z$-dependent solution for the heterotic string.
For the $z$-independent zero mode, the results translate into
\begin{eq}
    a=-\frac{6-p}{7}\frac{\rho_0}{r^{6-p}}\mcomma \qquad b=\frac{p+1}{7}\frac{\rho_0}{r^{6-p}}\mcomma \qquad d=0\mcomma \qquad \varphi=0\mperiod
\end{eq}

\subsubsection{Form fields}

The linearized equation for the potential ${\cal B}$ is
\begin{eq}
    {\cal B}_{zz}+{\cal B}_{rr}+\frac{7-p}{r}{\cal B}_r-\left[2\beta_p \phi + 2(p-3)\Omega\right]_z{\cal B}_z=0\mperiod
\end{eq}
Following the previous cases, one separates variables
\begin{eq}
    {\cal B}(z,r)={\cal B}_1(r){\cal B}_2(z)\mcomma
\end{eq}
obtaining two ordinary differential equations:
\begin{eq}\label[pluralequation]{eq:B_two_eq}
    {{\cal B}_1}_{rr}&+\frac{7-p}{r}{{\cal B}_1}_r=\mathfrak{m}^2 {{\cal B}_1}\mcomma \\
    {{\cal B}_2}_{zz}&-\left[2\beta_p\phi_z+2(p-3)\Omega_z\right]{{\cal B}_2}_z=- \mathfrak{m}^2 {{\cal B}_2}\mperiod
\end{eq}
Letting
\begin{eq}
    {{\cal B}_2}=\tilde{\psi} e^{\beta_p \phi + (p-3)\Omega}\mcomma
\end{eq}
the second of \cref{eq:B_two_eq} takes the Schr\"odinger form
\begin{eq}
    \left[-\frac{d}{dz}+\left(\beta_p\phi_z+(p-3)\Omega_z\right)\right]\left[\frac{d}{dz}+\left(\beta_p\phi_z+(p-3)\Omega_z\right)\right]\tilde{\psi}=\mathfrak{m}^2 \tilde{\psi}\mperiod
\end{eq}

For the two non-supersymmetric orientifolds, the potential takes the same form as in \cref{eq:schrodinger_potential} with
\begin{eq}
    \mu=\frac{|p-2|}{2}\mcomma \qquad\tilde{\mu}=\frac{|p-5|}{4}\mcomma
\end{eq}
and, referring to \cref{eq:muneq0} and \cref{eq:mu=0}, normalizability sets some of the $C_i$ to zero. For example, $C_4$ must vanish when $p=1$ and $C_2$ must vanish when $p=5$. For $p=3$, which is relevant for type 0'B string theory, all the $C_i$'s are allowed.
The normalizable zero modes with $\mathfrak{m}^2=0$ are
\begin{eq}\label{eq:B_Apsi=0}
    \tilde{\psi}=e^{-\beta_p\phi-(p-3)\Omega}
\end{eq}
for $p=3$, while for $p=1$ and $p=5$ \cref{eq:B_Apsi=0} is not normalizable, but one can still find a normalizable zero mode using the Wronskian method:
\begin{eq}
    p=1 \ &: \ \tilde{\Psi}(z)=\tilde{\psi}(z)\int_{z_m}^z \frac{dz'}{|\tilde{\psi}(z')|^2}\mcomma \\
    p=5 \ &: \ \tilde{\Psi}(z)=\tilde{\psi}(z)\int_{0}^z \frac{dz'}{|\tilde{\psi}(z')|^2}\mcomma
\end{eq}
where $\tilde{\psi}$ is defined in \cref{eq:B_Apsi=0} for the two values of $p$.
Taking into account that zero modes exhibit power-like $r^{p-6}$ behavior for ${\cal B}_1$, the field strengths of the three relevant cases become
\begin{eq}
    F_3 & = -\frac{Q_1}{r^5}\left[\frac{5}{r}\int_{z}^{z_m}dz' e^{-4\Omega(z')-\phi(z')}dr+ e^{-4\Omega(z)-\phi(z)} dz\right]\wedge dx^0 \wedge dx^1\mcomma \\
    F_7 & = \frac{Q_5}{r}\left[-\frac{1}{r}\int_0^z dz'e^{4\Omega(z')+\phi(z')}dr+ e^{4\Omega(z)+\phi(z)}dz\right]\wedge dx^0 \wedge \ldots \wedge dx^5 \mcomma
\end{eq}
for $p=1$ and $p=5$ in both orientifold models, and 
\begin{eq}
    F_5=\frac{Q}{\sqrt{2}}\left[\frac{1}{r^4}dx^0\wedge \ldots \wedge dx^3 \wedge dr+ dz\wedge {\text{vol}}_{S^4}\right]\mcomma
\end{eq}
for the $p=3$ self-dual case of type 0'B.
Note that only $p=3$ leads to a zero mode that is independent of $z$. This is consistent with the analysis of \cref{sec:spont_cpt}, of which this section provides the extension with $z$-dependence.

In the heterotic case, the limiting behavior of the Schr\"odinger potential is characterized by
\begin{eq}
    \mu=\frac{|p-1|}{4}\mcomma \qquad \tilde{\mu}=\frac{|p-5|}{4}\mcomma
\end{eq}
where $p=1$ and $p=5$ are the two relevant cases. For both $p=1$ and $p=5$, the normalizable zero mode can be extracted using the Wronskian method, and the final results are
\begin{eq}
    F_{3}&=-\frac{Q_1}{r^5}\left[\frac{5}{r}\int_z^{z_m}dz' e^{-4\Omega(z')+\phi(z')}dr+ e^{-4\Omega(z)+\phi(z)}dz\right]\wedge dx^0\wedge dx^1\mcomma \\
    F_7 &=\frac{Q_5}{r}\left[-\frac{1}{r}\int_0^z dz' e^{4\Omega(z')-\phi(z')}dr + e^{4\Omega(z)-\phi(z)} dz\right] \wedge dx^0\wedge\ldots\wedge dx^5 \mcomma
\end{eq}
which are consistent with the analysis of \cref{sec:spont_cpt}.

The other type of form field, $F_{p+3}$ of \cref{eq:brane_p+3_form}, is automatically solved by \cref{eq:h_eq}, so that no further linearized analysis is required.

\subsection{On the two approaches}

The results of this section show that it is possible to find linear perturbations of the Dudas-Mourad background that indicate the presence of both uncharged and charged brane solutions.
Nothing guarantees that the linear results can be completed to full brane-like solutions, but in~\cite{Mourad:2024mpg} we checked that the linear modes can be treated as small deformations of the background solution, albeit with subtleties regarding how to take the limits $z\to 0$ and $z\to z_m$. Even more importantly, there is no rationale for identifying the stable charged and uncharged branes of the three ten-dimensional non-supersymmetric strings, which would replace the BPS property of supersymmetric setups. 

These considerations lead to an interesting possibility: since there is no restriction on the global form of the background---it need not preserve any supercharge, for instance---the spacetime profiles of branes could mix the two approaches, that of \cref{sec:brane_isometries} with the usual brane isometries, ISO$(p,1)\times$SO$(9-p)$, and that of \cref{sec:branes} with the condition that branes be embedded in a vacuum solution.
A brane could be captured by the profiles of \cref{sec:brane_isometries} close to its core, while reducing to the linearized perturbations of \cref{sec:branes} far from it.

Alternatively, one of the two cases may apply depending on the hierarchy of lengths involved: on the one hand, the length of the Dudas-Mourad interval $l_{DM}$, and on the other hand, the typical length of the brane horizon, which is not captured by the two-derivative action and is therefore less clearly identified.
This is reminiscent of the usual arguments about whether the UV cutoff of an effective field theory allows the inclusion of a given object in the effective description. The tension of the brane would determine whether the solution detects the internal interval direction or not.

%% file: chapters/conclusion.tex

\chapter{Conclusion}
\label{conclusion}

The emergence of tadpoles is the string theory version of the vacuum energy (cosmological constant) problem in quantum field theory; however, while the latter raises issues about naturalness and backreaction, the former drags physics off-shell.
It is fascinating to see how the issue changes dramatically when moving to the more adequate language of UV-finite string theory.

As I have stressed in this thesis, the potential of \cref{eq:tadpole_potential_generic} signals the instability of a flat Minkowski background, opening up a runaway direction in parameter space. In regimes in which the analysis can be trusted, there is no chance to stabilize the solution without including additional ingredients, thus leading to the profiles of \cref{sec:vacuum_solutions}: spaces with singular boundaries or singular cosmologies.
These have the vexing problem of containing regions outside the regime of validity of the quantum effective action, and our current ignorance about singular boundaries obstructs a comprehensive understanding of these vacua.
The absence of procedures that can distinguish physical boundaries (those that become smooth in the complete theory) from non-physical ones is a crucial problem that requires further investigation, not least because boundaries can evade several important no-go theorems~\cite{Gibbons:1984kp,Maldacena:2000mw,Gautason:2015tig,Basile:2020mpt} and sum rules~\cite{Gibbons:2000tf,Leblond:2001xr}.
At the same time, the natural emergence of a time direction with potential roles in string-derived cosmology, while having similar problems with singularities, may provide a phenomenologically interesting picture of string-scale supersymmetry breaking.

Alternative approaches to obtain vacua, with more complicated manifolds or including additional ingredients, as in \cref{chapter:landscape}, are all hampered by the lack of efficient strategies to engineer solutions without spacetime supersymmetry. This is perhaps the most important spacetime issue that I can identify.

The worldsheet formulation is equally not exempt from problems. Although the discussion of \cref{sec:FS_mechanism} captures the leading-order physics, the current version of the Fischler-Susskind mechanism is embarrassingly primitive.
With no clear scheme for extracting higher-order tadpoles from vacuum amplitudes and no control over scattering amplitudes computed in the shifted vacuum, one is left with a partial grasp of quantum gravity without supersymmetry.
This is a clear drawback of the geometric on-shell definition of string theory. After all, the Fischler-Susskind mechanism is based on an on-shell formulation and attempts to control an off-shell scenario: this is already beyond expectations.

Two related questions are of utmost importance. First, it is not clear whether the full exponentiated tadpole potential gives the most relevant contribution to the scalar potential. In fact, in the absence of a protecting principle, integration over massive string modes may introduce other effective self-interactions for the dilaton that are comparable to \cref{eq:tadpole_potential_generic}, in a way that resembles open-string tachyon condensation. 
The second question concerns the endpoint of the Fischler-Susskind procedure.
If the shifted vacuum has non-vanishing massless tadpoles, the tadpole subtraction procedure should be repeated to avoid divergences. What seems to follow from this is that either the true vacuum has no massless scalars (which is good news if string theory ought to describe our nature) or it has massless scalars, but all tadpoles vanish.
The latter option is not simple to engineer without spacetime supersymmetry, and this resonates with~\cite{Sen:2015uoa}, where the endpoints of the flow were found to restore supersymmetry. Note the analogy with the spacetime arguments of~\cite{Ooguri:2016pdq}, arguing that stable vacua require spacetime supersymmetry.

Alternatively, the endpoint of the Fischler-Susskind mechanism may not be a perturbative string theory. Then, nothing guarantees that in the shifted vacuum gravity remains dynamical, although I would be surprised by such a possibility. The ultimate reason why string theory is so relevant to physics---the natural inclusion of gravity---may not hold without the protection of spacetime supersymmetry. 
In view of this, the honest conclusion is that we do not quite understand why apples fall.
